\documentclass[twocolumn,aps,prr,groupeaddress,showpacs,longbibliography]{revtex4-2}

\usepackage[english]{babel}
\usepackage{amsmath,amssymb,amsfonts,float,graphics,epsfig,epstopdf,color,verbatim,tabularx,bm,multirow,hyperref}
\usepackage{amsmath}
\usepackage{amssymb}
\usepackage{mathtools}
\usepackage{graphicx}
\usepackage{bm}
\usepackage{hyperref}
\usepackage{url}
\usepackage[utf8]{inputenc}
\usepackage{subfigure}
\usepackage{slashed,bbm}
\usepackage{graphics,psfrag,epsfig}
\usepackage{dsfont}
\usepackage{setspace}
\usepackage{wasysym}
\usepackage{slashed}
\usepackage{lipsum}
\usepackage{braket}
\usepackage{physics}
\usepackage{minitoc}

\usepackage{hyperref}
\usepackage{xcolor}
\definecolor{dark-red}{rgb}{0.4,0.15,0.15}
\definecolor{dark-blue}{rgb}{0.15,0.15,0.4}
\definecolor{medium-blue}{rgb}{0,0,0.5}
\hypersetup{colorlinks, linkcolor={blue},citecolor={blue},urlcolor={blue}}
\usepackage{ulem}

\newcommand{\be}{\begin{equation}}
\newcommand{\ee}{\end{equation}}
\newcommand{\bea}{\begin{eqnarray}}
\newcommand{\eea}{\end{eqnarray}}

\newcommand{\I}{i}
\newcommand{\e}{e}

\begin{document}

\title{Critical phase induced by Berry phase and dissipation in a spin chain}

\author{Simon Martin}
\author{Tarun Grover}
\address{Department of Physics, University of California at San Diego, La Jolla, California 92093, USA}

\begin{abstract}

Motivated by experiments on spin chains embedded in a metallic bath, as well as closed quantum systems described by long-range interacting Hamiltonians, we study a critical SU$(N)$ spin chain perturbed by dissipation, or equivalently, after space-time rotation, long-range spatial interactions. The interplay of dissipation and the Wess-Zumino (Berry phase) term results in a rich phase diagram with multiple renormalization-group fixed points. For a range of the exponent that characterizes the dissipative bath, we find  a second-order phase transition between the fixed point that describes an isolated critical spin chain and a dissipation-induced-ordered phase. More interestingly, for a different range of the exponent, we find a stable, gapless, nonrelativistic phase of matter whose existence necessarily requires coupling to the dissipative bath. Upon tuning the exponent, we find that the fixed point corresponding to this gapless, stable phase ``annihilates'' the fixed point that describes the transition out of this phase to the ordered phase. We also study a relativistic version of our model, and we identify a new critical point. We discuss  the implications of our work for Kondo lattice systems and engineered long-range interacting quantum systems.
\end{abstract}

\maketitle

\section{Introduction}

Two recurring themes in many-body quantum physics, especially in the context of quantum phases and phase transitions, are Berry phase effects and long-range interactions induced by coupling to gapless modes. For example, Berry phase effects can lead to critical states in systems where one might naively expect a gap to excitations \cite{Haldane1983}, while coupling to gapless modes can effectively generate nonlocal interactions that can influence the nature of quantum criticality \cite{Hertz1976,Millis1993}, and can also help circumvent the Mermin-Wagner-Hohenberg theorem \cite{Hohenberg1967,Mermin1966} for systems with local interactions \cite{Chakravarty1982,Bray1982,Leggett1987,Castro1997,Werner2005a,Werner2005b,Laflorencie2005,Cazalilla2006,Lobos2012,Sperstad2012,Yan2018,Weber2022,Danu2022,Cuomo2023}. In this paper, we will revisit the problem of one-dimensional dissipative quantum systems, which, in the special case of dissipative Luttinger liquids, has been extensively studied in the past \cite{Castro1997,Cazalilla2006,Lobos2012,Weber2022,Danu2022}. One common feature of various setups for dissipative Luttinger liquids is the possibility of long-range order in one dimension and the associated order-disorder transition. Here we will show that in a class of one-dimensional systems with a non-Abelian symmetry, an interplay of Berry-phase effects and dissipation can result in a new possibility: a stable, dissipative phase with power-law correlations in both space and time, and which has no counterpart in a one-dimensional, nondissipative system with short-range interactions. We will also demonstrate the phenomena of fixed-point annihilation in this system which is reminiscent of that seen in a zero-dimensional quantum impurity coupled to a dissipative bath \cite{Cuomo2022,Nahum2022,Beccaria2022,Hu2022,Weber2023}.

It is well known that long-range interactions can lead to new critical points that are neither mean-field, nor related to critical points in short-ranged interacting systems \cite{Fisher1972,Sak1973,Sak1977,Bhattacharjee1982,Paulos2016,Behan2017a,Behan2017b,Defenu2017,Slade2018,Gubser2019,Defenu2020,Chakraborty2021,Chai2021,Chai2022}. Previous studies in this context have predominantly focused on ``classical models'', i.e., models whose Euclidean action is real. Here we will focus on models whose action contains a Berry phase term, and the resulting critical points do not necessarily have a classical statistical mechanics interpretation. From an experimental perspective, long-range interactions similar to the present work can arise in ``hybrid-dimensionality'' Kondo lattice systems such as Yb${}_2$Pt${}_2$Pb \cite{Wu2016,Classen2018,Gannon2019}, and  engineered Kondo lattice systems \cite{Toskovic2016,Choi2017,Moro2019,Choi2019,Danu2019}. In such systems, local moments effectively live in a lower dimension compared to the conduction electrons. In the limit of weak Kondo coupling, one may integrate out the conduction electrons resulting in  long-range interactions between the local moments along the time direction \cite{Hertz1976,Millis1993,Lobos2012,Weber2022,Danu2022}. Yb${}_2$Pt${}_2$Pb in particular exhibits signatures of one-dimensional spinon-like excitations \cite{Wu2016,Classen2018,Gannon2019}, and it is natural to ask whether the fractionalized excitations seen here are identical to those in an isolated spin chain, or if they could be a signature of new physics where the coupling with the surrounding metal is crucial. A different setup relevant to our discussion is that of \textit{nondissipative} systems where spatially long-range interactions arise due to cavity-mediated interactions, or due to dipole-dipole interactions \cite{Richerme2014,Jurcevic2014,Britton2012,Neyenhuis2017,Liu2019}. The relation between these two different classes of systems, namely, dissipative spin chains and spatially long-range interacting spin chains is space-time rotation -- e.g., Ohmic dissipation maps to $1/r^2$ interaction after space-time rotation.


Our focus in this work will be on (1+1)-D  SU$(N)_k$ Wess–Zumino–Witten (WZW) CFTs \cite{Wess1971,Novikov1981,Witten1983,Witten1984,Polyakov1983}  perturbed by a dissipative term that can arise in models of solid-state systems \cite{Affleck1985,Affleck1986,Affleck1987,Affleck1989,Francesco1997,Nielsen2011,Bondesan2015}. Further, as discussed below, the RG analysis for this problem can be controlled using a large-$k$ expansion, similar to the nondissipative case \cite{Witten1984}. Recent work on (0+1)-$D$ dissipative spin impurities has shown the presence of multiple fixed points due to the interplay of Berry phase and dissipation \cite{Cuomo2022,Nahum2022,Beccaria2022,Weber2023,Hu2022}, and it is natural to wonder about the fate of models in higher dimensions where both dissipation and Berry phase effects exist. Lastly, analogous to the long-range Ising or O$(N)$ models \cite{Paulos2016}, a relativistic version of our model (which we also study) can potentially lead to an infinite number of new conformal field theories labeled by $(N,k)$.

\begin{figure}[h]
	\centering
	\includegraphics[width=0.47\textwidth]{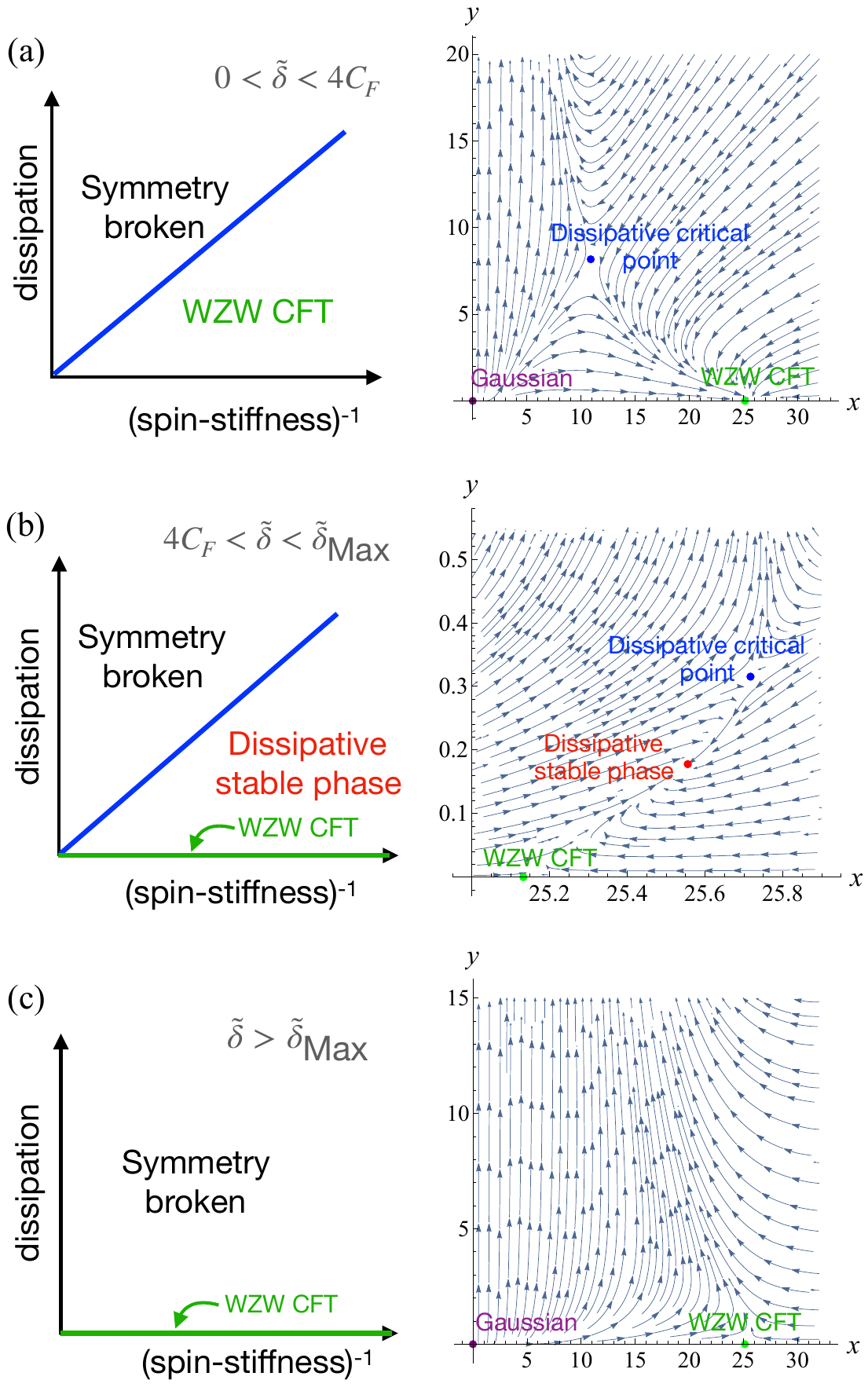}
	\caption{Schematic phase diagrams (left) and associated RG flows (right) in terms of the inverse ``spin stiffness'' ($\Tilde{\lambda} \sim x$) and the dissipation strength ($\Tilde{\gamma} \sim y$) for three different regimes, parameterized by $\Tilde{\delta} = k \delta$. All the RG flows have been obtained using $N=2$, and thus $4C_F = 3$, $\Tilde{\delta}_{\text{Max}}\approx 3.0429$. a) For $0 < \Tilde{\delta} < 4C_F$, a dissipative, critical fixed point separates the WZW CFT from a symmetry-broken (ordered) phase ($\tilde{\gamma} \gg 1$). The RG flow is plotted for $\Tilde{\delta} = 1$. b) For $4C_F < \Tilde{\delta} <  \Tilde{\delta}_{\text{Max}}$, one obtains a stable, dissipative  phase which is separated from the ordered phase by a dissipative critical point. For visualization purposes, we have zoomed on the interesting region of the RG flow, and plotted $\frac{1}{8}\beta(x)$, $4\beta(y)$. The RG flow has been obtained by setting $\Tilde{\delta}=3.04$. At $\Tilde{\delta} = \Tilde{\delta}_{\text{Max}}$, the two dissipative fixed points annihilate each other. c) For $\Tilde{\delta} > \Tilde{\delta}_{\text{Max}}$, there is no dissipative fixed point and the WZW CFT is unstable towards the broken symmetry (ordered) phase. The RG flow is plotted for $\Tilde{\delta} = 4$.}
	\label{fig:rglow_nonrel}
\end{figure}

\section{Model and its symmetries}
		
We will first consider a nonrelativistic setup where  dissipation induces interactions that are nonlocal only in time, analogous to the standard Hertz-Millis theory for antiferromagnets  \cite{Hertz1976,Millis1993} (the induced nonlocality in space due to dissipative bath is assumed to be subleading compared to the spatial kinetic energy term, and hence neglected \cite{Danu2022}). We consider a system which in the absence of dissipation is described by the 1+1-D SU$(N)_k$ WZW CFT \cite{Wess1971,Novikov1981,Witten1983,Witten1984}. The (Euclidean) action is
 
\begin{equation} \label{Eq:fullaction_nonrel}
S[g] = S_{\text{Grad}}[g] + S_{\text{WZ}}[g] + S_{\text{Dis}}[g].
\end{equation} 
In this equation

\begin{equation}
S_{\text{Grad}}[g] = \frac{1}{\lambda}\int d\tau dx \, \tr \left( \frac{1}{c^2} \partial_{\tau} g \partial_{\tau} g^{-1} + \partial_x g \partial_x g^{-1} \right)
\end{equation}  
		  
\noindent is the standard kinetic energy term for the matrix-valued field $g \in \text{SU}(N)$, transforming in the bifundamental represenation of SU$(N)_L$ $\otimes$ SU$(N)_R$. $c$ is a velocity which will run under RG as discussed below. Next,
 
\begin{equation} 
S_{\text{WZ}}[g] = \frac{\I k}{12\pi} \int_{B^3} d\tau\,dx\,du \, \epsilon^{ijk} \tr \left( \tilde{g}^{-1} \partial_i \tilde{g} \, \tilde{g}^{-1} \partial_j \tilde{g} \,\tilde{g}^{-1} \partial_k \tilde{g} \right)
\end{equation}

\noindent is the Wess-Zumino (WZ) Berry-phase term, defined in terms of $\tilde{g}(\tau,x,u)$  which is an  extension of the field $g(\tau,x)$ to a three-ball $B^3$ so that $\tilde{g}(\tau,x,u=0) = g_0$ is any chosen reference value, and $\tilde{g}(\tau,x,u=1) = g(\tau,x)$ is the physical value of $g$ at $(\tau,x)$ (= boundary $S^2$ of $B^3$). Finally,
 
\begin{equation} 
S_{\text{Dis}}[g] = k^2 \gamma \int d\tau d\tau' dx \, K(\tau-\tau') \, \tr [\mathds{1} - g(\tau,x) g^{-1}(\tau',x)] \, , 
\end{equation} 

\noindent where $\gamma > 0$, is the dissipation term. The kernel $K$ is defined as $K(\tau-\tau') = \frac{A}{|\tau-\tau'|^{3-\delta}}$ with the normalization $\quad A = \frac{(\delta-2)}{16\pi \Gamma(\delta-1) \cos(\pi \delta/2)}$ chosen so that the Fourier transform $\Tilde{K}(\omega)$ of $K(\tau)$ has a simple form suited for our RG analysis. We restrict $\delta$ to the range $0 < \delta < 2$ so that the Fourier transform $\tilde{K}(\omega)$ of $K(\tau)$ goes to zero as $\omega \rightarrow 0$, and $1/A$ is not divergent. The global continuous symmetry of this model is SU$(N)_L$ $\otimes$ SU$(N)_R$ where under SU$(N)_L$, $g \rightarrow U g$, and under SU$(N)_R$, $g \rightarrow g V$, where $U, V$ are arbitrary SU$(N)$ matrices. Since  $S_{\text{Grad}}[g] + S_{\text{WZ}}[g] $ is Lorentz invariant, after interchanging space and imaginary time, the action $S[g]$ describes a nondissipative closed system with long-range spatial interactions (a dynamical exponent $z$ in the dissipative system corresponds to a dynamic exponent $1/z$ in its space-time interchanged counterpart). 

The exponent $3-\delta$ for the kernel $K(\tau)$ is chosen so that  $\delta = \tilde{\delta}/k \ll 1$, with $\tilde{\delta}$ an $\mathcal{O}(1)$ number, allows for a controlled $1/k$ expansion. Relatedly, the couplings $\lambda$ and $\gamma$ will be of the order $1/k$ at all the RG fixed points, which implies that the three terms in the action $S[g]$ all scale as $k$. It will be useful to introduce the $\mathcal{O}(k^0)$ couplings $\Tilde{\lambda} = k \lambda$ and $\Tilde{\gamma} = k \gamma$. The dynamical exponent $z$ will be defined as part of the RG scheme, and will deviate from unity only by $\mathcal{O}(1/k)$, and therefore we also introduce an  $\mathcal{O}(k^0)$ variable $\Tilde{z}$ such that $z = 1 + \frac{\Tilde{z}}{k}$.

\section{Renormalization Group} \label{sec:rg}

To set up our RG calculation, we decompose the matrix-valued field $g$ as $g = g_s e^{W}$, where $g_s$ denotes ``slow'' variables, and $W$ denotes ``fast'' variables \cite{Witten1984}. The renormalization of $\Tilde{\lambda},\Tilde{\gamma}$ and $c$ is induced by integrating out the fast variables. At the leading order in $1/k$ (i.e., one-loop Feynman diagrams), we obtain the following $\beta$ functions for $\Tilde{\lambda}$, $\Tilde{\gamma}$ and $c$ (see Appendix \ref{sec:appendixA} for a detailed derivation):

\begin{equation} \label{eq:beta_lt_non-relativistic}
\beta(\Tilde{\lambda}) = \frac{1}{k} \Bigg[ - \Tilde{z} \Tilde{\lambda} + \frac{N c \Tilde{\lambda}^2}{8\pi} \Bigg( w - \frac{c^2 \Tilde{\lambda}^2}{(8\pi)^2} w^3 \Bigg) \Bigg] \, ,
\end{equation}

\begin{equation} \label{eq:beta_gt_non-relativistic}
\beta(\Tilde{\gamma}) = \frac{1}{k} \Bigg[ (\Tilde{\delta} - \Tilde{z}) \Tilde{\gamma} - \frac{C_F}{2\pi} c \Tilde{\lambda} \Tilde{\gamma} w \Bigg] \, ,
\end{equation}

\begin{align} \label{eq:beta_c}
\begin{split}
\beta(c) &= \frac{1}{k} \Bigg[ \Tilde{z} c - \frac{N c^2 \Tilde{\lambda}}{16 \pi} \Bigg( 1 + \frac{c^2 \Tilde{\lambda}^2}{(8\pi)^2} \Bigg) (w - w^3) \\ &\hspace{0.2cm} - \frac{C_F}{32\pi^2} c^4 \Tilde{\lambda}^2 \Tilde{\gamma} w + \frac{N}{(8\pi)^2} \Bigg( 1 + \frac{c^2 \Tilde{\lambda} \Tilde{\gamma}}{16\pi} \Bigg) c^4 \Tilde{\lambda}^2 \Tilde{\gamma} w^3 \Bigg] \, ,
\end{split}
\end{align}

\begin{figure}[t]
	\centering
	\includegraphics[width=1.0\hsize]{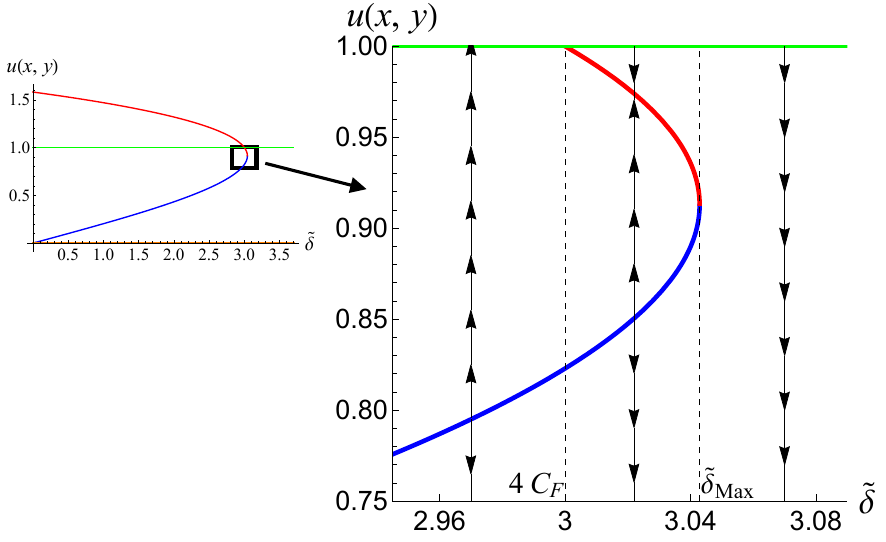}
	\caption{Two physical solutions of the cubic equation for $u(x,y)$ in terms of $\Tilde{\delta}$. The blue curve is the solution associated with the unstable dissipative critical point, while the red curve is the solution associated with the stable dissipative phase. Physical solutions must satisfy $0 \leq u(x,y) \leq 1$ (See Appendix \ref{sec:fp_analysis_solving_fp_A}). One of the three solutions is always negative and is thus not shown. Moreover, for $0 < \Tilde{\delta} < 4C_F$, the dissipative stable phase is located in $y<0$ and is thus also unphysical. The WZW fixed point (green curve) and the broken symmetry (ordered) phase (orange curve along $u(x,y)=0$ in the left figure) are also shown. The right part of the figure is a zoom on the interesting regime containing the two dissipative fixed points, with the stability of the various fixed points represented by arrows. This figure was obtained by setting $N=2$ and similar plots are obtained for other values of $N$.}
	\label{fig:cubicplot}
\end{figure}

\noindent where $w = \Big( 1 + \frac{1}{8\pi} c^2 \Tilde{\lambda} \Tilde{\gamma} \Big)^{-1/2}$ and $C_F = \frac{N^2-1}{2N}$ is the quadratic Casimir for SU$(N)$ in the fundamental representation. The main outcomes of these RG equations are as follows: 

\begin{enumerate}
	\item  When $0 < \Tilde{\delta} < 4 C_F$ (Fig. \ref{fig:rglow_nonrel}(a)), the WZW CFT fixed point is perturbatively stable against dissipation, which can also be deduced using the scaling dimension $\Delta_g \approx 2 C_F/k$ of the primary field $g$ at the WZW fixed point at large $k$. In this range of $\Tilde{\delta}$, as the magnitude $\Tilde{\gamma}$ of dissipation increases, the system eventually undergoes a single-parameter tuned second-order phase transition beyond which  $\Tilde{\gamma}$  flows to infinity. Based on energetical considerations, we expect that at large $\tilde{\gamma}$, the field $g$ acquires a non-zero expectation value, so that the SU$(N)_L$ $\otimes$ SU$(N)_R$ symmetry is spontaneously broken to  diagonal SU$(N)$, akin to the chiral symmetry-broken phase in QCD with massless quarks \cite{Peskin1995}, and we make this assumption in drawing the phase diagram in Fig. \ref{fig:rglow_nonrel}. Qualitatively, this scenario is similar to the one discussed in Ref.\cite{Laflorencie2005} for Heisenberg chain perturbed by long-range interactions (whose action can be thought of as a space-time rotated version of our action, Eq.\eqref{Eq:fullaction_nonrel}).
	
	Writing $g \sim e^{i \sum_a \pi_a T_a}$ where $\pi_a$ are the Goldstone modes, and $T_a$ are the SU$(N)$  generators, the low energy theory in the ordered phase is given by $\mathcal{L} = |\pi_a(k,\omega)|^2 (k^2 + \omega^{2-\delta}) + ...$, where ``$...$'' denotes interactions between the Goldstone modes. These interactions are irrelevant at low-energy, and spontaneous symmetry breaking stable, precisely due to long-range interactions that lead to the aforementioned nonrelativistic dispersion for the Goldstone modes (this is ultimately related to the fact that the integral $\int dk d\omega (k^2 + \omega^{2-\delta})^{-1}$ for $\delta > 0$ converges in the infra-red) \cite{Castro1997,Werner2005a,Werner2005b,Laflorencie2005,Cazalilla2006,Lobos2012,Sperstad2012,Yan2018,Weber2022,Danu2022,Cuomo2023}. In contrast, for a relativistic theory in 1+1-D with short-range interactions, Goldstone modes interact strongly and destabilize spontaneous symmetry breaking \cite{Hohenberg1967,Mermin1966,Polyakov1975}. The universal properties of the critical point separating the WZW CFT and the symmetry-broken phase are further discussed below. 
	
	\item When $\Tilde{\delta} > \Tilde{\delta}_{\text{Max}} = \frac{2}{3\sqrt{3}} \sqrt{ \frac{(4C_F+N)^3}{N}}$ (Fig. \ref{fig:rglow_nonrel}(c)), the WZW fixed point is unstable towards the aforementioned ordered phase for infinitesimal $\Tilde{\gamma}$.
	
	\item 	Most interestingly, in the intermediate regime, namely, when $4 C_F < \Tilde{\delta} < \Tilde{\delta}_{\text{Max}} $, the WZW CFT is unstable towards a \textit{nonrelativistic, dissipative, critical phase} which has no relevant perturbations if we only allow terms that respect the SU$(N)_L$ $\otimes$ SU$(N)_R$ symmetry (Fig. \ref{fig:rglow_nonrel}(b)). This phase is separated from the ordered phase by a single-parameter-tuned phase transition. At $\Tilde{\delta} = \Tilde{\delta}_{\text{Max}}$, one encounters a fixed-point annihilation between the fixed point corresponding to this stable phase and the fixed point corresponding to the phase transition out of this phase to the ordered phase. 
	
\end{enumerate}

The aforementioned analytical expression for  $\Tilde{\delta}_{\text{Max}}$ follows from solving $\beta(\Tilde{\lambda}) = \beta(\Tilde{\gamma}) = 0$, which leads to the following cubic equation for the variables $x = c \Tilde{\lambda}$ and $y = c \Tilde{\gamma}$: 

\begin{equation}
\label{Eq:cubiceq} 
N u^3(x,y) - (4C_F + N) u(x,y) + \Tilde{\delta} = 0
\end{equation}

\noindent where $u(x,y) = \frac{x}{8\pi} \frac{1}{\sqrt{1 + \frac{1}{8\pi} xy}}$. This cubic equation has three (one) real solutions for $u(x,y)$ when its discriminant is positive (negative), and the change of sign of the discriminant precisely corresponds to the fixed-point annihilation. As shown in  Appendix \ref{sec:fp_analysis_solving_fp_A}, physical solutions must respect $0\leq u(x,y) \leq 1$. Since one of the three solutions always has $u(x,y) < 0$, it can be dropped and is thus not shown in Fig. \ref{fig:cubicplot}. Furthermore, in the regime $(0 < \Tilde{\delta} < 4C_F)$, the solution associated with the stable dissipative phase has $u(x,y) > 1$ and is also unphysical.

\begin{figure}[h]
	\centering
	\includegraphics[width=1.0\hsize]{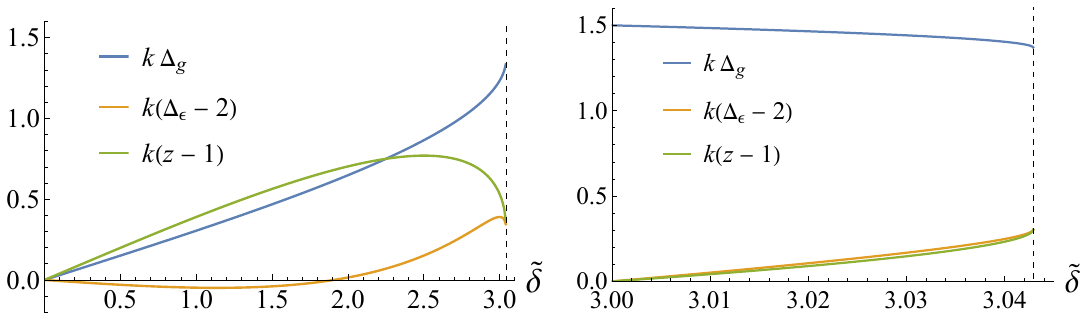}
	\caption{Coefficients of the $\mathcal{O}(1/k)$ contribution to the scaling dimensions $\Delta_g$ and $\Delta_{\epsilon}$ of the primary field $g$ and the energy density operator $\epsilon$ as well as the dynamical critical exponent $z$ at the dissipative critical point (left) and the dissipative critical phase (right) in terms of $\Tilde{\delta}$. The plots have been obtained using $N=2$, for which $4C_F = 3$ and $\Tilde{\delta}_{\text{Max}} \approx 3.0429$, which is represented by the vertical dashed line. }
	\label{fig:scaling_dim_nr}
\end{figure}

By adding a ``magnetic field'' term to the action,

\begin{equation}
S_h =  h \int d\tau\,dx\, \tr(g + g^{-1}) \, ,
\end{equation}

\noindent we obtain the $\beta$ function for $h$ (see Appendix \ref{sec:fp_analysis_magnetic_field_A} for the derivation): $\beta(h) = e_h h$ where $e_h = \left( 2 + \frac{\Tilde{z}}{k} \right) - \frac{C_F}{4\pi k} c \Tilde{\lambda}  \, w + \mathcal{O}(1/k^2)$ is the RG eigenvalue associated with $h$. The scaling dimension $\Delta_g$ of the primary field at a given fixed point is therefore given by $ 1+z - e^{*}_h$ where $e_h^{*}$ is evaluated at that fixed point. One may also extract the scaling dimension $\Delta_\epsilon$ of the energy density operator  $\epsilon=\tr \left( \frac{1}{c^2} \partial_{\tau} g \partial_{\tau} g^{-1} + \partial_x g \partial_x g^{-1} \right)$ using the RG equations. We numerically solve the RG equations for the fixed points, and we plot the dynamical exponent $z$ and the scaling dimensions $\Delta_g, \Delta_\epsilon$ at the two dissipative fixed points in terms of $\Tilde{\delta}$ in Fig. \ref{fig:scaling_dim_nr}. Moreover, by using the RG equations for $h$ and $\Tilde{\gamma}$, one can show that at either of these fixed points, the following equality holds: $\Tilde{z} = \Tilde{\delta} - 2k\Delta_g$, which corresponds to the expansion at order $\mathcal{O}(1/k)$ of $z = \frac{2-\eta}{2-\delta}$ where $\eta$ is the anomalous dimension of $g$ (see Appendix \ref{sec:relation_eta_z_A}). This relation can be argued to hold on the general ground that an RG transformation leaves the nonlocal term $\int d\tau d\tau\int dx K(\tau-\tau') \, \tr [g(\tau,x) g^{-1}(\tau',x)]$ invariant \cite{Nahum2022} and has also been seen in previous studies on nonrelativistic quantum criticality \cite{Gamba1999,Pankov2004,Sperstad2012}. Note that at either of the dissipative fixed points, the two-point correlation function $\ev{\tr\left(g(\tau,x) g^{-1}(0,0)\right)}$ has a non-trivial scaling behavior  both along space and time, with equal-time, unequal-space correlations decaying as $1/x^{2 \Delta_g}$, and unequal-time, equal-space correlations decaying as $1/\tau^{2\Delta_g/z}$.

We note a technical subtlety about our RG calculation: the total action $S[g]$ respects the discrete symmetry $g(\tau,x) \rightarrow g^{-1}(\tau,-x)$ which rules out terms such as $\int d\tau dx \tr \left(\partial_{\tau} g \partial_x g^{-1}\right)$. However, the aforementioned decomposition $g = g_s e^W$ ``fractionalizes'' the action of this discrete symmetry, and integrating out $W$ can and does generate an unphysical term 
$\int d\tau dx \tr \left(\partial_{\tau} g_s \partial_x g_s^{-1}\right)$ which should be discarded on symmetry grounds. One way to keep the symmetry manifest is by defining a symmetrized effective action $S_{\text{Eff}}[g_s] = \frac{1}{2} \big( S_{\text{Eff}}^{(1)}[g_s] + S_{\text{Eff}}^{(2)}[g_s] \big)$, where $(1)$ and $(2)$ correspond to the following two decompositions: $g = g_s e^W$ and $g = e^W g_s$. The two decompositions yield exactly the same RG for all the physical (i.e. symmetry allowed) terms, while the aforementioned unphysical term has a relative opposite sign. Due to this, $S_{\text{Eff}}[g_s]$ only contains terms allowed by symmetries.


\section{A Relativistic version}

As mentioned in the introduction, we also study a relativistic-invariant version of our model. The kinetic energy term and the WZW term are unchanged (we set $c=1$), while the dissipation is now chosen as Lorentz invariant

\begin{equation}
S_{\text{Dis}} = k^2 \gamma \int d^2\vb*{r} d^2\vb*{r'} \, K(|\vb*{r}-\vb*{r'}|) \tr \left( \mathds{1} - g(\vb*{r}) g^{-1}(\vb*{r'}) \right) \, ,
\end{equation}

\noindent where $\vb*{r} = (\tau,x)$ denotes Euclidean space-time, and the kernel is now $K(r) = \frac{B}{r^{4-\delta}}$ with $B = -\frac{1}{2^{1+\delta} \pi^2} \frac{\Gamma(2-\delta/2)}{\Gamma\big(\frac{\delta}{2} - 1\big)}$ and $r = |\vb*{r}|$. The normalization of the kernel is such that its Fourier transform is $\Tilde{K}(p) = -\frac{1}{8\pi} |p|^{2-\delta}$, with $p = |\vb*{p}|$, $\vb*{p} = (\omega,q)$. 

The RG analysis can be carried out using a scheme similar to that for the nonrelativisitic case (See Appendix \ref{sec:appendixB}).  It will again be useful to introduce $\mathcal{O}(k^0)$ couplings $\Tilde{\lambda} = k \lambda$ and $\Tilde{\gamma} = k \gamma$.   The corresponding $\beta$ functions to the leading order in $1/k$ are:

\begin{align} \label{eq:betalt_relativistic_main}
\begin{split}
\beta(\Tilde{\lambda}) &= \frac{1}{k} \Bigg[ \frac{N \Tilde{\lambda}^2}{8\pi} \Bigg( 1 - \frac{\Tilde{\lambda}^2}{(8\pi)^2} \Bigg) F^2(\Tilde{\lambda} \Tilde{\gamma}) - \frac{C_F}{16\pi^2} \Tilde{\lambda}^3 \Tilde{\gamma} F(\Tilde{\lambda} \Tilde{\gamma}) \\
& \hspace{0.5cm} + \frac{N}{(8\pi)^3} \Tilde{\lambda}^4 \Tilde{\gamma}^2 F^2(\Tilde{\lambda} \Tilde{\gamma}) + \frac{N}{32\pi^2} \Tilde{\lambda}^3 \Tilde{\gamma} F^2(\Tilde{\lambda} \Tilde{\gamma}) \Bigg] \, ,
\end{split}
\end{align}

\begin{equation} \label{eq:betagt_relativistic_main}
\beta(\Tilde{\gamma}) = \frac{1}{k} \Bigg[ \Tilde{\delta} \Tilde{\gamma} - \frac{C_F}{2\pi} \Tilde{\lambda} \Tilde{\gamma} F(\Tilde{\lambda} \Tilde{\gamma}) \Bigg] \, ,
\end{equation}

\noindent with $F(\Tilde{\lambda} \Tilde{\gamma}) = \frac{1}{1 + \frac{1}{8\pi} \Tilde{\lambda} \Tilde{\gamma}}$.

\begin{figure}[h]
	\centering
	\includegraphics[width=1\hsize]{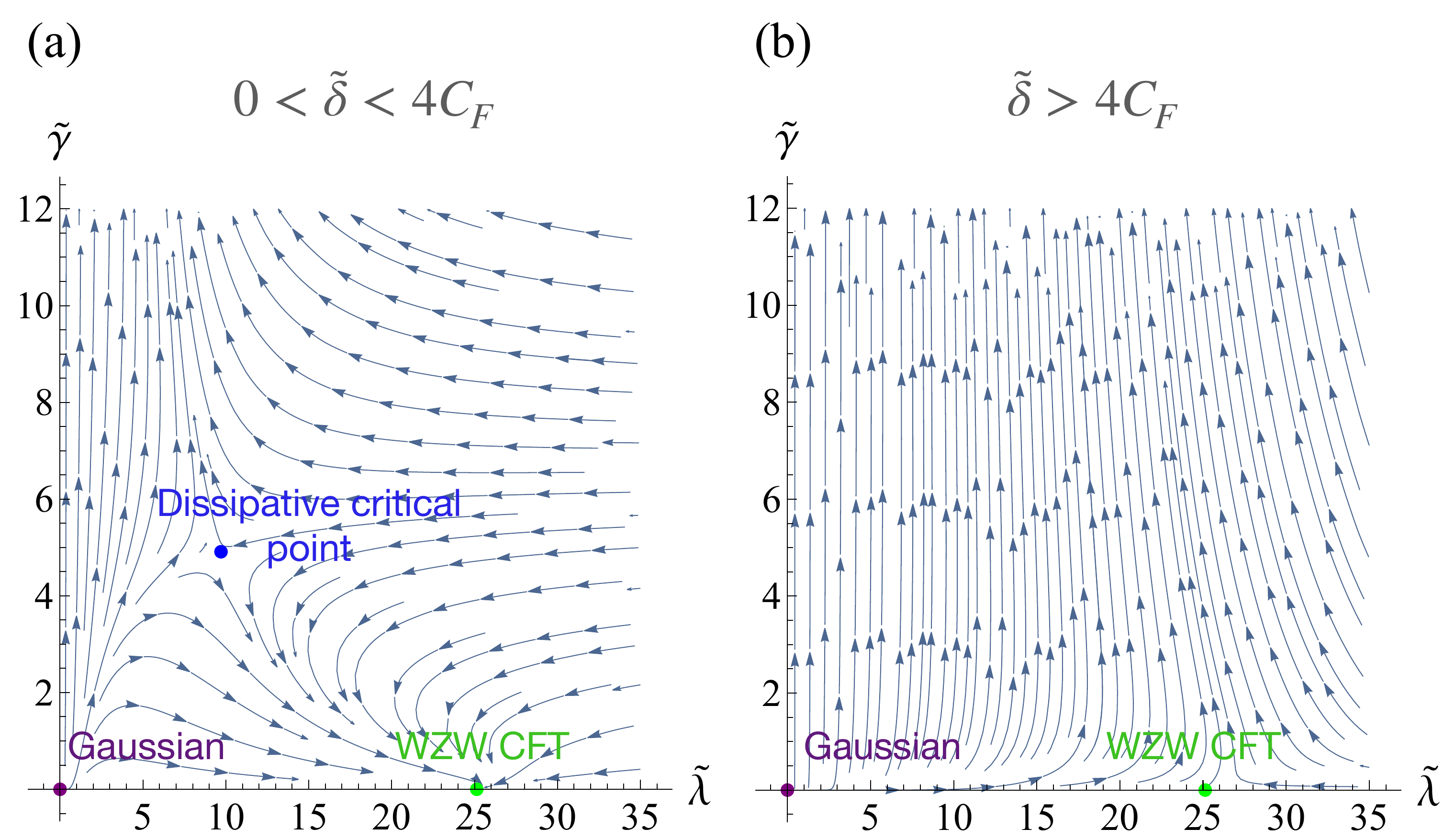}
	\caption{RG flows for the relativistic theory for the two different regimes parametrized by $\tilde{\delta} = k \delta$. For both cases, $N=2$ has been used, which means that $4C_F = 3$. a) For $\Tilde{\delta} < 4C_F$, the WZW CFT is separated from the dissipation-induced ordered phase by a dissipative critical point. The plot has been obtained using $\Tilde{\delta} = 0.4$. b) For $\Tilde{\delta} > 4C_F$, there is no dissipative fixed point and the WZW CFT is unstable to infinitely small dissipation. The plot has been obtained using $\Tilde{\delta} = 6$.}
	\label{fig:rglow_rel}
\end{figure}

In contrast to the nonrelativistic case, we now find only two qualitatively different phase diagrams as a function of $\Tilde{\delta}$, as illustrated by the RG flows in Fig. \ref{fig:rglow_rel}: when $\Tilde{\delta} < 4 C_F$, the WZW CFT is stable against dissipation and is separated from the large $\tilde{\gamma}$ fixed point (which presumably again corresponds to the symmetry-broken phase) by a single-parameter tuned quantum phase transition, while when $\Tilde{\delta} > 4 C_F$, the WZW fixed point is unstable towards the large $\tilde{\gamma}$ fixed point at infinitesimal dissipation. Furthermore, we find the following scaling dimensions for the primary field $g$ and the energy density operator $\epsilon$ at the dissipative fixed point: $\Delta_g = \frac{\Tilde{\delta}}{2k}, \Delta_{\epsilon} = 2 + \frac{\Tilde{\delta}}{64 C_F^3 k} \Bigg[ N \Tilde{\delta}^2 - \sqrt{N (1024 C_F^5 - 64 C_F^3 \Tilde{\delta}^2 + N \Tilde{\delta}^4)} \Bigg]$. The scaling dimensions at the WZW fixed point of course match the known exact results in the large-$k$ limit, namely, $\Delta_g = 2C_F/k, \Delta_{\epsilon} = 2 + 2N/k$. Analogous to the long-range Ising or O($N$) models \cite{Paulos2016}, it will be interesting to explore whether these theories potentially correspond to an infinite number of new conformal field theories labeled by the integers $(N,k)$.


\section{Summary and discussion}
	
We carried out an RG study of a class of (1+1)-D CFTs perturbed by long-range interactions along space and/or time, and we identified several RG fixed points (see Fig. \ref{fig:rglow_nonrel}). For a range of the exponent $\delta$ that characterizes  long-range interactions, we found that the CFT becomes unstable towards a stable, gapless dissipative phase that exhibits non-trivial scaling both along space and time. Upon tuning $\delta$, one encounters a fixed-point annihilation between the fixed point corresponding to the aforementioned stable, gapless phase, and  another dissipative fixed point with one relevant direction. Compared to relativistic systems with long-range interactions and no WZW term \cite{Fisher1972,Sak1973,Sak1977,Bhattacharjee1982,Paulos2016,Behan2017a,Behan2017b,Defenu2017,Slade2018,Gubser2019,Defenu2020,Chakraborty2021,Chai2021,Chai2022}, the novelty here is the presence of an intermediate coupling stable phase. We characterized this critical phase via the scaling dimensions of a few prominent operators and the dynamical critical exponent. We also studied a relativistic version of our theory that shows a novel quantum critical point between the WZW CFT and a dissipation-induced symmetry-broken phase (see Fig. \ref{fig:rglow_rel}). 

It is important to ask what lattice models may give rise to the non-trivial intermediate coupling dissipative phase (Fig. \ref{fig:rglow_nonrel}(b)). In our analysis we assumed  SU$(N)_L \otimes$ SU$(N)_R$ symmetry at low energies which may be difficult to achieve starting from a lattice model. Although one can certainly find fine-tuned lattice models that realize SU$(N)_k$ CFTs for any $N, k$ \cite{Babujian1982,Babujian1983,Takhtajan1982},  when $k > 1$  there generically exist relevant terms that explicitly break the SU$(N)_L \otimes$ SU$(N)_R$ symmetry down to the diagonal SU$(N)$ \cite{Affleck1987}. A natural way to realize SU$(N)_L \otimes$ SU$(N)_R$ without any fine-tuning is to consider a spin chain that corresponds to the edge mode of a 2D symmetry protected topological (SPT) phase \cite{Liu2013}. Further, 1+1-D models with $k = 1$ for any $N$ are also stable (assuming translation symmetry) since anomaly-based arguments imply that under RG flow the parity of the level cannot change \cite{Gepner1986,Furuya2017,Yao2019}. On that note, for a single impurity coupled to a  dissipative bath, one also finds a phase diagram broadly similar to our problem \cite{Cuomo2022,Nahum2022, Beccaria2022,Weber2023,Hu2022}, and although the corresponding calculation is justified only in a semiclassical limit somewhat analogous to ours (large spin $S$ for a single impurity versus large level $k$ for WZW CFT), numerical studies have shown that the qualitative aspects carry over even to spin-1/2 impurities \cite{Weber2023,Hu2022}. Therefore, it will be interesting to explore the effect of long-range interactions on (1+1)-D lattice models corresponding to SU$(N)_k$ CFTs even at $k = 1$ using quantum Monte Carlo (QMC) \cite{Werner2005a,Laflorencie2005,Sperstad2012,Weber2022,Song2023,Zhao2023}, or in engineered systems \cite{Toskovic2016,Choi2017,Moro2019,Choi2019,Richerme2014,Jurcevic2014,Britton2012,Neyenhuis2017,Liu2019}. Another direction worth exploring is the potential relation to models of deconfined quantum critical points that also have WZW terms and show fixed-point annihilation in fractional dimensions \cite{Nahum2020,Ma2020}.

Returning to the topic of hybrid-dimensionality Kondo lattice models, we speculate that the dissipative phase can potentially be a novel ``fractionalized Fermi liquid'' with a small Fermi surface. This is because the physics of Kondo singlet formation, and relatedly, that of a ``large Fermi surface'' heavy Fermi liquid phase  \cite{Oshikawa2000} where  local moments contribute to the Fermi surface volume, is non-perturbative in the Kondo coupling $J_K$ with an effective energy scale $e^{-c/J_K}$, where $c$ is a constant. If one imagines that our action $S[g]$ was obtained by integrating out a fermionic bath, then such physics is likely not operative in the dissipative phase since the Kondo coupling $J_K$ appears only perturbatively (with dissipation $\gamma \sim J^2_K$). In contrast to the ``conventional'' small Fermi surface fractionalized phases  \cite{Senthil2003, Senthil2004}, in such a dissipation-induced non-Fermi-liquid, here the electrons and spins do not completely decouple at low energies since non-zero dissipation must imply non-trivial entanglement between spins and electrons. At the same time, one can still inquire whether the fixed points we obtained are \textit{perturbatively} stable against flow to a large Fermi surface phase. For example, as discussed in Ref.\cite{Danu2020}, for a spin chain embedded in a Dirac semi-metal, the electronic bath completely decouples from the spin chain at weak Kondo coupling, resulting in a hybrid-dimensionality small-Fermi-surface fractionalized Fermi liquid \cite{Senthil2003,Senthil2004}. Another example is provided by ``Fermi-Bose Kondo impurity'' models \cite{Smith1999,Sengupta2000,Vojta2000,Zhu2002,Zarand2002}, where one finds an  intermediate dissipation fixed point which is again stable against Kondo singlet formation with the fermionic bath  \cite{Hu2022}. The existence of either of these fixed points can be inferred solely using a dissipative bosonic bath similar to our calculation  \cite{Cuomo2022,Nahum2022,Beccaria2022,Weber2023}.  In a similar vein, we expect that the WZW CFT fixed point (Fig. \ref{fig:rglow_nonrel}(a)), and more interestingly, the stable, dissipative fixed point (Fig. \ref{fig:rglow_nonrel}(b)) are also both stable against flow towards a large Fermi surface phase. The heuristic reasoning behind this expectation is that perturbatively, the dissipation coefficient $\gamma$ is proportional to $J^2_K$, and since the RG flow at either of these fixed points is attractive along the $\gamma$ direction, one expects that it will be attractive along the $J_K$ direction as well. We leave the further exploration of this topic to the future.

\begin{acknowledgments}
\emph{Acknowledgments:} The authors are grateful to John McGreevy, Adam Nahum, Masaki Oshikawa, Filip Ronning, Qimiao Si and Matthias Vojta for helpful discussions. TG is supported by the National Science Foundation under Grant No. DMR-1752417. This research was supported in part by the National Science Foundation under Grant No. NSF PHY-1748958.
	
\end{acknowledgments}



\onecolumngrid
\appendix
	
\section{RG analysis of the nonrelativistic theory} \label{sec:appendixA}

This Appendix presents the detailed RG calculation of the nonrelativistic theory presented in the main text.

	
\subsection{Expanding in slow/fast modes} \label{sec:expand_A}

The RG calculation is performed by splitting $g$ into slow and fast degrees of freedom: $g(\tau,x) = g_s(\tau,x) g_f(\tau,x)$, where $g_s$ is a slow-varying background field, while $g_f$ constitutes fast fluctuations about $g_s$ \cite{Witten1984}. The goal is to obtain the effective action for the slow fields $g_s$ due to the integration of the fast modes $g_f$. $g_f$ is thus expanded to quadratic order with the following decomposition:

\begin{equation}
g_f = \e^{W} \approx \mathds{1} + W + \frac{W^2}{2} + ... \, ,
\end{equation}

\noindent with $W(\tau,x) = \I T^a \phi^a(\tau,x)$, where $T^a$ are the $N^2-1$ generators of $\text{SU}(N)$ in the fundamental representation, which respect the algebra $[T^a,T^b] = \I f^{abc} T^c$ and are normalized according to $\tr (T^a T^b) = \frac{1}{2} \delta^{ab}$, while $\phi^a$ are $N^2-1$ real scalar fields. Below we analyze the three terms in the action $S[g] = S_{\text{Grad}}[g] + S_{\text{WZ}}[g] + S_{\text{Dis}}[g]$ with such a decomposition. Overall, the main simplification in the large $k$ limit is that at each order in $1/k$, there are only a finite number of Feynman diagrams that contribute to the RG flow, as explained in Appendix \ref{sec:integration_fast_A}.

\subsubsection{Gradient term} \label{sec:expand_grad_A}

Let us start with the gradient term. For $\mu = \tau$ or $\mu = x$ (no sum over $\mu$), we have

\begin{align}
\begin{split}
\tr \Big( \partial_{\mu} g \partial_{\mu} g^{-1} \Big) &= \tr \Big( \partial_{\mu} (g_s g_f) \partial_{\mu} (g_f^{-1} g_s^{-1}) \Big) \\ &= \tr \Big( \partial_{\mu} g_s g_f \partial_{\mu} g_f^{-1} g_s^{-1} + \partial_{\mu} g_s g_f g_f^{-1} \partial_{\mu} g_s^{-1} + g_s \partial_{\mu} g_f \partial_{\mu} g_f^{-1} g_s^{-1} + g_s \partial_{\mu} g_f g_f^{-1} \partial_{\mu} g_s^{-1} \Big) \\ &= \tr \Big( \partial_{\mu} g_s \partial_{\mu} g_s^{-1} \Big) + \tr \Big( \partial_{\mu} g_f \partial_{\mu} g_f^{-1} \Big) + 2 \tr \Big( g_s^{-1} \partial_{\mu} g_s g_f \partial_{\mu} g_f^{-1} \Big) \, ,
\end{split}
\end{align}

\noindent where we have used the fact that $g_s g_s^{-1} = g_f g_f^{-1} = \mathds{1}$, which implies that $\partial_{\mu} g_{s} g_{s}^{-1} = - g_{s} \partial_{\mu} g_{s}^{-1}$ (same thing for $g_f$). Expanding the second term to quadratic order in $W$ yields

\begin{equation}
\tr \big[ \partial_{\mu} g_f \partial_{\mu} g_f^{-1} \big] = - \tr \big[ \partial_{\mu} W \partial_{\mu} W \big] + \mathcal{O}(W^3) \, .
\end{equation}

\noindent For the third term, we get

\begin{align}
\begin{split}
2 \tr \Big( g_s^{-1} \partial_{\mu} g_s g_f \partial_{\mu} g_f^{-1} \Big) &= 2 \tr \Big[ g_s^{-1} \partial_{\mu} g_s \Big( \mathds{1} + W + \frac{W^2}{2} \Big) \partial_{\mu} \Big( \mathds{1} - W + \frac{W^2}{2} \Big) \Big] + ... \\ &= 2 \tr \Big[ g_s^{-1} \partial_{\mu} g_s \Big( \frac{1}{2} W \partial_{\mu} W + \frac{1}{2} \partial_{\mu} W W - W \partial_{\mu} W \Big) \Big] + \text{Terms linear in } W + \mathcal{O}(W^2) \\ &= \tr \Big( g_s^{-1} \partial_{\mu} g_s [\partial_{\mu}W,W] \Big) + ... \, ,
\end{split}
\end{align}

\noindent where the terms linear in $W$ can be dropped, since these will yield vanishing contributions when computing loop diagrams over fast modes (no momentum exchange between slow and fast modes is compatible with momentum conservation).

Therefore, using the results derived above, the gradient term becomes 

\begin{align}
\begin{split}
S_{\text{Grad}}[g_s g_f] = S_{\text{Grad}}[g_s] + S_{\text{Grad}}^{(2)}[W] + S_{\text{Int,Grad}}^{(2)}[g_s,W] \, ,
\end{split}
\end{align}

\noindent with

\begin{equation}
S_{\text{Grad}}[g_s] = \frac{1}{\lambda} \int d\tau dx \, \tr \Bigg( \frac{1}{c^2} \partial_{\tau} g_s \partial_{\tau} g_s^{-1} + \partial_x g_s \partial_x g_s^{-1} \Bigg) \, ,
\end{equation}

\begin{align}
\begin{split}
S_{\text{Grad}}^{(2)}[W] &= - \frac{1}{\lambda}\int d\tau dx \, \tr \Bigg( \frac{1}{c^2} \partial_{\tau} W \partial_{\tau} W + \partial_x W \partial_x W \Bigg) \\ &= \frac{1}{2} \int \frac{d\omega dq}{(2\pi)^2} \Tilde{\phi}^a(\omega,q) \Pi^{-1}(\omega,q) \Tilde{\phi}^a(-\omega,-q) \, , \quad \Pi(\omega,q) = \frac{\lambda}{\frac{\omega^2}{c^2} + q^2} \, ,
\end{split}
\end{align}

\begin{equation}
S_{\text{Int,Grad}}^{(2)}[g_s,W] = \frac{1}{\lambda} \int d\tau dx \, \tr \Bigg( \frac{1}{c^2} g_s^{-1}\partial_{\tau} g_s [\partial_{\tau} W,W] + g_s^{-1}\partial_x g_s [\partial_x W,W] \Bigg) \, .
\end{equation}

\noindent Note that the second term has been written in Fourier space, after having taken the trace over the generators. This term will contribute to the fast propagator.

\subsubsection{WZ term} \label{sec:expand_WZ_A}

Let us now split the degrees of freedom in the WZ term. To do so, note that

\begin{equation}
g^{-1} dg = g_f^{-1} g_s^{-1} d (g_s g_f^{-1}) = g_f^{-1} g_s^{-1} d g_s g_f + g_f^{-1} d g_f \, .
\end{equation}

\noindent Therefore, the trace becomes

\begin{align}
\begin{split}
&\tr \Big[ g^{-1} dg \wedge g^{-1}dg \wedge g^{-1} dg \Big] \\ &= \tr \Big[ \Big( g_f^{-1} g_s^{-1} d g_s g_f + g_f^{-1} d g_f \Big) \wedge \Big( g_f^{-1} g_s^{-1} d g_s g_f + g_f^{-1} d g_f \Big) \wedge \Big( g_f^{-1} g_s^{-1} d g_s g_f + g_f^{-1} d g_f \Big) \Big] \, .
\end{split}
\end{align}

\noindent Expanding this expression yields eight terms, which can be combined to give

\begin{align}
\begin{split}
\tr \Big[ g^{-1} dg \wedge g^{-1}dg \wedge g^{-1} dg \Big] &= \tr \Big[ g_s^{-1} dg_s \wedge g_s^{-1}dg_s \wedge g_s^{-1} dg_s \Big] + 3 \tr \Big[ dg_s^{-1} \wedge d g_s \wedge g_f d g_f^{-1} \Big] \\ &\hspace{0.5cm} - 3 \tr \Big[ g_s^{-1} dg_s \wedge d g_f \wedge d g_f^{-1} \Big] + \mathcal{O}(W^3) \, .
\end{split}
\end{align}

\noindent Expanding the second term to quadratic order in $W$ yields

\begin{align}
\begin{split}
3 \tr \Big[ dg_s^{-1} \wedge d g_s \wedge g_f d g_f^{-1} \Big] &\approx 3 \tr \Big[ dg_s^{-1} \wedge d g_s \wedge \Big( \mathds{1} + W + \frac{W^2}{2} \Big) d \Big( \mathds{1} - W + \frac{W^2}{2} \Big) \Big] \\ &= 3 \tr \Big( dg_s^{-1} \wedge dg_s \wedge \frac{1}{2} [dW,W] \Big) + \text{Linear term in } W + \mathcal{O}(W^3) \, ,
\end{split}
\end{align}

\noindent while we get for the third term

\begin{align}
\begin{split}
- 3 \tr \Big[ g_s^{-1} dg_s \wedge d g_f \wedge d g_f^{-1} \Big] = 3 \tr \Big( g_s^{-1} d g_s \wedge dW \wedge dW \Big) + \mathcal{O}(W^3) \, .
\end{split}
\end{align}

\noindent Hence, combining everything leads to

\begin{align}
\begin{split}
\tr \Big[ g^{-1} dg \wedge g^{-1}dg \wedge g^{-1} dg \Big] &= \tr \Big[ g_s^{-1} dg_s \wedge g_s^{-1}dg_s \wedge g_s^{-1} dg_s \Big] + \frac{3}{2} \tr \Big( dg_s^{-1} \wedge dg_s \wedge [dW,W] \Big) \\ &\hspace{0.5cm} + 3 \tr \Big( g_s^{-1} d g_s \wedge dW \wedge dW \Big) \\ &= \tr \Big[ g_s^{-1} dg_s \wedge g_s^{-1}dg_s \wedge g_s^{-1} dg_s \Big] + \frac{3}{2} \tr \, d \Big( g_s^{-1} d g_s \wedge [dW,W] \Big) \, ,
\end{split}
\end{align}

\noindent where the second and the third terms have been combined in a total derivative in the last step. Hence, applying Stoke's theorem, the WZ action becomes

\begin{align}
\begin{split}
S_{\text{WZ}}[g_s g_f] &= S_{\text{WZ}}[g_s] +  S_{\text{Int,WZ}}^{(2)}[g_s,W] \\ &= S_{\text{WZ}}[g_s] + \frac{\I k}{8\pi} \int d\tau dx \, \epsilon_{\mu \nu} \tr \Big( g_s^{-1} \partial_{\mu} g_s [\partial_{\nu}W,W] \Big) \, .
\end{split}
\end{align}

\noindent The relativistic notation $\mu = (\tau,x)$ is used here.

\subsubsection{Dissipation term} \label{sec:expand_dis_A}

Finally, we focus on the dissipation term. The trace becomes

\begin{align}
\begin{split}
\tr \Big(\mathds{1} - g g^{\prime \, -1}\Big) &= \tr \big( \mathds{1} - g_s g_f g_f^{\prime \, -1} g_s^{\prime \, -1} \big) \\ &\approx \tr \Bigg[ \mathds{1} - g_s \Big( \mathds{1} + W + \frac{W^2}{2} \Big) \Big( \mathds{1} - W' + \frac{W'^2}{2} \Big) g_s^{\prime \, -1} \Bigg] \\ &= \tr \Big( \mathds{1} - g_s^{\prime \, -1} g_s \Big) - \tr \Bigg( \frac{W^2}{2} + \frac{W'^2}{2} - W W' \Bigg) \\ &\hspace{0.5cm} + \tr \Bigg[ \Big( \mathds{1} - g_s^{\prime \, -1} g_s \Big) \Bigg( \frac{W^2}{2} + \frac{W'^2}{2} - W W' \Bigg) \Bigg] + \mathcal{O}(W^3) \, ,
\end{split}
\end{align}

\noindent where a prime means evaluated at $(\tau',x)$. Once again, the linear terms in $W$ are dropped. In this case, the dissipation action takes the following form

\begin{equation}
S_{\text{Dis}}[g_s g_f] = S_{\text{Dis}}[g_s] + S_{\text{Dis}}^{(2)}[W] + S_{\text{Int,Dis}}^{(2)}[g_s,W] \, ,
\end{equation}

\noindent with 

\begin{equation}
S_{\text{Dis}}[g_s] = k^2 \gamma \int d\tau d\tau' dx \, K(\tau-\tau') \, \tr \Big(\mathds{1} - g_s(\tau,x) g_s^{-1}(\tau',x)\Big) \, ,
\end{equation}

\begin{align}
\begin{split}
S_{\text{Dis}}^{(2)}[W] &= -k^2 \gamma \int d\tau d\tau' \int dx K(\tau-\tau') \tr \Bigg( \frac{W^2}{2} + \frac{W^{\prime \, 2}}{2} - W W' \Bigg) \, ,
\end{split}
\end{align}

\begin{align}
\begin{split}
S_{\text{Int,Dis}}^{(2)}[g_s,W] = k^2 \gamma \int d\tau d\tau' \int dx K(\tau-\tau') \, \tr \Bigg[ \Big(\mathds{1}-g_s^{\prime \, -1} g_s\Big) \Bigg( \frac{W^2}{2} + \frac{W^{\prime \, 2}}{2} - W W' \Bigg) \Bigg] \, .
\end{split}
\end{align}

\noindent The second term (purely fast part) can be written in Fourier space

\begin{align}
\begin{split}
S_{\text{Dis}}^{(2)}[W] &= \frac{k^2 \gamma}{2} \int d\tau d\tau' dx \int_{\omega,\omega',\omega''} \int_{q,q'} \Tilde{K}(\omega'') \Tilde{\phi}^a(\omega,q) \Tilde{\phi}^a(\omega',q') \e^{\I \omega''(\tau-\tau')} \\ &\hspace{0.5cm}\times \Bigg[ \frac{1}{2} \e^{\I(\omega \tau + q x)} \e^{\I (\omega' \tau+q' x)} + \frac{1}{2} \e^{\I(\omega \tau' + q x)} \e^{\I (\omega' \tau'+q' x)} - \e^{\I(\omega \tau + q x)} \e^{\I (\omega' \tau'+q' x)} \Bigg] \\ &= \frac{k^2 \gamma}{2} \int_{\omega,q} \Big( \Tilde{K}(0) - \Tilde{K}(-\omega) \Big) \Tilde{\phi}^a(\omega,q) \Tilde{\phi}^a(-\omega,-q) \, ,
\end{split}
\end{align}

\noindent where $\int_{\omega} = \int \frac{d\omega}{2\pi}$, $\int_q = \int \frac{dq}{2\pi}$. In the first equality, the trace over the generators has been performed, while in the second equality, integrals over momentum/frequency $\delta$ functions have been carried out. The Fourier transform of the kernel is obtained using the general formula

\begin{equation}
\int d^dx \frac{\e^{-\I p\cdot x}}{|x|^{\beta}} = \frac{\Gamma\Big( \frac{d}{2}-\frac{\beta}{2} \Big)}{\pi^{d/2} 2^{\beta} \Gamma(\beta/2)} (2\pi)^d \frac{1}{|p|^{d-\beta}} \, ,
\end{equation}

\noindent for $d$ Euclidean dimensions. In our case, $d=1$ and $\beta = 3-\delta$ for the Fourier transform of the kernel, which yields $\Tilde{K}(\omega) = - \frac{1}{8\pi} |\omega|^{2-\delta}$. This shows that $\Tilde{K}(0) = 0$ and the fast part of the dissipation action thus becomes

\begin{align}
\begin{split}
S_{\text{Dis}}^{(2)}[W] = - \frac{k^2 \gamma}{2} \int_{\omega,q} \Tilde{K}(\omega) \Tilde{\phi}^a(\omega,q) \Tilde{\phi}^a(-\omega,-q) = - \frac{k^2 \gamma}{2} \int_{\omega,q} \Bigg( -\frac{1}{8\pi} |\omega|^{2-\delta} \Bigg) \Tilde{\phi}^a(\omega,q) \Tilde{\phi}^a(-\omega,-q)
\end{split}
\end{align}

\subsubsection{Recap} \label{sec:expand_recap_A}

As a recap, the action expanded at quadratic order in $W$ can be grouped in three terms: $S[g_s g_f] = S[g_s] + S^{(2)}[W] + S_{\text{Int}}^{(2)}[g_s,W]$. The first term is simply the initial action evaluated at $g = g_s$

\begin{align}
\begin{split}
S[g_s] &= S_{\text{Grad}}[g_s] + S_{\text{WZ}}[g_s] + S_{\text{Dis}}[g_s] \\ &= \frac{1}{\lambda} \int d\tau dx \, \tr \Bigg( \frac{1}{c^2} \partial_{\tau} g_s \partial_{\tau} g_s^{-1} + \partial_x g_s \partial_x g_s^{-1} \Bigg) + \frac{\I k}{12 \pi} \int_{B^3} \tr \Big( g_s^{-1} dg_s \wedge g_s^{-1} dg_s \wedge g_s^{-1} dg_s \Big) \\ &\hspace{0.5cm}+ k^2 \gamma \int d\tau d\tau' dx \, K(\tau-\tau') \, \tr \Big(\mathds{1} - g_s(\tau,x) g_s^{-1}(\tau',x)\Big) \, .
\end{split}
\end{align}

\noindent It contributes to the $\beta$ functions only via the final rescaling step. The second contribution to the expanded action regroups the two terms which only contain fast fields:

\begin{align}
\begin{split}
S^{(2)}[W] &= S_{\text{Grad}}^{(2)}[W] + S_{\text{Dis}}^{(2)}[W] \\ &= - \frac{1}{\lambda}\int d\tau dx \, \tr \Bigg( \frac{1}{c^2} \partial_{\tau} W \partial_{\tau} W + \partial_x W \partial_x W \Bigg) -k^2 \gamma \int d\tau d\tau' \int dx K(\tau-\tau') \tr \Bigg( \frac{W^2}{2} + \frac{W^{\prime \, 2}}{2} - W W' \Bigg) \\ &= \frac{1}{2} \int \frac{d\omega dq}{(2\pi)^2} \Tilde{\phi}^a(\omega,q) \Big( \Pi^{-1}(\omega,q) - k^2 \gamma \Tilde{K}(\omega) \Big) \Tilde{\phi}^a(-\omega,-q) \\ &= \frac{1}{2} \int \frac{d\omega dq}{(2\pi)^2} \Tilde{\phi}^a(\omega,\phi) \Tilde{G}^{-1}(\omega,q) \Tilde{\phi}^a(-\omega,-q) \, ,
\end{split}
\end{align}

\noindent where we have identified the fast propagator

\begin{equation}
\Tilde{G}(\omega,q) = \frac{\lambda}{q^2 + \frac{\omega^2}{c^2} + \frac{k^2}{8\pi} \lambda \gamma |\omega|^{2-\delta}} \, .
\end{equation}

\noindent Finally, the last piece contains all the terms mixing slow and fast modes, which are denoted as interaction terms

\begin{align}
\begin{split}
S_{\text{Int}}^{(2)}[g_s,W] &= S_{\text{Int,Grad}}^{(2)}[g_s,W] + S_{\text{Int,WZ}}^{(2)}[g_s,W] + S_{\text{Int,Dis}}^{(2)}[g_s,W] \\ &= \frac{1}{\lambda} \int d\tau dx \, \tr \Bigg( \frac{1}{c^2} g_s^{-1}\partial_{\tau} g_s [\partial_{\tau} W,W] + g_s^{-1}\partial_x g_s [\partial_x W,W] \Bigg) + \frac{\I k}{8\pi} \int d\tau dx \, \epsilon_{\mu \nu} \tr \Big( g_s^{-1} \partial_{\mu} g_s [\partial_{\nu}W,W] \Big) \\ & \hspace{0.5cm}+ k^2 \gamma \int d\tau d\tau' \int dx K(\tau-\tau') \, \tr \Bigg[ \Big(\mathds{1}-g_s^{\prime \, -1} g_s\Big) \Bigg( \frac{W^2}{2} + \frac{W^{\prime \, 2}}{2} - W W' \Bigg) \Bigg] \, .
\end{split}
\end{align}

\noindent The first two terms can be combined into a ``WZW interaction term'' $S_{\text{Int,WZW}}^{(2)}$:

\begin{align}
\begin{split}
S_{\text{Int,WZW}}^{(2)}[g_s,W] = S_{\text{Int,Grad}}^{(2)}[g_s,W] + S_{\text{Int,WZ}}^{(2)}[g_s,W] = \int d\tau dx \, \tr \Big( \Phi_{\mu}(\tau,x) [\partial_{\mu}W,W] \Big) \, ,
\end{split}
\end{align}

\noindent where

\begin{align}
\begin{split}
\Phi_{\tau}(\tau,x) = g_s^{-1} \Big( \frac{1}{c^2 \lambda} \partial_{\tau} - \frac{i k}{8\pi} \partial_x \Big) g_s \, , \qquad \Phi_x(\tau,x) = g_s^{-1} \Big( \frac{1}{\lambda} \partial_x + \frac{i k}{8\pi} \partial_{\tau} \Big) g_s \, .
\end{split}
\end{align}


\subsection{Fourier representation of interaction terms} \label{sec:Fourier_rep_A}

We now express the interaction terms, which we will average over with respect to the fast propagator, in Fourier space.

\subsubsection{WZW interaction term} \label{sec:Fourier_rep_WZW_A}

\begin{align}
\begin{split}
S_{\text{Int,WZW}}^{(2)}[g_s,W] &=  \int d\tau dx \, \tr \Big( \Phi_{\mu}(\tau,x) [\partial_{\mu}W,W] \Big) \\ &= \I \int d\tau dx \int_{p_s} \int_{p, p'} \e^{\I (p+p'+p_s)\cdot x} (p_{\mu} - p_{\mu}') \tr \Big[ \Tilde{\Phi}_{\mu}(p_s) \Tilde{W}(p) \Tilde{W}(p') \Big] \\ &= \I \int_{p_s} \int_p (2p_{\mu} + p_{s \, \mu}) \tr \Big[ \Tilde{\Phi}_{\mu}(p_s) \Tilde{W}(p) \Tilde{W}(-p-p_s) \Big] \, ,
\end{split}
\end{align}

\noindent where $p=(\omega,q)$ is a fast 2-momentum and $p_s=(\omega_s,q_s)$ is a slow 2-momentum.

\subsubsection{Dissipation interaction term} \label{sec:Fourier_rep_dis_A}

To treat the dissipation interaction term $S_{\text{Int,Dis}}^{(2)}$, let us define

\begin{equation}
D_s(\tau,\tau',x) = \mathds{1} - g_s^{-1}(\tau',x) g_s(\tau,x) = \int \frac{d \omega_s}{2\pi} \frac{d \omega'_s}{2\pi} \int \frac{dq_s}{2\pi} \tilde{D}_s(\omega_s,\omega'_s,q_s) \e^{\I (\omega_s \tau + \omega'_s \tau' + q_s x)} \, .
\end{equation}

\noindent Therefore, by Fourier transforming, we get

\begin{align}
\begin{split}
S_{\text{Int,Dis}}^{(2)}[g_s,W] &= k^2 \gamma \int d\tau d\tau' dx \int_{\omega_s,\omega_s',q_s} \int_{\omega,\omega',\Omega} \int_{q,q'} \, \Tilde{K}(\Omega) \e^{\I \Omega(\tau-\tau')} \, \tr \Bigg[ \Tilde{D}_s(\omega_s,\omega_s',q_s) \e^{\I(\omega_s \tau+\omega_s' \tau'+q_s x)} \\ &\hspace{0.5cm}\times \Bigg( \frac{1}{2} \Tilde{W}(\omega,q) \Tilde{W}(\omega',q') \e^{\I(\omega \tau+q x)} \e^{\I(\omega' \tau+q' x)} + \frac{1}{2} \Tilde{W}(\omega,q) \Tilde{W}(\omega',q') \e^{\I(\omega \tau'+q x)} \e^{\I(\omega' \tau'+q' x)} \\ &\hspace{1cm} - \Tilde{W}(\omega,q) \Tilde{W}(\omega',q') \e^{\I(\omega \tau+q x)} \e^{\I(\omega' \tau'+q' x)} \Bigg)\Bigg] \, ,
\end{split}
\end{align}

\noindent where frequencies and momenta with a subscript ``$s$'' are slow modes, while the others are fast modes, except for $\Omega$ which is unspecified for now. The space and time integrals yield $\delta$ functions over frequencies and momenta. Performing them, we arrive at 

\begin{equation}
S_{\text{Int,Dis}}^{(2)}[g_s,W] = T_1 + T_2 + T_3 \, ,  \label{eq:diss_t1t2t3}
\end{equation}

\noindent where

\begin{align}
\begin{split}
T_1 = \frac{k^2\gamma}{2} \int_{\omega_s, \omega'_s,q_s} \int_{\omega, q} \tilde{K}(\omega'_s) \, \tr \Big[ \tilde{D}_s(\omega_s,\omega'_s,q_s) \tilde{W}(\omega,q) \tilde{W}(-\omega_s-\omega'_s-\omega,-q-q_s) \Big] \, ,
\end{split}
\end{align}

\begin{align}
\begin{split}
T_2 = \frac{k^2\gamma}{2} \int_{\omega_s, \omega'_s,q_s} \int_{\omega, q} \tilde{K}(\omega_s) \, \tr \Big[ \tilde{D}_s(\omega_s,\omega'_s,q_s) \tilde{W}(\omega,q) \tilde{W}(-\omega_s-\omega'_s-\omega,-q-q_s) \Big] \, ,
\end{split}
\end{align}

\begin{align}
\begin{split}
T_3 = -k^2\gamma \int_{\omega_s, \omega'_s,q_s} \int_{\omega, q} \tilde{K}(\omega_s+\omega) \, \tr \Big[ \tilde{D}_s(\omega_s,\omega'_s,q_s) \tilde{W}(\omega,q) \tilde{W}(-\omega_s-\omega'_s-\omega,-q-q_s) \Big] \, .
\end{split}
\end{align}

\subsubsection{Diagrammatic representation} \label{sec:Fourier_rep_diagrams}

\begin{figure}[H]
	\centering
	\includegraphics[width=0.9\hsize]{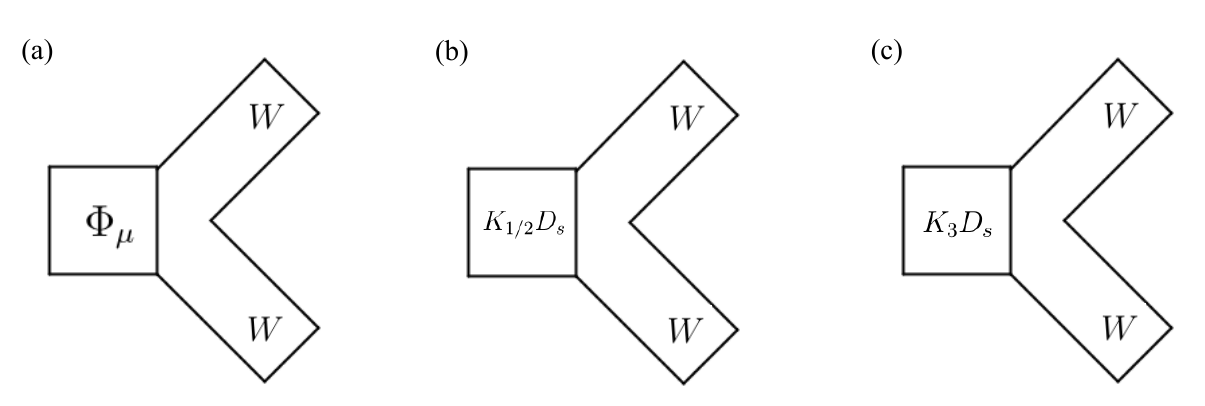}
	\caption{Diagrammatic representation of the three different types of terms in the interaction action. a) Representation of $S_{\text{Int,WZW}}^{(2)}[g_s,W]$, where the square corresponds to the slow object $\Phi_{\mu}$. b) Representation of $T_1$ and $T_2$. These two terms are essentially the same, since the Fourier transform of their kernel only contains slow modes. The square depicts an insertion of the kernel from either $T_1$ or $T_2$, times the purely slow object $D_s$. c) Representation of $T_3$. The square corresponds to an insertion of the kernel in $T_3$, which mixes slow and fast modes, times $D_s$. }
	\label{fig:vertex}
\end{figure}

The interaction terms presented in the two previous sections can be represented diagrammatically in terms of the vertices presented in Fig. \ref{fig:vertex}.

\noindent In each vertex, the square represents the part of the interaction action containing slow modes. Since the action has been expanded to quadratic order in $W$, each vertex contains two $W$ insertions, represented as double lines, which can be seen as the two matrix indices of $W$.


\subsection{Integration of fast modes}\label{sec:integration_fast_A}

We are now in a position to integrate the fast modes. To do so, we proceed with a cumulant expansion.

\begin{equation}
S_{\text{Eff}}[g_s] \approx S[g_s] + \ev{S_{\text{Int}}^{(2)}[g_s,W]}_f - \frac{1}{2} \ev{(S^{(2)}_{\text{Int}}[g_s,W])^2}_f^c + ... \, ,
\end{equation}

\noindent where the expectation value is taken with respect to the fast modes, while $c$ stands for connected correlation function. We perform the RG calculation at one-loop, which is controlled using a large-$k$ expansion. This requires the couplings $\lambda$ and $\gamma$ to be of order $1/k$ as well as $\delta$, which justifies the introduction of the $\mathcal{O}(k^0)$ parameters $\Tilde{\lambda} = k \lambda$, $\Tilde{\gamma} = k \gamma$ and $\Tilde{\delta} = k \delta$.

With $\e^{W}$ expanded to quadratic order in $W$, only one-loop diagrams are generated, as we can see from the vertices of Fig. \ref{fig:vertex}. Moreover, it is clear that diagrams at order $n$ in the cumulant expansion contain $n$ vertices. Two-loop diagrams can be obtained by expanding to higher powers in $W$. However, these terms will be suppressed with additional powers of $1/k$. This comes from the fact that every vertex is of order $k$, but the propagator is of order $1/k$. Hence, the order in $1/k$ of a diagram is given by $n_{p} - n_{v}$ (respectively the number of propagators and the number of vertices). However, $n_{p} - n_{v} = n_{l} - 1$, where $n_l$ is the number of loops in a given diagram. Therefore, the order in $1/k$ of a diagram is directly related to the number of loops it has.

\subsubsection{Order 1 in interaction action}\label{sec:integration_fast_order1_A}

Let us start by evaluating the first expectation value

\begin{equation}
\ev{S_{\text{Int}}^{(2)}[g_s,W]}_f = \ev{S_{\text{Int,Dis}}^{(2)}[g_s,W]}_f + \ev{S_{\text{Int,WZW}}^{(2)}[g_s,W]}_f \, .
\end{equation}

\vspace{0.5cm}
\noindent\underline{\textbf{Dissipation term:}} The expectation value of the dissipation term is separated into the expectation value of its three pieces (see Eq. (\ref{eq:diss_t1t2t3}) above)

\begin{equation}
\ev{S_{\text{Int,Dis}}^{(2)}[g_s,W]}_f = \ev{T_1}_f + \ev{T_2}_f + \ev{T_3}_f \, ,
\end{equation}

\noindent which can be represented by the following three Feynman diagrams

\begin{figure}[H]
	\centering
	\includegraphics[width=0.9\hsize]{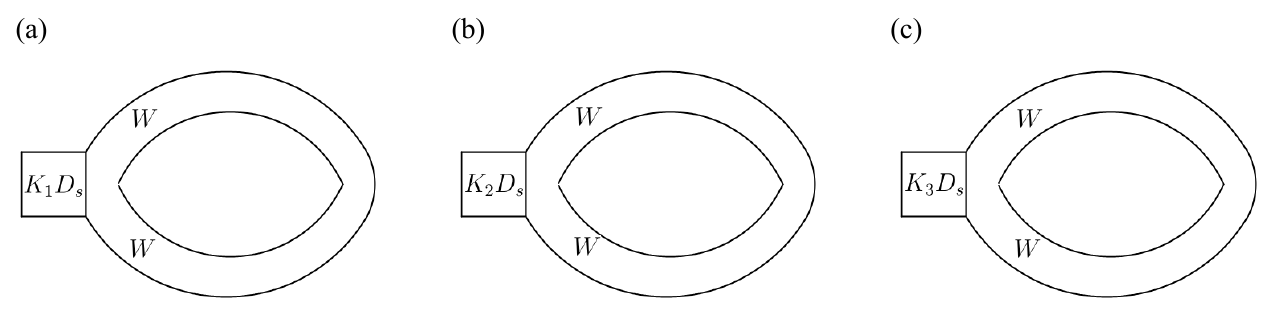}
	\caption{one-loop Feynman diagrams contributing to $\ev{S_{\text{Int,Dis}}^{(2)}[g_s,W]}_f$. (a), (b) and (c) correspond respectively to $\ev{T_1}_f$, $\ev{T_2}_f$ and $\ev{T_3}_f$.}
	\label{fig:diagrams_O(1)_1}
\end{figure}

\noindent For the first term, we have

\begin{align}
\begin{split}
\ev{T_1}_f &= \frac{k^2 \gamma}{2} \int_{\omega_s \omega_s' q_s} \int_{\omega,q} \tilde{K}(\omega_s') \tr \Big[ 
\tilde{D}_s(\omega_s,\omega_s',q_s) \ev{\tilde{W}(\omega,q) \tilde{W}(-\omega-\omega_s-\omega_s',-q-q_s)}_f \Big] \\ &= i^2 \frac{k^2 \gamma}{2} \int_{\omega_s \omega_s' q_s} \int_{\omega,q} \tilde{K}(\omega_s') \tr \Big[ 
\tilde{D}_s(\omega_s,\omega_s',q_s) T^a T^b\Big] \ev{\tilde{\phi}^a(\omega,q) \tilde{\phi}^b(-\omega-\omega_s-\omega_s',-q-q_s)}_f \, .
\end{split}
\end{align}

\noindent The expectation value yields a single Wick contraction

\begin{equation}
\ev{\tilde{\phi}^a(\omega,q) \tilde{\phi}^b(-\omega-\omega_s-\omega_s',-q-q_s)}_f = \delta^{ab} \tilde{G}(\omega,q) (2\pi)^2 \delta(\omega_s+\omega_s') \delta(q_s) \, ,
\end{equation}

\noindent from which we get

\begin{align}
\begin{split}
\ev{T_1}_f &= - \frac{k^2 \gamma}{2} \int_{\omega,q} \tilde{G}(\omega,q) \int_{\omega_s} \tilde{K}(\omega_s) \tr \Big[ 
\Tilde{D}_s(\omega_s,-\omega_s,0) T^a T^a\Big] \\ &= - \frac{k^2 \gamma}{4} \Big( N - \frac{1}{N} \Big) \int_{\omega,k} \tilde{G}(\omega,k) \int_{\omega_s} \tilde{K}(\omega_s) \, \tr \Big[ \Tilde{D}_s(\omega_s,-\omega_s,0) \Big] \\ &= - \frac{k^2\gamma}{2} C_F \, I_1 \int d\tau d\tau' \int dx \, K(\tau-\tau') \, \tr \Big( \mathds{1} - g_s^{\prime \, -1} g_s \Big) \, ,
\end{split}
\end{align}

\noindent where the trace has been simplified using the SU$(N)$ completeness relation $T^a_{ij} T^a_{kl} = \frac{1}{2} \Big( \delta_{il} \delta_{jk} - \frac{1}{N} \delta_{ij} \delta_{kl} \Big)$. We have also defined the SU$(N)$ quadratic Casimir in the fundamental representation $C_F = \frac{N^2-1}{2N}$ and the fast integral

\begin{equation}
I_1 = \int \frac{d\omega dq}{(2\pi)^2} \Tilde{G}(\omega,q) = \int \frac{d\omega dq}{(2\pi)^2} \frac{\lambda}{q^2 + \frac{\omega^2}{c^2} + \frac{k^2}{8\pi} \lambda \gamma |\omega|^{2-\delta}} \, . 
\end{equation}

\noindent In the last step, the following inverse Fourier transform has been employed

\begin{align}
\begin{split}
\int_{\omega_s} \tilde{K}(\omega_s) \, \tr \Big[ \Tilde{D}_s(\omega_s,-\omega_s,0) \Big] &= \int d\tau d\tau' d\tau'' dx \int_{\omega_s} K(\tau'') \tr \Big( D_s(\tau,\tau',x) \Big) \e^{\I \omega_s(\tau'-\tau-\tau'')} \\ &= \int d\tau d\tau' dx K(\tau-\tau') \tr \Big( D(\tau,\tau',x) \Big) \\ &= \int d\tau d\tau' \int dx \, K(\tau-\tau') \, \tr \Big( \mathds{1} - g_s^{\prime \, -1} g_s \Big)
\end{split}
\end{align}

\noindent By performing a very similar calculation, one can show that $\ev{T_2}_f = \ev{T_1}_f$. For $T_3$, using the above result for the expectation value of the fast modes, we get

\begin{align}
\begin{split}
\ev{T_3}_f &= \frac{k^2 \gamma}{2} \Big( N - \frac{1}{N} \Big) \int_{\omega,q} \tilde{G}(\omega,q) \int_{\omega_s} \tilde{K}(\omega+\omega_s) \, \tr \Big[ \Tilde{D}_s(\omega_s,-\omega_s,0) \Big] \, .
\end{split}
\end{align}

\noindent The kernel is now expanded to quadratic order in $\omega_s$

\begin{equation}
\Tilde{K}(\omega+\omega_s) = -\frac{1}{8\pi} \Bigg( |\omega|^{2-\delta} + (2-\delta) \frac{\omega}{|\omega|^{\delta}} \omega_s + \frac{1}{2}(2-\delta)(1-\delta) \frac{\omega_s^2}{|\omega|^{\delta}} \Bigg) + \mathcal{O}(\omega_s^3) \, ,
\end{equation}

\noindent Clearly, the contribution from the second term vanishes since the fast integrand is odd under $\omega \rightarrow -\omega$. Moreover, the contribution from the first term can also be shown to vanish

\begin{align}
\begin{split}
\int_{\omega_s} \tr \Big( \Tilde{D}_s(\omega_s,-\omega_s,0) \Big) &= \int d\tau d\tau' dx \int_{\omega_s} \tr \Big( D_s(\tau,\tau',x) \Big) \e^{\I \omega_s (\tau'-\tau)} \\ &= \int d\tau dx \tr \Big( D_s(\tau,\tau,x) \Big) \\ &= 0 \, ,
\end{split}
\end{align}

\noindent since $D_s(\tau,\tau,x) = \mathds{1} - g_s(\tau,x)g^{-1}_s(\tau,x) = 0$. Hence, only the quadratic term in $\omega_s$ survives. Therefore

\begin{align}
\begin{split}
\ev{T_3}_f &= -\frac{k^2\gamma C_F}{16\pi} (2-\delta) (1-\delta) \int_{\omega,q} \frac{\Tilde{G}(\omega,q)}{|\omega|^{\delta}} \int_{\omega_s} \omega_s^2 \tr \Big( \Tilde{D}_s(\omega_s,-\omega_s,0) \Big) \, .
\end{split}
\end{align}

\noindent This is proportional to $\int d\tau dx \tr \Big( \partial_{\tau} g_s \partial_{\tau} g_s^{-1} \Big)$, as can be seen from the following manipulations

\begin{align}
\begin{split}
\int_{\omega_s} \omega_s^2 \tr \Big( \Tilde{D}_s(\omega_s,-\omega_s,0) \Big) &= \int d\tau d\tau' dx \int_{\omega_s} \tr \Big( D_s(\tau,\tau',x) \Big) \omega_s^2 \e^{\I \omega_s(\tau'-\tau)} \\ &= - \int d\tau d\tau' dx \int_{\omega_s} \tr \Big( D_s(\tau,\tau',x) \Big) \partial_{\tau}^2 \e^{\I \omega_s(\tau'-\tau)} \\ &= - \int d\tau d\tau' dx \int_{\omega_s} \tr \Big( \partial_{\tau}^2 D_s(\tau,\tau',x) \Big) \e^{\I \omega_s(\tau'-\tau)} \\ &= - \int d\tau d\tau' dx \tr \Big( \partial_{\tau}^2 D_s(\tau,\tau',x) \Big) \delta(\tau'-\tau) \\ &= \int d\tau d\tau' dx \tr \Big( g_s^{-1}(\tau',x) \partial_{\tau}^2 g_s(\tau,x) \Big) \delta(\tau'-\tau) \\ &= - \int d\tau dx \tr \Big( \partial_{\tau} g_s \partial_{\tau} g_s^{-1} \Big) \, ,
\end{split}
\end{align}

\noindent where $\omega_s^2$ has been replaced by $-\partial_{\tau}^2$ acting on the exponential, while integration by parts has also been used twice. Therefore

\begin{align}
\begin{split}
\ev{T_3}_f &= \frac{k^2\gamma C_F}{16\pi} (2-\delta) (1-\delta) \int_{\omega,q} \frac{\Tilde{G}(\omega,q)}{|\omega|^{\delta}} \int d\tau dx \tr \Big( \partial_{\tau} g_s \partial_{\tau} g_s^{-1} \Big) \, .
\end{split}
\end{align}

\begin{figure}[H]
	\centering
	\includegraphics[width=0.4\hsize]{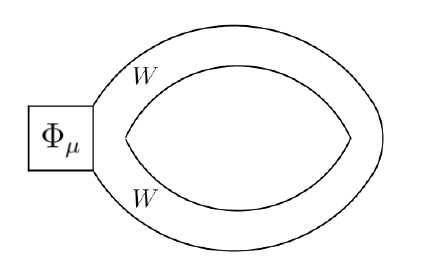}
	\caption{one-loop Feynman diagram contributing to $\ev{S_{\text{Int,WZW}}^{(2)}[g_s,W]}_f$.}
	\label{fig:diagrams_O(1)_2}
\end{figure}

\vspace{0.5cm}
\noindent\underline{\textbf{WZW term:}} We now move to the expectation value of the WZW interaction term, corresponding to the Feynman diagrams shown in Fig. \ref{fig:diagrams_O(1)_2}.

\noindent The calculation of the diagram yields

\begin{align}
\begin{split}
\ev{S_{\text{Int,WZW}}^{(2)}[g_s,W]}_f &= \I \int_{p_s} \int_{p} (2p_{\mu} + p_{s\, \mu}) \, \tr \Big( \ev{\Tilde{\Phi}_{\mu}(p_s) \Tilde{W}(p) \Tilde{W}(-p - p_s)}_f \Big) \\ &= -\I \int_{p_s} \int_p (2p_{\mu} + p_{s\, \mu}) \tr \Big( \Tilde{\Phi}_{\mu}(p_s) T^a T^b \Big) \ev{\Tilde{\phi}^a(p) \Tilde{\phi}^b(-p-p_s)} \\ &= -\I \int_{p_s} \int_p (2p_{\mu} + p_{s\, \mu}) \tr \Big( \Tilde{\Phi}_{\mu}(p_s) T^a T^a \Big) (2\pi)^2 \delta^{(2)}(p_s) \Tilde{G}(p) \\ &= -2\I \int_p p_{\mu} \Tilde{G}(p) \tr \Big( \Tilde{\Phi}_{\mu}(0) T^a T^a \Big) \\ &= 0 \, ,
\end{split}
\end{align}

\noindent where we still have $p=(\omega,q)$, $p_s = (\omega_s,q_s)$. The above expression vanishes for two reasons. First, the integral over fast modes vanishes due to an odd integrand. Secondly, simplifying the trace using the $\text{SU}(N)$ completeness relation yields a trace of $\Phi_{\mu}$, which vanishes. This can be shown by writing $g = v \mathds{1} + \I N^a T^a$, with $\Vec{n} = (v,\Vec{N})^T$, $\Vec{n} \cdot \Vec{n} = 1$. In this case, $\tr \Phi_{\mu} \sim \Vec{n} \cdot \partial_{\mu} \Vec{n} = 0$, since $\Vec{n}$ is perpendicular to its derivative.

\vspace{0.5cm}
\noindent\underline{\textbf{Recap:}} Therefore, the expectation value of the interaction action is

\begin{align}
\begin{split}
\ev{S_{\text{Int}}^{(2)}[g_s,W]}_f &= \ev{T_1}_f + \ev{T_2}_f + \ev{T_3}_f \\ &= -k^2\gamma C_F \, I_1 \int d\tau d\tau' \int dx \, K(\tau-\tau') \, \tr \Big( \mathds{1} - g_s^{\prime \, -1} g_s \Big) \\ &\hspace{0.5cm}+\frac{k^2\gamma C_F}{16\pi} (2-\delta) (1-\delta) \int_{\omega,q} \frac{\Tilde{G}(\omega,q)}{|\omega|^{\delta}} \int d\tau dx \tr \Big( \partial_{\tau} g_s \partial_{\tau} g_s^{-1} \Big) \, .
\end{split}
\end{align}

\subsubsection{Order 2 in interaction action}\label{sec:integration_fast_order2_A}

We now move to the term quadratic in the interaction action in the cumulant expansion. There are three terms to consider

\begin{equation} \label{eq:S2Exp}
\ev{(S^{(2)}_{\text{Int}}[g_s,W])^2}_f^c = \ev{(S_{\text{Int,WZW}}^{(2)})^2}_f^c + \ev{(S_{\text{Int,Dis}}^{(2)})^2}_f^c + 2 \ev{S_{\text{Int,Dis}}^{(2)} S_{\text{Int,WZW}}^{(2)}}_f^c \, . 
\end{equation}

\begin{figure}[H]
	\centering
	\includegraphics[width=0.4\hsize]{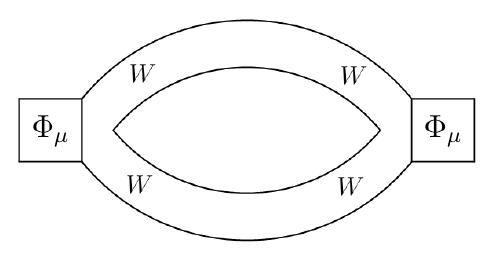}
	\caption{one-loop Feynman diagram contributing to $\ev{(S^{(2)}_{\text{Int,WZW}})^2}_f^c$.}
	\label{fig:diagrams_O(2)_2}
\end{figure}

\vspace{0.5cm}
\noindent\underline{\textbf{Squared WZW term:}} We start with the expectation value of the squared WZW interaction term, the diagrammatic representation of which is shown in Fig. \ref{fig:diagrams_O(2)_2}.

\noindent The diagram corresponds to

\begin{align}
\begin{split}
\ev{(S_{\text{Int,WZW}}^{(2)})^2}_f^c &= -\int_{p_s,p_s'} \int_{p,p'} (2p_{\mu} + p_{s\, \mu}) (2p'_{\nu} + p'_{s\, \nu}) \Big\langle\tr \Big( \tilde{\Phi}_{\mu}(p_s) \tilde{W}(p) \tilde{W}(-p-p_s) \Big) \\ &\hspace{4cm}\times \tr \Big( \tilde{\Phi}_{\nu}(p_s') \tilde{W}(p') \tilde{W}(-p'-p_s') \Big) \Big\rangle_f^c \\ &\approx -4 \int_{p_s,p_s'} \int_{p,p'} p_{\mu}p'_{\nu} \Big\langle\tr \Big( \tilde{\Phi}_{\mu}(p_s) \tilde{W}(p) \tilde{W}(-p-p_s) \Big) \\ &\hspace{4cm}\times \tr \Big( \tilde{\Phi}_{\nu}(p_s') \tilde{W}(p') \tilde{W}(-p'-p_s') \Big) \Big\rangle_f^c \, ,
\end{split}
\end{align}

\noindent where $p = (\omega,q)$, $p_s = (\omega_s,q_s)$, $p' = (\omega',q')$, $p'_s = (\omega'_s,q'_s)$. Note that $p_{s\, \mu}$ and $p'_{s\, \nu}$ have been dropped since the expression is already quadratic in derivatives (from the two $\Phi_{\mu}$). Slow modes, when expressed in real space, correspond to derivatives, which means even more irrelevant terms. Let us focus our attention on the expectation value

\begin{align} \label{eq:Expectation_value_WZW2}
\begin{split}
& \ev{\tr \Big( \tilde{\Phi}_{\mu}(p_s) \tilde{W}(p) \tilde{W}(-p-p_s) \Big) \tr \Big( \tilde{\Phi}_{\nu}(p_s') \tilde{W}(p') \tilde{W}(-p'-p_s') \Big)}_f^c \\ &= \ev{\tilde{\phi}^a(p) \tilde{\phi}^b(-p-p_s) \tilde{\phi}^c(p') \tilde{\phi}^d(-p'-p_s')}_f^c \tr \Big( \tilde{\Phi}_{\mu}(p_s) T^a T^b \Big) \tr \Big( \tilde{\Phi}_{\nu}(p_s') T^c T^d \Big) \, .
\end{split}
\end{align}

\noindent The expectation value is computed using Wick contractions. There are two connected pieces, denoted as $W_1$ and $W_2$. First, let us consider $W_1$,

\begin{align}
\begin{split}
W_1 &= (2\pi)^4 \delta^{(2)}(p+p') \delta^{(2)}(p+p'+p_s+p_s') \delta_{ac} \delta_{bd} \Tilde{G}(p) \Tilde{G}(p+p_s) \tr \Big( \tilde{\Phi}_{\mu}(p_s) T^a T^b \Big) \tr \Big( \tilde{\Phi}_{\nu}(p_s') T^c T^d \Big) \\ &= (2\pi)^4 \delta^{(2)}(p+p') \delta^{(2)}(p+p'+p_s+p_s') \Tilde{G}(p) \Tilde{G}(p+p_s) \tr \Big( \tilde{\Phi}_{\mu}(p_s) T^a T^b \Big) \tr \Big( \tilde{\Phi}_{\nu}(p_s') T^a T^b \Big) \, .
\end{split}
\end{align}

\noindent The traces are computed using the the completeness relation for the SU$(N)$ generators which leads to

\begin{align}
\begin{split}
\tr \Big( \tilde{\Phi}_{\mu}(p_s) T^a T^b \Big) \tr \Big( \tilde{\Phi}_{\nu}(p_s') T^a T^b \Big) &= \tilde{\Phi}^{ij}_{\mu}(p_s) \tilde{\Phi}^{lm}_{\nu}(p_s') \Big( \frac{1}{2} \delta_{jn} \delta_{km} - \frac{1}{2N} \delta_{jk} \delta_{mn} \Big) \Big( \frac{1}{2} \delta_{kl} \delta_{in} - \frac{1}{2} \delta_{ik} \delta_{nl} \Big) \\ &= -\frac{1}{2N} \tr \Big( \tilde{\Phi}_{\mu}(p_s) \tilde{\Phi}_{\nu}(p_s') \Big) + \frac{1}{4} \Big( 1 + \frac{1}{N^2} \Big) \tr \Big( \tilde{\Phi}_{\mu}(p_s) \Big) \, \tr \Big( \tilde{\Phi}_{\nu}(p_s') \Big) \\ &= -\frac{1}{2N} \tr \Big( \tilde{\Phi}_{\mu}(p_s) \tilde{\Phi}_{\nu}(p_s') \Big) \, ,
\end{split}
\end{align}

Next, consider $W_2$,

\begin{align}
\begin{split}
W_2 &= (2\pi)^4 \delta^{(2)}(p-p'-p_s') \delta^{(2)}(p'-p-p_s) \delta_{ad} \delta_{bc} \Tilde{G}(p+p_s) \Tilde{G}(p) \tr \Big( \tilde{\Phi}_{\mu}(p_s) T^a T^b \Big) \tr \Big( \tilde{\Phi}_{\nu}(p_s') T^c T^d \Big) \\ &= (2\pi)^4 \delta^{(2)}(p-p'-p_s') \delta^{(2)}(p'-p-p_s) \Tilde{G}(p+p_s) \Tilde{G}(p) \tr \Big( \tilde{\Phi}_{\mu}(p_s) T^a T^b \Big) \tr \Big( \tilde{\Phi}_{\nu}(p_s') T^b T^a \Big) \, .
\end{split}
\end{align}

\noindent Computing the traces leads to

\begin{align} \label{eq:trace2}
\begin{split}
\tr \Big( \tilde{\Phi}_{\mu}(p_s) T^a T^b \Big) \tr \Big( \tilde{\Phi}_{\nu}(p_s') T^b T^a \Big) &= \tilde{\Phi}_{\mu}^{ij}(p_s) \tilde{\Phi}^{lm}_{\nu}(p_s') \Big( \frac{1}{2} \delta_{jl} \delta_{kn} - \frac{1}{2N} \delta_{jk} \delta_{nl} \Big) \Big( \frac{1}{2} \delta_{kn} \delta_{im} - \frac{1}{2N} \delta_{ki} \delta_{mn} \Big) \\ &= \Big( \frac{N}{4} - \frac{1}{2N} \Big) \tr \Big( \tilde{\Phi}_{\mu}(p_s) \tilde{\Phi}_{\nu}(p_s') \Big) + \frac{1}{4N^2} \tr \Big( \tilde{\Phi}_{\mu}(p_s) \Big) \, \tr \Big( \tilde{\Phi}_{\nu}(p_s') \Big) \\ &= \Big( \frac{N}{4} - \frac{1}{2N} \Big) \tr \Big( \tilde{\Phi}_{\mu}(p_s) \tilde{\Phi}_{\nu}(p_s') \Big) \, .
\end{split}
\end{align}

\noindent Combining everything and integrating over the $\delta$ functions yields

\begin{align}
\begin{split}
\ev{(S_{\text{Int,WZW}}^{(2)})^2}_f^c &= -4 \int_{p_s} \int_p p_{\mu} \Tilde{G}(p) \Tilde{G}(p+p_s) \, \tr \Big( \Tilde{\Phi}_{\mu}(p_s) \Tilde{\Phi}_{\nu}(-p_s) \Big) \Bigg[ \frac{1}{2N} p_{\nu} + \Big( \frac{N}{4} - \frac{1}{2N} \Big) (p_{\nu} + p_{s\, \nu}) \Bigg] \\ &\approx -N \int_{p_s} \int_p p_{\mu} p_{\nu} \Tilde{G}^2(p) \, \tr \Big( \Tilde{\Phi}_{\mu}(p_s) \Tilde{\Phi}_{\nu}(-p_s) \Big) \\ &= -N \int \frac{d\omega dq}{(2\pi)^2} (\omega,q)_{\mu} (\omega,q)_{\nu} \Tilde{G}^2(\omega,q) \, \int d\tau dx \tr \Big( \Phi_{\mu}(\tau,x) \Phi_{\nu}(\tau,x) \Big) \, ,
\end{split}
\end{align}

\noindent where slow modes have once again been neglected compared to fast modes. Note that the fast integral vanishes if $\mu \neq \nu$. Therefore

\begin{align}
\begin{split}
\ev{(S_{\text{Int,WZW}}^{(2)})^2}_f^c &= -N \int \frac{d\omega dq}{(2\pi)^2} \omega^2 \Tilde{G}^2(\omega,q) \, \int d\tau dx \tr \Big( \Phi_{\tau}(\tau,x) \Phi_{\tau}(\tau,x) \Big) \\ &\hspace{0.5cm}-N \int \frac{d\omega dq}{(2\pi)^2} q^2 \Tilde{G}^2(\omega,q) \, \int d\tau dx \tr \Big( \Phi_{x}(\tau,x) \Phi_{x}(\tau,x) \Big) \\ &= -N I_2 \int d\tau dx \tr \Big( \Phi_{\tau}(\tau,x) \Phi_{\tau}(\tau,x) \Big) - N I_3 \int d\tau dx \tr \Big( \Phi_{x}(\tau,x) \Phi_{x}(\tau,x) \Big) \, ,
\end{split}
\end{align}

\noindent where we have defined the following fast integrals

\begin{align}
\begin{split}
I_2 &= \int \frac{d\omega dq}{(2\pi)^2} \omega^2 \Tilde{G}^2(\omega,q) = \int \frac{d\omega dq}{(2\pi)^2} \omega^2 \frac{\lambda^2}{(q^2 + \omega^2/c^2 + \frac{k^2}{8\pi} \lambda \gamma |\omega|^{2-\delta})^2} \\ I_3 &= \int \frac{d\omega dq}{(2\pi)^2} q^2 \Tilde{G}^2(\omega,q) = \int \frac{d\omega dq}{(2\pi)^2} q^2 \frac{\lambda^2}{( q^2 + \omega^2/c^2 + \frac{k^2}{8\pi} \lambda \gamma |\omega|^{2-\delta})^2} \, .
\end{split}
\end{align}

\noindent Using the expressions for $\Phi_{\tau}$ and $\Phi_x$ to simplify the traces and regrouping similar terms, we get

\begin{align}\label{eq:SIntWZW2Expectation_value}
\begin{split}
\ev{(S_{\text{Int,WZW}}^{(2)})^2}_f^c &= \frac{N}{c^4 \lambda^2} \Bigg( I_2 - \frac{k^2 c^4\lambda^2}{(8\pi)^2} I_3 \Bigg) \int d\tau dx \tr \Big( \partial_{\tau} g_s \partial_{\tau} g_s^{-1} \Big) \\ &\hspace{0.5cm}+ \frac{N}{\lambda^2} \Bigg( I_3 - \frac{k^2 \lambda^2}{(8\pi)^2} I_2 \Bigg) \int d\tau dx \tr \Big( \partial_x g_s \partial_x g_s^{-1} \Big) \\ &\hspace{0.5cm}+ N \frac{\I k}{4\pi} \Bigg( \frac{1}{\lambda} I_3 - \frac{1}{c^2\lambda} I_2 \Bigg) \int d\tau dx \tr \Big(\partial_{\tau} g_s \partial_x g_s^{-1}\Big) \, ,
\end{split}
\end{align}

\noindent The first two  terms contribute to the renormalization of the gradient term in the action. However, the third term is unphysical. Indeed, as pointed out in the main text, the operator $\tr \Big(\partial_{\tau} g_s \partial_x g_s^{-1}\Big)$ breaks a symmetry from the original action, since it is not invariant under $g(\tau,x) \rightarrow g^{-1}(\tau,-x)$. This term is generated due to the fact that when performing the splitting of the degrees of freedom using the decomposition $g = g_s g_f$, this symmetry is ``fractionalized'' between the slow and fast modes and is effectively lost when the latter are integrated out. For the purpose of the RG analysis, this unphysical term can thus be dropped from the effective action.

However, by doing the ``opposite'' decomposition, that is $g = g_f g_s$, one can easily show that the expanded action to quadratic order in $W$ is essentially the same as the one derived above, but with the important difference that the sign of  $S_{\text{Int,WZ}}^{(2)}[g_s,W]$ reverses, that is

\begin{align}
\begin{split}
S_{\text{WZ}}[g_f g_s] &= S_{\text{WZ}}[g_s] +  S_{\text{Int,WZ}}^{\prime \, (2)}[g_s,W] \\ &= S_{\text{WZ}}[g_s] - \frac{\I k}{8\pi} \int d\tau dx \, \epsilon_{\mu \nu} \tr \Big( g_s \partial_{\mu} g_s^{-1} [\partial_{\nu}W,W] \Big) \, ,
\end{split}
\end{align}

\noindent which is equivalent to the replacement $k \rightarrow -k$ (strictly speaking, there are also a few other minor differences, such as $g_s \partial_{\mu} g_s^{-1}$ instead of $g_s^{-1} \partial_{\mu} g_s$, but these do not affect the renormalization of any physical term). Hence, doing the RG with this new decomposition yields the same expression as Eq. (\ref{eq:SIntWZW2Expectation_value}), but with a relative negative sign in the third term. Therefore, by defining the symmetrized effective action $S_{\text{Eff}}[g_s] = \frac{1}{2} \big( S_{\text{Eff}}^{(1)}[g_s] + S_{\text{Eff}}^{(2)}[g_s] \big)$, where $ S_{\text{Eff}}^{(1)}[g_s]$ is obtained using $g = g_s g_f$ and $ S_{\text{Eff}}^{(2)}[g_s]$ comes from using $g = g_f g_s$, the unphysical terms cancel, leaving an effective action containing only terms allowed by symmetries.
%

\begin{figure}[H]
	\centering
	\includegraphics[width=0.9\hsize]{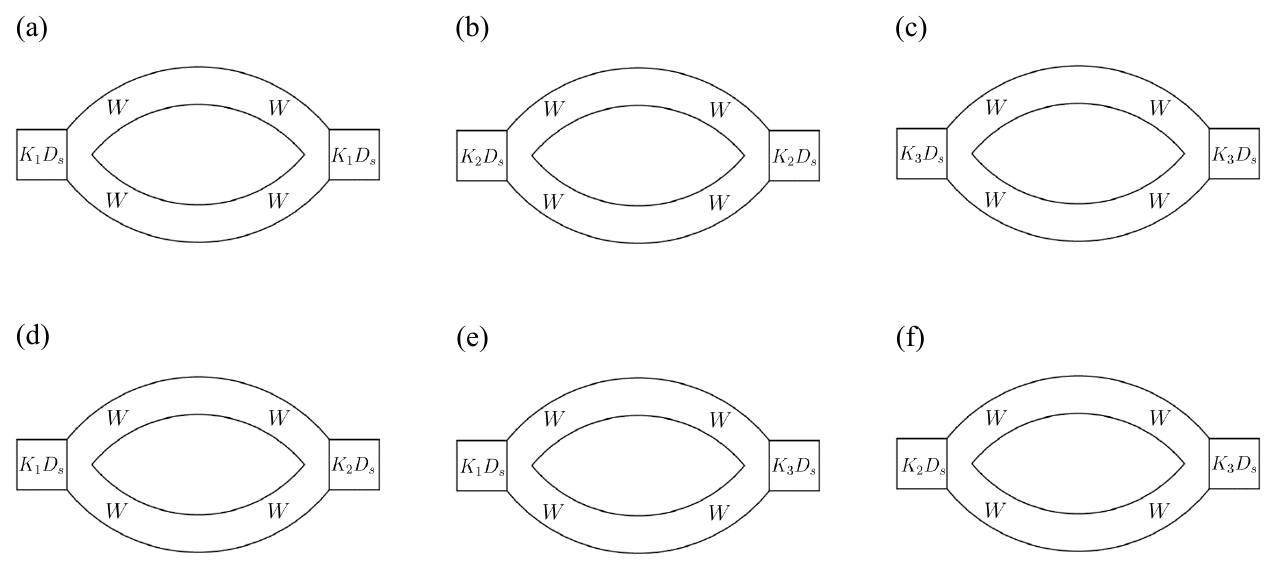}
	\caption{one-loop Feynman diagrams contributing to $\ev{(S_{\text{Int,Dis}}^{(2)})^2}_f^c$.}
	\label{fig:diagrams_O(2)_1}
\end{figure}

\vspace{0.5cm}
\noindent\underline{\textbf{Squared dissipation term:}} Let us next consider the square of the dissipation term,

\begin{align}
\begin{split}
\ev{(S_{\text{Int,Dis}}^{(2)})^2}_f^c &= \ev{(T_1+T_2+T_3)^2}_f^c \\ &= \ev{T_1^2}_f^c + \ev{T_2^2}_f^c + \ev{T_3^2}_f^c + 2 \ev{T_1 T_2}_f^c + 2 \ev{T_1 T_3}_f^c + 2 \ev{T_2 T_3}_f^c \, ,
\end{split}
\end{align}

\noindent which can be represented diagrammatically by Fig. \ref{fig:diagrams_O(2)_1}.

\noindent From the Fourier-space expressions of $T_1$ and $T_2$, we see that the first three terms will contain two slow kernels. Therefore, terms with three time integrals and two kernels will be generated. An example of such a term is

\begin{equation} \label{eq:T_1^2}
\int d\tau d\tau' d\tau'' dx K(\tau) K(\tau') \tr \Bigg[ \Big( \mathds{1} - g_s^{-1}(\tau''-\tau) g_s(\tau'') \Big) \Big( \mathds{1} - g_s^{-1}(\tau''+\tau') g_s(\tau'') \Big) \Bigg] \, ,
\end{equation}

\noindent where the fields' $x$-dependence is implicit. Let us now analyze the relevance of this term compared to the terms in the initial action. To do so, we apply the rescaling $x \rightarrow b x$, $\tau \rightarrow b^z \tau$, where $b>0$ and $z$ is the dynamical critical exponent. Using this, we have

\begin{align}
\begin{split}
&\int d\tau dx  \tr \Big( \partial_{\tau} g_s \partial_{\tau} g_s^{-1} \Big) \sim b^{1-z} \\ &\int d\tau dx  \tr \Big( \partial_{x} g_s \partial_{x} g_s^{-1} \Big) \sim b^{z-1} \\ &\int d\tau d\tau' dx \frac{1}{|\tau-\tau'|^{3-\delta}}  \tr \Big( \mathds{1} - g_s^{\prime \, -1} g_s \Big) \sim b^{1+z(\delta-1)} \, ,
\end{split}
\end{align}

\noindent (we can take the naive vanishing scaling dimension for the fields since $\Delta_g>0$ makes terms even more irrelevant). By performing the same rescaling for Eq. (\ref{eq:T_1^2}), we see that it goes as $b^{1+z(2\delta-3)}$. Therefore, for $\delta < 1$ (which is required for our controlled large-$k$ expansion), this term is less relevant then the terms in the initial action and thus can be neglected.

Let us now move on to the two terms $\ev{T_1 T_3}_f^c$ and $\ev{T_2 T_3}_f^c$. $T_1$ and $T_2$ contribute with a slow kernel, while $T_3$ gives a mixed kernel containing slow and fast modes. The mixed kernel needs to be expanded in powers of $\omega_s$ as in the calculation of $\ev{T_3}_f$. Therefore, the resulting contributions will be like the initial dissipation term, but with additional time derivatives. For example, at order $\omega_s^2$ (the first non-vanishing order), we would have something of the form

\begin{equation}
\int d\tau d\tau' dx \, \partial_{\tau}^2 K(\tau-\tau') \tr \Big( \mathds{1} - g_s(\tau',x) g_s(\tau,x) \Big) \, ,
\end{equation}

\noindent which is of course very irrelevant and can be dropped.

Finally, let us compute the expectation value of $T_3^2$,

\begin{align}
\begin{split}
\ev{T_3^2}_f^c &= k^4 \gamma^2 \int_{\omega_s,\omega'_s,q_s} \int_{\omega,q} \int_{\Omega_s,\Omega_s',l_s} \int_{\Omega,l} \Tilde{K}(\omega+\omega_s) \Tilde{K}(\Omega+\Omega_s) \\ &\hspace{0.5cm}\times \Big\langle\tr \Big( \Tilde{D}_s(\omega_s,\omega'_s,q_s) \Tilde{W}(\omega,q) \Tilde{W}(-\omega-\omega_s-\omega_s',-q-q_s) \Big) \\ &\hspace{1cm} \tr \Big( \Tilde{D}_s(\Omega_s,\Omega'_s,l_s) \Tilde{W}(\Omega,l) \Tilde{W}(-\Omega-\Omega_s-\Omega_s',-l-l_s) \Big) \Big\rangle_f^c \, ,
\end{split}
\end{align}

\noindent where $l$ and $l_s$ are respectively fast and slow momenta. Once again, we start by considering the expectation value

\begin{align}
\begin{split}
&\Big\langle\tr \Big( \Tilde{D}_s(\omega_s,\omega'_s,q_s) \Tilde{W}(\omega,q) \Tilde{W}(-\omega-\omega_s-\omega_s',-q-q_s) \Big) \tr \Big( \Tilde{D}_s(\Omega_s,\Omega'_s,l_s) \Tilde{W}(\Omega,l) \Tilde{W}(-\Omega-\Omega_s-\Omega_s',-l-l_s) \Big) \Big\rangle_f^c  \\ &= \ev{\tilde{\phi}^a(\omega,q) \tilde{\phi}^b(-\omega-\omega_s-\omega_s',-q-q_s) \tilde{\phi}^c(\Omega,l) \tilde{\phi}^d(-\Omega-\Omega_s-\Omega_s',-l-l_s)}_f^c \\ &\hspace{0.5cm} \tr \Big( \tilde{D}_{s}(\omega_s,\omega_s',q_s) T^a T^b \Big) \tr \Big( \tilde{D}_{s}(\Omega_s,\Omega_s',l_s) T^c T^d \Big) \, .
\end{split}
\end{align}

\noindent The calculation of this expectation value is quite similar to the one performed before (see Eqs. (\ref{eq:Expectation_value_WZW2})--(\ref{eq:trace2})). Let us denote the two connected pieces as $W_1$ and $W_2$, where

\begin{align}
\begin{split}
W_1 &= (2\pi)^4 \delta(\omega+\Omega) \delta(q+l) \delta(\omega+\omega_s+\omega_s'+\Omega+\Omega_s+\Omega_s') \delta(q+q_s+l+l_s) \Tilde{G}(\omega,q) \Tilde{G}(\omega+\omega_s+\omega_s',q+q_s) \\ &\hspace{1cm}\times \Bigg[ -\frac{1}{2N} \tr \Big( \Tilde{D}_s(\omega_s,\omega_s',q_s) \Tilde{D}_s(\Omega_s,\Omega_s',l_s) \Big) + \frac{1}{4} \Bigg( 1 + \frac{1}{N^2} \Bigg) \tr \Big( \Tilde{D}_s(\omega_s,\omega_s',q_s) \Big) \tr \Big( \Tilde{D}_s(\Omega_s,\Omega_s',l_s) \Big) \Bigg] \, ,
\end{split}
\end{align}

\noindent and

\begin{align}
\begin{split}
W_2 &= (2\pi)^4 \delta(\omega-\Omega-\Omega_s-\Omega_s') \delta(q-l-l_s) \delta(\Omega-\omega-\omega_s-\omega_s') \delta(l-q-q_s) \Tilde{G}(\omega,q) \Tilde{G}(\omega+\omega_s+\omega_s',q+q_s) \\ &\hspace{1cm}\times \Bigg[ \frac{1}{4} \Bigg( N - \frac{2}{N} \Bigg) \tr \Big( \Tilde{D}_s(\omega_s,\omega_s',q_s) \Tilde{D}_s(\Omega_s,\Omega_s',l_s) \Big) + \frac{1}{4N^2} \tr \Big( \Tilde{D}_s(\omega_s,\omega_s',q_s) \Big) \tr \Big( \Tilde{D}_s(\Omega_s,\Omega_s',l_s) \Big) \Bigg] \, .
\end{split}
\end{align}

\noindent By adding the two Wick contractions and integrating over the $\delta$ functions, we get

\begin{align}
\begin{split} \label{eq:T3toExpand}
\ev{T_3^2}_f^c &= k^4 \gamma^2 \int_{\omega_s,\omega_s',\Omega_s,q_s} \int_{\omega,q} \Tilde{K}(\omega+\omega_s) \Tilde{G}(\omega,q) \Tilde{G}(\omega+\omega_s+\omega_s',q+q_s) \\ &\hspace{0.5cm} \times \Bigg[ \Bigg( -\frac{1}{2N} \Tilde{K}(\omega-\Omega_s) + \frac{1}{4} \Big( N - \frac{2}{N} \Big) \Tilde{K}(\omega+\omega_s+\omega_s'+\Omega_s) \Bigg) \\ &\hspace{1cm} \times \tr \Big( \Tilde{D}_s(\omega_s,\omega_s',q_s) \Tilde{D}_s(\Omega_s,-\omega_s-\omega_s'-\Omega_s,-q_s) \Big) \\ &\hspace{1cm} + \Bigg( \frac{1}{4} \Big( 1 + \frac{1}{N^2} \Big) \Tilde{K}(\omega-\Omega_s) + \frac{1}{4N^2} \Tilde{K}(\omega+\omega_s+\omega_s'+\Omega_s) \Bigg) \\ &\hspace{1cm} \times \tr \Big( \Tilde{D}_s(\omega_s,\omega_s',q_s) \Big) \tr \Big( \Tilde{D}_s(\Omega_s,-\omega_s-\omega_s'-\Omega_s,-q_s) \Big)  \Bigg] \, .
\end{split}
\end{align}

\noindent We now need to expand the kernels as well as the second propagator in powers of the slow modes $\omega_s$, $\omega_s'$, $\Omega_s$ and $q_s$. Instead of expanding directly, which would yield a huge number of terms, let us analyze the various possible slow contributions that can be generated. We will only focus on the term which contains $ \tr \Big( \Tilde{D}_s(\omega_s,\omega_s',q_s) \Tilde{D}_s(\Omega_s,-\omega_s-\omega_s'-\Omega_s,-q_s) \Big)$, since the structure of the other term which is proportional to $\tr \Big( \Tilde{D}_s(\omega_s,\omega_s',q_s) \Big) \tr \Big( \Tilde{D}_s(\Omega_s,-\omega_s-\omega_s'-\Omega_s,-q_s) \Big)$ follows from a very similar analysis (in fact, it turns out that the contribution from this latter term vanishes as discussed below). At leading order in the slow mode expansion, the contribution from Eq. (\ref{eq:T3toExpand}) to the effective action for the slow field $g_s$ is proportional to 

\begin{align}
\begin{split}
&\int_{\omega_s,\omega_s',\Omega_s,q_s} \tr \Big( \Tilde{D}_s(\omega_s,\omega_s',q_s) \Tilde{D}_s(\Omega_s,-\omega_s-\omega_s'-\Omega_s,-q_s) \Big) \\ &= \int d\tau_1 d\tau_2 d\tau_3 d\tau_4 \int dx dy \int_{\omega_s,\omega_s',\Omega_s,q_s} \tr \Big( D_s(\tau_1,\tau_2,x) D_s(\tau_3,\tau_4,y) \Big) \\ &\hspace{0.5cm}\times \e^{-\I \omega_s \tau_1} \e^{-\I \omega_s' \tau_2} \e^{-\I q_s x} \e^{-\I \Omega_s \tau_s} \e^{\I (\omega_s+\omega_s'+\Omega_s) \tau_4} \e^{\I q_s y} \\ &= \int d\tau_1 d\tau_2 d\tau_3 d\tau_4 \int dx dy \tr \Big( D_s(\tau_1,\tau_2,x) D_s(\tau_3,\tau_4,y) \Big) \delta(\tau_4-\tau_1) \delta(\tau_4-\tau_2) \delta(\tau_4-\tau_3) \delta(y-x) \\ &= 0 \, ,
\end{split}
\end{align}

\noindent which vanishes since $D_s(\tau,\tau,x) = 0$. Next, at linear order in slow modes, all the contributions vanish, since these terms will also be linear in fast modes, which will yield an odd fast integrand. Therefore, to get a nonzero contribution, we must go to quadratic order in fast modes. There are various possible combinations. Let us analyze them. First, we could have a term with $\omega_s^2$. Its contribution to the effective action will be proportional to

\begin{align}
\begin{split}
&\int_{\omega_s,\omega_s',\Omega_s,q_s} \omega_s^2 \tr \Big( \Tilde{D}_s(\omega_s,\omega_s',q_s) \Tilde{D}_s(\Omega_s,-\omega_s-\omega_s'-\Omega_s,-q_s) \Big) \\ &= \int d\tau_1 d\tau_2 d\tau_3 d\tau_4 \int dx dy \int_{\omega_s,\omega_s',\Omega_s,q_s} \omega_s^2 \tr \Big( D_s(\tau_1,\tau_2,x) D_s(\tau_3,\tau_4,y) \Big) \\ &\hspace{0.5cm}\times \e^{-\I \omega_s \tau_1} \e^{-\I \omega_s' \tau_2} \e^{-\I q_s x} \e^{-\I \Omega_s \tau_s} \e^{\I (\omega_s+\omega_s'+\Omega_s) \tau_4} \e^{\I q_s y} \\ &= -\int d\tau_1 d\tau_2 d\tau_3 d\tau_4 \int dx dy \int_{\omega_s,\omega_s',\Omega_s,q_s} \tr \Big( D_s(\tau_1,\tau_2,x) D_s(\tau_3,\tau_4,y) \Big) \\ &\hspace{1cm}\times \partial_{\tau_1}^2 \e^{\I\omega_s(\tau_4-\tau_1)} \e^{\I\omega'_s(\tau_4-\tau_2)} \e^{\I\Omega_s(\tau_4-\tau_3)} \e^{\I q_s (y-x)} \\ &= -\int d\tau_1 d\tau_2 d\tau_3 d\tau_4 \int dx dy \tr \Big( \partial_{\tau_1}^2 D_s(\tau_1,\tau_2,x) D_s(\tau_3,\tau_4,y) \Big) \delta(\tau_4-\tau_1) \delta(\tau_4-\tau_2) \delta(\tau_4-\tau_3) \delta(y-x) \\ &= -\int d\tau_1 d\tau_2 d\tau_3 \int dx \tr \Big( \partial_{\tau_1}^2 D_s(\tau_1,\tau_2,x) D_s(\tau_3,\tau_3,x) \Big) \delta(\tau_3-\tau_1) \delta(\tau_3-\tau_2) \\ &= 0 \, ,
\end{split}
\end{align}

\noindent where integration by parts has been used. With an identical calculation, terms with $\omega_s'^2$ will be the same as above, except with $\partial_{\tau_2}^2$ instead of $\partial_{\tau_1}^2$, while terms with $\Omega_s^2$ will contain $\partial_{\tau_3}^2$. Clearly, these terms also vanish for the same reason as above. Therefore, we recognize a pattern here: a term with a $T_3$ vanishes if there is no time derivative that acts on the associated $D_s$. Hence, we see that terms with $\omega_s \omega_s'$ also vanish, since no derivatives will be acting on $D_s(\tau_3,\tau_4,x)$. 

Let us now look at the contribution from terms with $\omega_s \Omega_s$, which will contain $\partial_{\tau_1}$ and $\partial_{\tau_3}$. This will be proportional to

\begin{align} \label{eq:omegasOmegas}
\begin{split}
&\int_{\omega_s,\omega_s',\Omega_s,q_s} \omega_s \Omega_s \tr \Big( \Tilde{D}_s(\omega_s,\omega_s',q_s) \Tilde{D}_s(\Omega_s,-\omega_s-\omega_s'-\Omega_s,-q_s) \Big) \\ &= -\int d\tau_1 d\tau_2 d\tau_3 d\tau_4 \int dx dy \tr \Big( \partial_{\tau_1} D_s(\tau_1,\tau_2,x) \partial_{\tau_3} D_s(\tau_3,\tau_4,y) \Big) \delta(\tau_4-\tau_1) \delta(\tau_4-\tau_2) \delta(\tau_4-\tau_3) \delta(y-x) \\ &= -\int d\tau_1 d\tau_2 d\tau_3 d\tau_4 \int dx \tr \Big( g_s^{-1}(\tau_2,x) \partial_{\tau_1} g_s(\tau_1,x) g_s^{-1}(\tau_4,x) \partial_{\tau_3} g_s(\tau_3,x) \Big) \delta(\tau_4-\tau_1) \delta(\tau_4-\tau_2) \delta(\tau_4-\tau_3) \\ &= - \int d\tau dx \tr \Big( g_s^{-1} \partial_{\tau} g_s g_s^{-1} \partial_{\tau} g_s \Big) \\ &= \int d\tau dx \tr \Big( \partial_{\tau} g_s \partial_{\tau} g_s^{-1} \Big) \, .
\end{split}
\end{align}

\noindent The contribution from terms with $\omega_s' \Omega_s$ is quite similar

\begin{align}
\begin{split}
&\int_{\omega_s,\omega_s',\Omega_s,q_s} \omega_s' \Omega_s \tr \Big( \Tilde{D}_s(\omega_s,\omega_s',q_s) \Tilde{D}_s(\Omega_s,-\omega_s-\omega_s'-\Omega_s,-q_s) \Big) \\ &= -\int d\tau_1 d\tau_2 d\tau_3 d\tau_4 \int dx dy \tr \Big( \partial_{\tau_2} D_s(\tau_1,\tau_2,x) \partial_{\tau_3} D_s(\tau_3,\tau_4,y) \Big) \delta(\tau_4-\tau_1) \delta(\tau_4-\tau_2) \delta(\tau_4-\tau_3) \delta(y-x) \\ &= -\int d\tau_1 d\tau_2 d\tau_3 d\tau_4 \int dx \tr \Big( \partial_{\tau_2} g_s^{-1}(\tau_2,x) g_s(\tau_1,x) g_s^{-1}(\tau_4,x) \partial_{\tau_3} g_s(\tau_3,x) \Big) \delta(\tau_4-\tau_1) \delta(\tau_4-\tau_2) \delta(\tau_4-\tau_3) \\ &= - \int d\tau dx \tr \Big( \partial_{\tau} g_s^{-1} g_s g_s^{-1} \partial_{\tau} g_s \Big) \\ &= -\int d\tau dx \tr \Big( \partial_{\tau} g_s \partial_{\tau} g_s^{-1} \Big) \, .
\end{split}
\end{align}

\noindent All the other possible quadratic terms contain at least a momentum $q_s$. All of these terms will vanish, since $q_s$ will yield a space derivative, which does not prevent the two time coordinates in $D_s$ from being the same.

Therefore, we only need to keep track of the terms with $\omega_s \Omega_s$ and $\omega'_s \Omega_s$ in the expansion of Eq. (\ref{eq:T3toExpand}). However, since these two terms have an opposite sign, any contribution from the combination $(\omega_s+\omega_s')\Omega_s$ vanishes when expanding Eq. (\ref{eq:T3toExpand}). Knowing this, we can set $\omega_s+\omega_s'=q_s=0$ in $ \Tilde{G}(\omega+\omega_s+\omega_s',q+q_s)$ as well as $\omega_s+\omega_s'=0$ in  $\Tilde{K}(\omega+\omega_s+\omega_s'+\Omega_s)$. Moreover, for the remaining non-vanishing contributions, since each $D_s$ becomes $g_s^{-1} \partial_{\tau} g_s$ (up to an integration by parts), we see that the term proportional to $ \tr \Big( \Tilde{D}_s(\omega_s,\omega_s',q_s) \Big) \tr \Big( \Tilde{D}_s(\Omega_s,-\omega_s-\omega_s'-\Omega_s,-q_s) \Big)$ also vanishes, since as argued before, $\tr (g_s^{-1} \partial_{\mu} g_s) = 0$. Therefore, we are left with

\begin{align}
\begin{split}
\ev{T_3^2}_f^c &\approx k^4 \gamma^2 \int_{\omega_s,\omega_s',\Omega_s,q_s} \int_{\omega,q} \Tilde{K}(\omega+\omega_s) \Tilde{G}^2(\omega,q)  \Bigg[ \Bigg( -\frac{1}{2N} \Tilde{K}(\omega-\Omega_s) + \frac{1}{4} \Big( N - \frac{2}{N} \Big) \Tilde{K}(\omega+\Omega_s) \Bigg) \\ &\hspace{7cm} \times \tr \Big( \Tilde{D}_s(\omega_s,\omega_s',q_s) \Tilde{D}_s(\Omega_s,-\omega_s-\omega_s'-\Omega_s,-q_s) \Big)  \Bigg] \, .
\end{split}
\end{align}

\noindent Expanding the kernels yields

\begin{align}
\begin{split}
\Tilde{K}(\omega+\omega_s) \Tilde{K}(\omega-\Omega_s) &\approx - \frac{(2-\delta)^2}{(8\pi)^2} \frac{\omega^2}{|\omega|^{2\delta}} \omega_s \Omega_s + ... \, , \\ \Tilde{K}(\omega+\omega_s) \Tilde{K}(\omega+\Omega_s) &\approx \frac{(2-\delta)^2}{(8\pi)^2} \frac{\omega^2}{|\omega|^{2\delta}} \omega_s \Omega_s + ... \, .
\end{split}
\end{align}

\noindent Hence, by using Eq. (\ref{eq:omegasOmegas}), we finally get

\begin{align}
\begin{split}
\ev{(S_{\text{Int,Dis}}^{(2)})^2}_f^c &= \ev{T_3^2}_f^c + ... = \frac{N (2-\delta)^2}{4(8\pi)^2} k^4 \gamma^2 \int_{\omega,q} \frac{\omega^2}{|\omega|^{2\delta}} \Tilde{G}^2(\omega,q) \int d\tau dx \tr \Big( \partial_{\tau} g_s \partial_{\tau} g_s^{-1} \Big) + ... \, , 
\end{split}
\end{align}

\noindent where the ellipsis denote irrelevant terms.

\vspace{0.5cm}
\noindent\underline{\textbf{Mixed WZW-Dissipation term:}} Finally, we must compute the mixed WZW-dissipation contribution

\begin{align}
\begin{split}
2 \ev{S_{\text{Int,WZW}}^{(2)} S_{\text{Int,Dis}}^{(2)}}_f^c = 2 \ev{S_{\text{Int,WZW}}^{(2)}T_1}_f^c + 2 \ev{S_{\text{Int,WZW}}^{(2)} T_2}_f^c + 2 \ev{S_{\text{Int,WZW}}^{(2)} T_3}_f^c \, ,
\end{split}
\end{align}

\noindent which can be represented by

\begin{figure}[H]
	\centering
	\includegraphics[width=0.9\hsize]{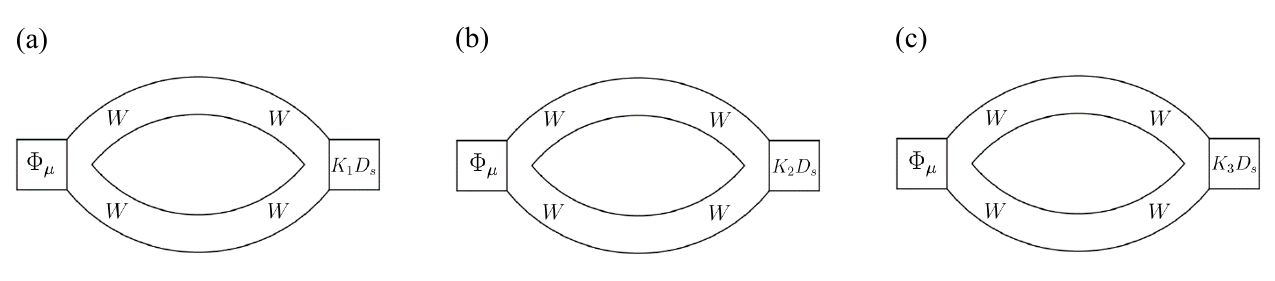}
	\caption{one-loop Feynman diagrams contributing to $\ev{S_{\text{Int,WZW}}^{(2)} S_{\text{Int,Dis}}^{(2)}}_f^c$.}
	\label{fig:diagrams_O(2)_3}
\end{figure}

\noindent Let us focus on the first term

\begin{align}
\begin{split}
2\ev{S_{\text{Int,WZW}}^{(2)}T_1}_f^c &= \I k^2 \gamma \int_{\omega_s,\omega_s',q_s} \int_{\omega,q} \int_{\Omega_s,l_s} \int_{\Omega,l} \Tilde{K}(\omega_s) (2\Omega+\Omega_s,2l+l_s)_{\mu} \\ &\hspace{0.5cm}\times \Big\langle\tr \Big( \Tilde{D}_s(\omega_s,\omega'_s,q_s) \Tilde{W}(\omega,q) \Tilde{W}(-\omega-\omega_s-\omega_s',-q-q_s) \Big) \\ &\hspace{1cm} \tr \Big( \Tilde{\Phi}_{\mu}(\Omega_s,l_s) \Tilde{W}(\Omega,l) \Tilde{W}(-\Omega-\Omega_s,-l-l_s) \Big) \Big\rangle_f^c \, .
\end{split}
\end{align}

\noindent The computation of the expectation value is quite similar to the one in $\ev{T_3^2}_f^c$, involving two connected Wick contractions. After integrating over the $\delta$ functions, we get

\begin{align}
\begin{split}
2\ev{S_{\text{Int,WZW}}^{(2)}T_1}_f^c &= \I \frac{N}{4} k^2 \gamma \int_{\omega_s,\omega_s',q_s} \int_{\omega,q} \Tilde{K}(\omega_s') (2\omega+\omega_s+\omega_s',2q+q_s)_{\mu} \Tilde{G}(\omega,q) \\ &\hspace{0.5cm}\times \Tilde{G}(\omega+\omega_s+\omega_s',q+q_s) \tr \Big( \Tilde{D}_s(\omega_s,\omega_s',q_s) \Tilde{\Phi}_{\mu}(-\omega_s-\omega_s',-q_s) \Big) \, .
\end{split}
\end{align}

\noindent Since there is a slow kernel and a derivative coming from $\Phi_{\mu}$, we can take the leading order term in the slow modes expansion

\begin{align}
\begin{split}
2\ev{S_{\text{Int,WZW}}^{(2)}T_1}_f^c &\approx \I \frac{N}{2} k^2 \gamma \int_{\omega_s,\omega_s',q_s} \int_{\omega,q} \Tilde{K}(\omega_s') (\omega,q)_{\mu}  \Tilde{G}^2(\omega,q) \tr \Big( \Tilde{D}_s(\omega_s,\omega_s',q_s) \Tilde{\Phi}_{\mu}(-\omega_s-\omega_s',-q_s) \Big) = 0 \, .
\end{split}
\end{align}

\noindent The expression vanishes due to the fact that the fast integrand is odd. Clearly, the exact same thing happens with $T_2$. Therefore, let us analyze the third term

\begin{align}
\begin{split}
2 \ev{S_{\text{Int,WZW}}^{(2)} T_3}_f^c &= -2\I k^2 \gamma \int_{\omega_s,\omega_s',q_s} \int_{\omega,q} \int_{\Omega_s,l_s} \int_{\Omega,l} \Tilde{K}(\omega+\omega_s) (2\Omega+\Omega_s,2l+l_s)_{\mu} \\ &\hspace{0.5cm}\times \Big\langle\tr \Big( \Tilde{D}_s(\omega_s,\omega'_s,q_s) \Tilde{W}(\omega,q) \Tilde{W}(-\omega-\omega_s-\omega_s',-q-q_s) \Big) \\ &\hspace{1cm} \tr \Big( \Tilde{\Phi}_{\mu}(\Omega_s,l_s) \Tilde{W}(\Omega,l) \Tilde{W}(-\Omega-\Omega_s,-l-l_s) \Big) \Big\rangle_f^c \, .
\end{split}
\end{align}

\noindent Computing the expectation value and the integrals over the $\delta$ functions yields

\begin{align}
\begin{split}\label{eq:SintT3}
2 \ev{S_{\text{Int,WZW}}^{(2)} T_3}_f^c &= -\I \frac{N}{2} k^2 \gamma \int_{\omega_s,\omega'_s,q_s} \int_{\omega,q} \Tilde{K}(\omega+\omega_s) (2\omega+\omega_s+\omega_s',2q+q_s)_{\mu} \Tilde{G}(\omega,q) \\ &\hspace{1cm} \times \Tilde{G}(\omega+\omega_s+\omega_s',q+q_s) \tr \Big( \Tilde{D}_s(\omega,\omega_s',q_s) \Tilde{\Phi}_{\mu}(-\omega_s-\omega_s',-q_s) \Big) \, .
\end{split}
\end{align}

\noindent This time, we need to expand to linear order in the various slow modes, since there is already a derivative in $\Phi_{\mu}$ (the leading order contribution of course vanishes). As we did before, let us look at the various possibilities one encounters when expanding Eq. (\ref{eq:SintT3}). First, linear terms in $\omega_s$ yield contributions to the effective action for $g_s$ proportional to

\begin{align}
\begin{split}\label{eq:omegas}
&\int_{\omega_s,\omega_s',q_s} \omega_s \tr \Big( \Tilde{D}_s(\omega_s,\omega_s',q_s) \Tilde{\Phi}_{\mu}(-\omega_s-\omega_s',-q_s) \Big) \\ &= \int d\tau_1 d\tau_2 d\tau_3 \int dx dy \int_{\omega_s,\omega_s',q_s} \omega_s \tr \Big( D_s(\tau_1,\tau_2,x) \Phi_{\mu}(\tau_3,y) \Big) \e^{-\I \omega_s \tau_1} \e^{-\I \omega_s' \tau_2} \e^{-\I q_s x} \e^{\I (\omega_s+\omega_s') \tau_3} \e^{\I q_s y} \\ &= \I \int d\tau_1 d\tau_2 d\tau_3 \int dx dy \int_{\omega_s,\omega_s',q_s} \tr \Big( D_s(\tau_1,\tau_2,x) \Phi_{\mu}(\tau_3,y) \Big) \partial_{\tau_1} \e^{\I \omega_s (\tau_3 - \tau_1)} \e^{\I \omega_s'(\tau_3-\tau_2)} \e^{\I q_s(y-x)} \\ &= -\I \int d\tau_1 d\tau_2 d\tau_3 \int dx dy \tr \Big( \partial_{\tau_1} D_s(\tau_1,\tau_2,x) \Phi_{\mu}(\tau_3,x) \Big) \delta(\tau_3-\tau_1) \delta(\tau_3-\tau_2) \delta(x-y) \\ &= \I \int d\tau_1 d\tau_2 d\tau_3 \int dx \tr \Big( g_s^{-1}(\tau_2) \partial_{\tau_1} g_s(\tau_1) \Phi_{\mu}(\tau_3,x) \Big) \delta(\tau_3-\tau_1) \delta(\tau_3-\tau_2) \\ &= -\I \int d\tau dx \tr \Big( \partial_{\tau} g_s^{-1} g_s \Phi_{\mu}(\tau,x) \Big) \, .
\end{split}
\end{align}

\noindent For linear terms in $\omega_s'$, the situation is identical, but with $\partial_{\tau_2}$ instead of $\partial_{\tau_1}$

\begin{align}
\begin{split}
&\int_{\omega_s,\omega_s',q_s} \omega_s' \tr \Big( \Tilde{D}_s(\omega_s,\omega_s',q_s) \Tilde{\Phi}_{\mu}(-\omega_s-\omega_s',-q_s) \Big) \\ &= \I \int d\tau_1 d\tau_2 d\tau_3 \int dx \tr \Big( \partial_{\tau_2} g_s^{-1}(\tau_2) g_s(\tau_1) \Phi_{\mu}(\tau_3,x) \Big) \delta(\tau_3-\tau_1) \delta(\tau_3-\tau_2) \\ &= \I \int d\tau dx \tr \Big( \partial_{\tau} g_s^{-1} g_s \Phi_{\mu}(\tau,x) \Big) \, .
\end{split}
\end{align}

\noindent Finally, it is clear that terms with $q_s$ vanish, since they will be proportional to $\tr \Big( \partial_x D_s(\tau,\tau,x) \Phi_{\mu}(\tau,x) \Big) = 0$. Hence, since the terms with $\omega_s$ and $\omega_s'$ have an opposite sign, Eq. (\ref{eq:SintT3}) becomes

\begin{align}
\begin{split}
2 \ev{S_{\text{Int,WZW}}^{(2)} T_3}_f^c &\approx -\I N k^2 \gamma \int_{\omega_s,\omega'_s,q_s} \int_{\omega,q} \Tilde{K}(\omega+\omega_s) (\omega,q)_{\mu} \Tilde{G}^2(\omega,q) \tr \Big( \Tilde{D}_s(\omega,\omega_s',q_s) \Tilde{\Phi}_{\mu}(-\omega_s-\omega_s',-q_s) \Big) + ... \, .
\end{split}
\end{align}

\noindent The expansion of the kernel at linear order in $\omega_s$ yields

\begin{equation}
\Tilde{K}(\omega+\omega_s) \approx - \frac{(2-\delta)}{8\pi} \frac{\omega}{|\omega|^{\delta}} \omega_s + ... \, ,
\end{equation}

\noindent from which we get, using Eq. (\ref{eq:omegas})

\begin{align}
\begin{split}
2 \ev{S_{\text{Int,WZW}}^{(2)} T_3}_f^c &= \frac{N(2-\delta)}{8\pi} k^2 \gamma \int_{\omega,q} \frac{\omega}{|\omega|^{\delta}}(\omega,q)_{\mu} \Tilde{G}^2(\omega,q) \int d\tau dx \, \tr \Big( \partial_{\tau} g_s^{-1} g_s \Phi_{\mu}(\tau,x) \Big) \\ &= \frac{N(2-\delta)}{8\pi} k^2 \gamma \int_{\omega,q} \frac{\omega^2}{|\omega|^{\delta}} \Tilde{G}^2(\omega,q) \int d\tau dx \, \tr \Big( \partial_{\tau} g_s^{-1} g_s \Phi_{\tau}(\tau,x) \Big) \, ,
\end{split}
\end{align}

\noindent where in the second equality, the fast integral is only nonzero if $\mu = \tau$. Using the expression for $\Phi_{\tau}$, this becomes

\begin{align}
\begin{split}
2 \ev{S_{\text{Int,WZW}}^{(2)} T_3}_f^c &= \frac{N(2-\delta)}{8\pi} \frac{k^2 \gamma}{c^2 \lambda} \int_{\omega,q} \frac{\omega^2}{|\omega|^{\delta}} \Tilde{G}^2(\omega,q) \int d\tau dx \, \tr \Big( \partial_{\tau} g_s \partial_{\tau} g_s^{-1} \Big) \\ &\hspace{0.5cm} - \I \frac{N(2-\delta)}{(8\pi)^2} k^3 \gamma  \int_{\omega,q} \frac{\omega^2}{|\omega|^{\delta}} \Tilde{G}^2(\omega,q) \int d\tau dx \, \tr \Big( \partial_{\tau} g_s \partial_{x} g_s^{-1} \Big) \, .
\end{split}
\end{align}

\noindent Once again, an unphysical term with mixed partial derivatives is generated. It can again be ignored for the rest of the RG calcualtion since it drops out from a symmetrized version of the RG (see the discussion right after Eq. (\ref{eq:SIntWZW2Expectation_value})).

\vspace{0.5cm}
\noindent\underline{\textbf{Recap:}} Combining all the contributions we found above, the expectation value of the squared interaction action is

\begin{align}
\begin{split}
\ev{(S^{(2)}_{\text{Int}}[g_s,W])^2}_f^c &= \frac{N}{c^4 \lambda^2} \Bigg( I_2 - \frac{k^2 c^4\lambda^2}{(8\pi)^2} I_3 \Bigg) \int d\tau dx \tr \Big( \partial_{\tau} g_s \partial_{\tau} g_s^{-1} \Big) \\ &\hspace{0.5cm}+ \frac{N}{\lambda^2} \Bigg( I_3 - \frac{k^2 \lambda^2}{(8\pi)^2} I_2 \Bigg) \int d\tau dx \tr \Big( \partial_x g_s \partial_x g_s^{-1} \Big) \\ &\hspace{0.5cm}+ \frac{N (2-\delta)^2}{4(8\pi)^2} k^4 \gamma^2 \int_{\omega,q} \frac{\omega^2}{|\omega|^{2\delta}} \Tilde{G}^2(\omega,q) \int d\tau dx \tr \Big( \partial_{\tau} g_s \partial_{\tau} g_s^{-1} \Big) \\ &\hspace{0.5cm}+ \frac{N(2-\delta)}{8\pi} \frac{k^2 \gamma}{c^2 \lambda} \int_{\omega,q} \frac{\omega^2}{|\omega|^{\delta}} \Tilde{G}^2(\omega,q) \int d\tau dx \, \tr \Big( \partial_{\tau} g_s \partial_{\tau} g_s^{-1} \Big) + ... \, ,
\end{split}
\end{align}

\noindent where the ellipsis denote the unphysical terms with mixed partial derivatives (which are neglected, as justified above).

\subsubsection{Higher order terms in the cumulant expansion} \label{sec:integration_fast_higher_terms_A}

Higher order terms in the cumulant expansion, that is, expectation values of higher powers of the interaction action, will yield other one-loop contributions. However, only irrelevant terms with more derivatives and kernels will be generated, and we can then stop at quadratic order in the interaction action.

\subsubsection{Effective action full expression} \label{sec:integration_fast_effective_action_A}

Therefore, by collecting all potentially relevant terms that have been computed above, the effective action is thus

\begin{align}
\begin{split}
S_{\text{Eff}}[g_s] &= \frac{1}{\lambda} \int d\tau dx \, \tr \Bigg( \frac{1}{c^2} \partial_{\tau} g_s \partial_{\tau} g_s^{-1} + \partial_x g_s \partial_x g_s^{-1} \Bigg) \\ &\hspace{0.5cm} + \frac{\I k}{12 \pi} \int_{B^3} \tr \Big( g_s^{-1} dg_s \wedge g_s^{-1} dg_s \wedge g_s^{-1} dg_s \Big) \\ &\hspace{0.5cm}+ k^2 \gamma \int d\tau d\tau' dx \, K(\tau-\tau') \, \tr \Big(\mathds{1} - g_s(\tau,x) g_s^{-1}(\tau',x)\Big) \\ &\hspace{0.5cm}- k^2 \gamma \, C_F \, I_1 \, \int d\tau d\tau' dx \, K(\tau-\tau') \tr \Big(\mathds{1} - g_s(\tau,x) g_s^{ -1}(\tau',x)\Big) \\ &\hspace{0.5cm}+ \frac{k^2 \gamma C_F}{16\pi} (2-\delta) (1-\delta) \, \int_{\omega,q} \frac{\Tilde{G}(\omega,q)}{|\omega|^{\delta}} \int d\tau dx \tr \Big(\partial_{\tau} g_s \partial_{\tau} g_s^{-1}\Big) \\ &\hspace{0.5cm} - \frac{N}{2 c^4 \lambda^2} \Bigg( I_2 - \frac{k^2 c^4 \lambda^2}{(8\pi)^2} I_3 \Bigg) \int d\tau dx \tr \Big(\partial_{\tau} g_s \partial_{\tau} g_s^{-1}\Big) \\ &\hspace{0.5cm} - \frac{N}{2\lambda^2} \Bigg( I_3 - \frac{k^2 \lambda^2}{(8\pi)^2} I_2 \Bigg) \int d\tau dx \tr \Big(\partial_{x} g_s \partial_{x} g_s^{-1}\Big) \\ &\hspace{0.5cm} - N \frac{(2-\delta)^2}{8(8\pi)^2} k^4 \gamma^2 \, \int_{\omega,q} \frac{\omega^2}{|\omega|^{2\delta}} \Tilde{G}^2(\omega,q) \int d\tau dx \tr \Big(\partial_{\tau} g_s \partial_{\tau} g_s^{-1}\Big) \\ &\hspace{0.5cm} - \frac{N(2-\delta)}{16\pi} \frac{k^2 \gamma}{c^2 \lambda} \int_{\omega,q} \frac{\omega^2}{|\omega|^{\delta}} \Tilde{G}^2(\omega,q) \int d\tau dx \tr\Big(\partial_{\tau} g_s \partial_{\tau} g_s^{-1}\Big)
\end{split}
\end{align}


\subsection{$\beta$ functions calculation} \label{sec:beta_functions_A}

Having obtained the effective action, we are now in a position to compute the $\beta$ functions. There will be three of these, from the three terms that are getting renormalized: $\partial_{\tau} g \partial_{\tau}g^{-1}$, $\partial_{x} g \partial_{x}g^{-1}$ and the dissipation $K(\tau-\tau') \tr \big( \mathds{1} - g(\tau,x) g^{-1}(\tau',x) \big)$. Note that the WZ term does not get renormalized, as expected, since its coefficient $k$ is quantized to be an integer.

The $\beta$ functions are obtained by rescaling space and time according to

\begin{equation}
x \rightarrow b x = \e^{dl} x \, , \qquad \tau \rightarrow b^z \tau = \e^{z dl} \tau \, ,
\end{equation}

\noindent where $b = \e^{dl}$, with $dl$ an infinitesimal positive quantity and $z$ is the dynamical critical exponent. As we will see eventually, all the terms containing fast integrals (obtained from the one-loop analysis) will be proportional to $dl$, so we only need to rescale terms coming from $S[g_s]$ in the effective action. From a simple power-counting, the following rescaling factors are deduced for the three $\beta$ functions:

\vspace{0.3cm}

\noindent Spatial derivatives term:

\begin{equation}
S_{\text{Grad, spatial}} \sim \int d\tau dx \partial_x^2 \implies \text{Factor of} \hspace{0.5cm} b^{z-1} \approx 1 + (z-1) dl \, ,
\end{equation}

\noindent Time derivatives term:

\begin{equation}
S_{\text{Grad, time}} \sim \int d\tau dx \partial_{\tau}^2 \implies \text{Factor of} \hspace{0.5cm} b^{1-z} \approx 1 + (1-z) dl \, ,
\end{equation}

\noindent Dissipation term:

\begin{equation}
S_{\text{Dis}} \sim \int d\tau d\tau' \int dx \frac{1}{|\tau-\tau'|^{3-\delta}} \implies \text{Factor of} \hspace{0.5cm} b^{1+(\delta-1)z} \approx 1 + [1+(\delta-1)z] dl \, .
\end{equation}

\noindent Therefore, after applying the rescaling, comparing the effective action with the initial action yields the following renormalized couplings:

\begin{align}
\begin{split}
\frac{1}{\lambda_R} = \frac{1}{\lambda} + \frac{z-1}{\lambda} dl - \frac{N}{2\lambda^2} \Bigg( I_3 - \frac{k^2 \lambda^2}{(8\pi)^2} I_2 \Bigg) \, ,
\end{split}
\end{align}

\begin{align}\label{eq:c2l}
\begin{split}
\frac{1}{(c^2 \lambda)_R} &= \frac{1}{c^2 \lambda} + \frac{1-z}{c^2 \lambda} dl + \frac{k^2 \gamma C_F}{16\pi} (2-\delta) (1-\delta) \, \int_{\omega,q} \frac{\Tilde{G}(\omega,q)}{|\omega|^{\delta}} - \frac{N}{2 c^4 \lambda^2} \Bigg( I_2 - \frac{k^2 c^4 \lambda^2}{(8\pi)^2} I_3 \Bigg) \\ &\hspace{1cm} - N \frac{(2-\delta)^2}{8(8\pi)^2} k^4 \gamma^2 \, \int_{\omega,q} \frac{\omega^2}{|\omega|^{2\delta}} \Tilde{G}^2(\omega,q) - \frac{N(2-\delta)}{16\pi} \frac{k^2 \gamma}{c^2 \lambda} \int_{\omega,q} \frac{\omega^2}{|\omega|^{\delta}} \Tilde{G}^2(\omega,q) \, ,
\end{split}
\end{align}

\begin{align}
\begin{split}
k^2 \gamma_R = k^2 \gamma + k^2 [1+(\delta-1)z] \gamma dl - k^2 \gamma \, C_F \, I_1 \, .
\end{split}
\end{align}

\noindent The $\beta$ function of a a given coupling $g$ is then defined to be $\beta(g) = \frac{g_R-g}{dl} = \frac{dg}{dl}$. Therefore, the next step is to evaluate the fast integrals over a frequency/momentum-shell. 

However, before doing that, the above expressions can be greatly simplified in the context of the $1/k$ expansion. Indeed, at large-$k$, all the fixed points should be located at values of $\lambda$ and $\gamma$ of order $1/k$, which is why we introduced the $\mathcal{O}(k^0)$ couplings $\Tilde{\lambda}$ and $\Tilde{\gamma}$. Since the Gaussian fixed point ($\lambda=\gamma=0$) is relativistic, it has $z=1$. Therefore, all the non-trivial fixed points should have $z = 1 + \frac{\Tilde{z}}{k}$. Knowing this, we see that all the terms on the right-hand-side of the above three equations, except the first one in each case, are all of order $k^0$ (recalling that $\Tilde{G}(\omega,q) \sim \lambda \sim 1/k$). Hence, since $\delta = \frac{\Tilde{\delta}}{k}$, we can set $\delta = 0$ in all the three fast integrals $I_1$, $I_2$ and $I_3$ as well as in all the prefactors appearing in Eq. (\ref{eq:c2l}). Keeping $\delta$ would simply add corrections of higher power in $1/k$ to the $\beta$ functions. In this case, the second equation reduces to

\begin{align}\label{eq:c2l_simplified}
\begin{split}
\frac{1}{(c^2 \lambda)_R} = \frac{1}{c^2 \lambda} + \frac{1-z}{c^2 \lambda} dl + \frac{k^2 \gamma C_F}{8\pi} \, I_1 - \frac{N}{2 c^4 \lambda^2} \Bigg( I_2 - \frac{k^2 c^4 \lambda^2}{(8\pi)^2} I_3 \Bigg) - \frac{N}{2(8\pi)^2} k^4 \gamma^2 \, I_2 - \frac{N}{8\pi} \frac{k^2 \gamma}{c^2 \lambda} I_2 \, .
\end{split}
\end{align}

\subsubsection{Evaluation of the fast integrals} \label{sec:beta_functions_fast_integrals_A}

Let us now evaluate the fast integrals. We have for $I_1$

\begin{equation}
I_1 = \int_{\omega,q} \Tilde{G}(\omega,q) = \int \frac{d\omega dq}{(2\pi)^2} \frac{\lambda}{q^2 + \frac{\omega^2}{c^2} + \frac{k^2}{8\pi} \lambda \gamma \omega^2} + \mathcal{O}(\delta) \, .
\end{equation}

\noindent We now rescale $\omega \rightarrow c \, \omega$, which means that the integral becomes

\begin{equation}
I_1 = c \lambda \int \frac{d\omega dq}{(2\pi)^2} \frac{1}{q^2 + \omega^2 + \frac{k^2}{8\pi} c^2 \lambda \gamma \omega^2} \, .
\end{equation}

\noindent This integral is performed using polar coordinates $(\omega,q) = p (\cos\theta,\sin\theta)$ over the shell $b^{-1} = \e^{-dl} < p < 1$. Hence

\begin{align}
\begin{split}
I_1 &= \frac{c \lambda}{4\pi^2} \int_0^{2\pi} d\theta \int_{\e^{-dl}}^1 dp \frac{p}{p^2 + \frac{k^2}{8\pi} c^2 \lambda \gamma p^2 \cos^2\theta} \\ &= \frac{c \lambda}{4\pi^2} dl \int_0^{2\pi} d\theta \frac{1}{1+\frac{k^2}{8\pi} c^2 \lambda \gamma \cos^2\theta} \\ &= \frac{c \lambda}{2\pi} \frac{dl}{\sqrt{1+\frac{k^2}{8\pi} c^2 \lambda \gamma}} \\ &= \frac{c\lambda}{2\pi} w(c^2\lambda \gamma) dl \, ,
\end{split}
\end{align}

\noindent where we have introduced the quantity $w(c^2\lambda \gamma)=\frac{1}{\sqrt{1+\frac{k^2}{8\pi} c^2 \lambda \gamma}}$. The two other integrals are computed using the same method

\begin{align}
\begin{split}
I_2 &= \lambda^2 \int \frac{d\omega dq}{(2\pi)^2} \frac{\omega^2}{\big( q^2 + \frac{\omega^2}{c^2} + \frac{k^2}{8\pi} c^2 \lambda \gamma \omega^2 \big)^2} + \mathcal{O}(\delta) \\ &= \frac{c^3 \lambda^2}{4\pi^2} \int_0^{2\pi} d\theta \int_{\e^{-dl}}^1 dp \frac{p^3 \cos^2\theta}{\big( p^2 + \frac{k^2}{8\pi} c^2 \lambda \gamma p^2 \cos^2\theta \big)^2} \\ &= \frac{c^3 \lambda^2}{4\pi} \frac{dl}{\big( 1 + \frac{k^2}{8\pi} c^2 \lambda \gamma \big)^{3/2}} \\ &= \frac{c^3 \lambda^2}{4\pi} w^3(c^2\lambda\gamma) dl \, ,
\end{split}
\end{align}

\begin{align}
\begin{split}
I_3 &= \lambda^2 \int \frac{d\omega dq}{(2\pi)^2} \frac{q^2}{\big( q^2 + \frac{\omega^2}{c^2} + \frac{k^2}{8\pi} c^2 \lambda \gamma \omega^2 \big)^2} + \mathcal{O}(\delta) \\ &= \frac{c \lambda^2}{4\pi^2} \int_0^{2\pi} d\theta \int_{\e^{-dl}}^1 dp \frac{p^3 \sin^2\theta}{\big( p^2 + \frac{k^2}{8\pi} c^2 \lambda \gamma p^2 \cos^2\theta \big)^2} \\ &= \frac{c \lambda^2}{4\pi} \frac{dl}{\sqrt{ 1 + \frac{k^2}{8\pi} c^2 \lambda \gamma}} \\ &= \frac{c \lambda^2}{4\pi} w(c^2\lambda\gamma) dl \, .
\end{split}
\end{align}

\subsubsection{$\beta$ functions} \label{sec:beta_functions_beta_functions_A}

Using the results of the previous section, we thus get

\begin{align}
\begin{split}
\beta\Big( \frac{1}{\lambda} \Big) = \frac{z-1}{\lambda} - \frac{N c}{8\pi} \Bigg( w - \frac{k^2 c^2 \lambda^2}{(8\pi)^2} w^3 \Bigg) \, ,
\end{split}
\end{align}

\begin{align}
\begin{split}
\beta\Big( \frac{1}{c^2 \lambda} \Big) = \frac{1-z}{c^2 \lambda} + \frac{C_F}{16\pi^2} k^2 c \lambda \gamma w - \frac{N}{8\pi c} \Bigg( w^3 - \frac{k^2 c^2 \lambda^2}{(8\pi)^2} w \Bigg) - \frac{N}{(8\pi)^3} k^4 c^3 \lambda^2 \gamma^2 w^3 - \frac{N}{32\pi^2} k^2 c \lambda \gamma w^3 \, ,
\end{split}
\end{align}

\begin{equation}
\beta(\gamma) = [1+(\delta-1)z]\gamma - \frac{C_F}{2\pi} c \lambda \gamma w \, .
\end{equation}

\noindent Using the chain rule, one can write $\beta(\lambda) = -\lambda^2 \beta\Big(\frac{1}{k}\Big)$ and $\beta(c) = - \frac{c^3 \lambda}{2} \beta\Big( \frac{1}{c^2 \lambda} \Big) + \frac{c \lambda}{2} \beta\Big( \frac{1}{\lambda} \Big)$. By using the $\mathcal{O}(k^0)$ variables introduced previously, we finally get the three $\beta$ functions in their final form

\begin{align}
\begin{split}
\beta(\Tilde{\lambda}) = \frac{1}{k} \Bigg[ -\Tilde{z} \Tilde{\lambda} + \frac{N c \Tilde{\lambda}^2}{8\pi} \Bigg( w - \frac{c^2 \Tilde{\lambda}^2}{(8\pi)^2} w^3 \Bigg) \Bigg] + \mathcal{O}(1/k^2) \, ,
\end{split}
\end{align}

\begin{align}
\begin{split}
\beta(c) = \frac{1}{k} \Bigg[ \Tilde{z} c - \frac{N c^2 \Tilde{\lambda}}{16 \pi} \Bigg( 1 + \frac{c^2 \Tilde{\lambda}^2}{(8\pi)^2} \Bigg) (w - w^3) - \frac{C_F}{32\pi^2} c^4 \Tilde{\lambda}^2 \Tilde{\gamma} w + \frac{N}{2(8\pi)^3} c^6 \Tilde{\lambda}^3 \Tilde{\gamma}^2 w^3 + \frac{N}{(8\pi)^2} c^4 \Tilde{\lambda}^2 \Tilde{\gamma} w^3 \Bigg] + \mathcal{O}(1/k^2) \, ,
\end{split}
\end{align}

\begin{align}
\begin{split}
\beta(\Tilde{\gamma}) = \frac{1}{k} \Bigg[ (\Tilde{\delta}-\Tilde{z}) \Tilde{\gamma} - \frac{C_F}{2\pi} c \Tilde{\lambda} \Tilde{\gamma} w \Bigg] + \mathcal{O}(1/k^2) \, .
\end{split}
\end{align}

Apparently, we have four unknowns to solve for, namely, the fixed point(s) values of $ \Tilde{\lambda}, \Tilde{\gamma}, c$ and $\Tilde{z}$ and only three equations. However, the fixed point value of the velocity $c$ is not a universal characteristic of a fixed point, and in fact, each fixed point should be thought of as a line of fixed points labeled by a different value of the velocity $c$. This is similar to the renormalization group in other systems, e.g., see Refs.\cite{Gamba1999,Lee2007}. The fact that universal exponents do not depend on $c$ can be seen by introducing the variables $x = c \Tilde{\lambda}$ and $y = c \Tilde{\gamma}$. Their respective $\beta$ function is then $\beta(x) = c \beta(\Tilde{\lambda}) + \Tilde{\lambda} \beta(c)$, $\beta(y) = c \beta(\Tilde{\gamma}) + \Tilde{\gamma} \beta(c)$

\begin{align}
\begin{split}
\beta(x) &= \frac{1}{k} \Bigg[ \frac{N x^2}{16\pi} \Bigg( 1 - \frac{x^2}{(8\pi)^2} \Bigg) \Big( w(x y) + w^3(x y) \Big) - \frac{C_F}{32\pi^2} x^3 y w(xy) \\ &\hspace{1cm}+ \frac{N}{2(8\pi)^3} x^4 y^2 w^3(xy) + \frac{N}{(8\pi)^2} x^3 y w^3(x y) \Bigg] + \mathcal{O}(1/k^2) \, ,
\end{split}
\end{align}

\begin{align}
\begin{split}
\beta(y) &= \frac{1}{k} \Bigg[ \Tilde{\delta} y - \frac{C_F}{2\pi} x y w(xy) - \frac{N}{16 \pi} x y \Bigg( 1 + \frac{x^2}{(8\pi)^2} \Bigg) \Big( w(xy) - w^3(xy) \Big) \\ &\hspace{1cm} - \frac{C_F}{32\pi^2} x^2 y^2 w(xy) + \frac{N}{2(8\pi)^3} x^3 y^3 w^3(xy) + \frac{N}{(8\pi)^2} x^2 y^2 w^3(xy) \Bigg] + \mathcal{O}(1/k^2) \, ,
\end{split}
\end{align}

\noindent while $\beta(c)$ is unchanged. $\beta(x)$ and $\beta(y)$ are now independent of $c$ and $\Tilde{z}$ and can thus be plotted in the $x-y$ plane to locate the fixed points.


\subsection{Fixed point analysis} \label{sec:fp_analysis_A}

\subsubsection{Solving for fixed points} \label{sec:fp_analysis_solving_fp_A}

Let us now find the fixed points of the RG flow equations. Consider first the relativistic case, where $\Tilde{\gamma} = 0$. Since the theory is relativistic, $\Tilde{z}=0$. For this case, it is more illuminating to work with the three $\beta$ functions $\beta(\Tilde{\lambda})$, $\beta(c)$ and $\beta(\Tilde{\gamma})$. We need to solve $\beta(\Tilde{\lambda}) = \beta(c) = \beta(\Tilde{\gamma}) = 0$.  The last two $\beta$ functions vanish, while the condition from the first $\beta$ function becomes

\begin{equation}
0 = \frac{N c \Tilde{\lambda}^2}{8\pi} \Bigg( 1-\frac{c^2\Tilde{\lambda}^2}{(8\pi)^2} \Bigg) \, .
\end{equation}

\noindent There are thus two relativistic fixed points, the first one being the trivial Gaussian fixed point in $\Tilde{\lambda} = 0$. There is also a non-trivial fixed point in $\Tilde{\lambda} = \frac{8\pi}{c}$. This is in fact a line of fixed points, as argued previously. This is nothing less than the WZW fixed point, which can easily be seen by setting $c=1$.

We now move on to the case of nonrelativistic fixed points, for which $\Tilde{\gamma} > 0$ and $\Tilde{z}\neq 0$. Note that in this case, there is no fixed point for $\Tilde{\lambda}=0$. Therefore, we get the expression for $\Tilde{z}$ from $\beta(\Tilde{\lambda}) = 0$

\begin{align}\label{eq:zt}
\begin{split}
\Tilde{z} = \frac{N c \Tilde{\lambda}}{8\pi} \Bigg( w(c^2 \Tilde{\lambda} \Tilde{\gamma}) - \frac{c^2 \Tilde{\lambda}^2}{(8\pi)^2} w^3(c^2 \Tilde{\lambda} \Tilde{\gamma}) \Bigg) = \frac{N x}{8\pi} \Bigg( w(x y) - \frac{x^2}{(8\pi)^2} w^3(x y) \Bigg) \, .
\end{split}
\end{align}

\noindent By replacing the expression for $\Tilde{z}$ in $\beta(\Tilde{\gamma}) = 0$, we get the cubic equation presented in the main text, namely,

\begin{align}
\begin{split} \label{eq:Cubicu}
0 = \Tilde{\delta} - \frac{N x}{8\pi} \Bigg( w(x y) - \frac{x^2}{(8\pi)^2} w^3(x y) \Bigg) - \frac{C_F}{2\pi} x w(x y) = \Tilde{\delta} - (4C_F + N) u(x,y) + N u^3(x,y) \, ,
\end{split}
\end{align}

\noindent where we have introduced $u(x,y) = \frac{x}{8\pi} w(xy) = \frac{x}{8\pi} \Big( 1 + \frac{1}{8\pi} xy \Big)^{-1/2}$. By setting $\beta(c)=0$ and using the expression for $\Tilde{z}$, we get the following second equation

\begin{align}
\begin{split}
0 = \frac{N}{16 \pi} \Bigg( 1 - \frac{x^2}{(8\pi)^2} \Bigg)\Big( w(x y) + w^3(x y) \Big) - \frac{C_F}{32\pi^2} x y w(x y) + \frac{N}{2(8\pi)^3} x^2 y^2 w^3(x y) + \frac{N}{(8\pi)^2} x y w^3(x y) \, .
\end{split}
\end{align}

\noindent Hence, fixed points are solutions of the above two equations. This system of equations does not have a compact solution, and therefore, we obtain the positions of the fixed point(s) numerically (in principle, one may obtain analytical expressions for the fixed-point values of $x$ and $y$, but they are very long and not particularly illuminating).

Let us now focus our attention on the second equation. By writing $x y = 8\pi \Big( 
\frac{1}{w^2} - 1 \Big)$, $x^2 = (8\pi)^2 \frac{u^2}{w^2}$, $u(x,y)$ can be expressed solely in terms of $w(x y)$

\begin{equation}
u(x,y) = \sqrt{1 - \frac{4C_F}{N} \frac{1-w^2(x y)}{1+w^2(x y)}} \, .
\end{equation}

\noindent Since $x,y \geq 0$, $w(xy)$ respects $0 \leq w(x y) \leq 1$. From the above expression, we then see that $u(x,y)$ also respects $0 \leq u(x,y) \leq 1$, which puts constraints on the three solutions of Eq. (\ref{eq:Cubicu}). Again, the closed-form expressions are not very illuminating and therefore we don't write them down explicitly. Nevertheless, one can easily see that one solution is always negative and is thus unphysical. The two other solutions are always non-negative, as we can see from Figure \ref{fig:cubicplot} in the main text, and correspond to the two possible dissipative fixed points. However, there are three different regimes, depending on the value of $\Tilde{\delta}$: (i) For $0 <\Tilde{\delta} < 4C_F$, one of the solutions has $u(x,y)>1$, and is therefore unphysical since the associated fixed point has $y<0$. This corresponds to the regime with only a dissipative critical point. (ii) When $\Tilde{\delta} > \Tilde{\delta}_{\text{Max}} = \frac{2}{3\sqrt{3}} \sqrt{ \frac{(4C_F+N)^3}{N}}$, the two solutions of the cubic equations are complex and there are thus no dissipative fixed points. $ \Tilde{\delta}_{\text{Max}}$ is the value of $\Tilde{\delta}$ where the discriminant of the cubic equation vanishes and where the fixed point annihilation occurs. (iii) Finally, for $4C_F < \Tilde{\delta} < \Tilde{\delta}_{\text{Max}}$, the two solutions of the cubic equation are physical, which corresponds to the regime with two dissipative fixed points: the unstable dissiaptive critical point and a new stable dissipative phase.

\subsubsection{Adding a magnetic field} \label{sec:fp_analysis_magnetic_field_A}

Our goal is now to compute universal quantities at the aforementioned fixed points. To obtain the scaling dimension of the primary field $g$, a ``magnetic field'' term is added to the action 

\begin{equation}
S_h[g] = h \int d\tau dx \, \tr \Big( g + g^{-1} \Big) \, ,
\end{equation}

\noindent which breaks the SU$(N)_L$ $\otimes$ SU$(N)_R$ symmetry down to it's diagonal SU$(N)$ subgroup. Splitting slow and fast modes and expanding to quadratic order in $W$, it is easy to see that

\begin{equation}
S_h[g] = S_h[g_s] + S_{\text{Int,}h}^{(2)}[g_s,W] = h \int d\tau dx \tr \Big( g_s+g_s^{-1} \Big) + \frac{h}{2} \int d\tau dx \tr \Big((g_s + g_s^{-1}) W^2\Big) \, .
\end{equation}

\noindent Writing the interaction action in Fourier space yields

\begin{equation}
S_{\text{Int,}h}^{(2)}[g_s,W] = \frac{h}{2} \int_{p_s} \int_p \tr \Big( \Tilde{B}_s(p_s) \Tilde{W}(p) \Tilde{W}(-p-p_s) \Big) \, ,
\end{equation}

\noindent where $B_s(\tau,x) = g_s + g_s^{-1}$. Let us then find the renormalization equation for $h$. At one-loop, we have

\begin{equation}
S_{h\text{,Eff}}[g_s] = S_h[g_s] + \ev{S_{\text{Int,}h}^{(2)}[g_s,W]}_f + ... \, .
\end{equation}

\noindent The computation of the expectation value is straightforward

\begin{align}
\begin{split}
\ev{S_{\text{Int,}h}^{(2)}[g_s,W]}_f &= - \frac{h}{2} \int_{p_S} \int_p \tr \Big( \Tilde{B}_s(p_s) T^a T^b \Big) \ev{\Tilde{\phi}^a(p) \Tilde{\phi}^b(-p-p_s)}_f \\ &= -\frac{h}{2} \int_p \Tilde{G}(p) \tr \Big( \Tilde{B}_s(0) T^a T^a \Big) \\ &= -\frac{h}{2} C_F \, I_1 \int d\tau dx \tr \Big( g_s + g_s^{-1} \Big) \, .
\end{split}
\end{align}

\noindent After rescaling by $b^{z+1} \approx 1 + (1+z) dl$ and using the expression for $I_1$ derived before, we get the following $\beta$ function for $h$

\begin{align}\label{eq:betah}
\begin{split}
\beta(h) = \Big( 2 + \frac{\Tilde{z}}{k} \Big) h - \frac{C_F}{4\pi k} c \Tilde{\lambda} h \, w(c^2 \Tilde{\lambda} \Tilde{\gamma}) + \mathcal{O}(1/k^2) = \Big( 2 + \frac{\Tilde{z}}{k} \Big) h - \frac{C_F}{4\pi k} x h \, w(x y) + \mathcal{O}(1/k^2)
\end{split}
\end{align}

\subsubsection{Dynamical critical exponent and scaling dimensions} \label{sec:fp_analysis_scaling_dim_A}

We are now in a position to compute universal quantities at the different fixed points. We will focus on the dynamical critical exponent $z$, the scaling dimension of $g$, $\Delta_g$ and the scaling dimension of the energy density operator $\epsilon = \tr \Big( \frac{1}{c^2} \partial_{\tau}g \partial_{\tau}g^{-1} + \partial_{x}g \partial_{x}g^{-1} \Big)$, $\Delta_{\epsilon}$.

First, the dynamical critical exponent $z = 1 + \frac{\Tilde{z}}{k}$ is obtained directly using Eq. (\ref{eq:zt}), evaluated at the various fixed points. Next, to compute $\Delta_g$, we need the eigenvalue $e_h$, which is computed using $\beta(h)$. By replacing the expression for $\tilde{z}$ in Eq. (\ref{eq:betah}), we get

\begin{equation}
e_h = 2 + \frac{1}{k} \Bigg[ \frac{N x}{8\pi} \Bigg( w(xy) - \frac{x^2}{(8\pi)^2} w^3(xy) \Bigg) - \frac{C_F x}{4\pi} w(xy) \Bigg] + \mathcal{O}(1/k^2) \, ,
\end{equation}

\noindent which needs to be evaluated at the various fixed points. The scaling dimension $\Delta_g$ is then given by $\Delta_g = 1 + z - e_h = \frac{1}{k} (\Tilde{z} - \Tilde{e}_h)$, where we have defined $e_h = 2 + \frac{\Tilde{e}_h}{k}$. Finally, the calcualtion of $\Delta_{\epsilon}$ requires the diagonalization of the following $2\times2$ matrix

\begin{equation}
M_{xy} = \begin{pmatrix} \partial_x \beta(x) & \partial_y \beta(x) \\ 
\partial_x \beta(y) & \partial_y \beta(y) \end{pmatrix}\Big|_{(x,y)=(x^*,y^*)} \, .
\end{equation}

\noindent In general, this matrix does not have vanishing entries, which means that the energy density operator $\epsilon$ (associated with coupling $x$) and the dissipation operator (associated with coupling $y$) mix among themselves. Therefore, the energy density operator is a linear combination of the two scaling operators $\mathcal{O}_+$ and $\mathcal{O}_-$ (eigenvectors of the above matrix), which have an associated eigenvalue $e_+$ and $e_-$ respectively, where $e_+ > e_-$. Following \cite{Cardy1996}, the scaling dimension of the energy density operator is then given by $\Delta_{\epsilon} = 1 + z - e_+ = 2 + \frac{1}{k} (\Tilde{z} - \Tilde{e}_+)$ whith $e_+ = \frac{\Tilde{e}_+}{k}$.

Let us compute these quantities at the various fixed points. We start with the trivial Gaussian fixed point, which has $x^* = y^* = 0$. Since it is relativistic, $z = 1$ ($\Tilde{z} = 0$). For the scaling dimensions, we get $\Delta_g = 0$ and $\Delta_{\epsilon} = 2$. We now move to the WZW fixed point, located at $x=8\pi$, $y=0$. It is also a relativistic fixed point, thus $z=1$. The scaling dimensions are $\Delta_g = \frac{2C_F}{k} = \frac{N^2-1}{N k}$ and $\Delta_{\epsilon} = 2 + \frac{2N}{k}$. These two results of course agree with the large-$k$ expansion of the exact expressions, $\Delta_g = \frac{N^2-1}{N(N+k)}$ and $\Delta_{\epsilon} = \frac{4N+2k}{N+k}$, as they should \cite{Witten1984}. Moreover, note that for these two relativistic fixed points, the energy density operator is a scaling operator.

Finally, we must proceed numerically for the two dissipative fixed points since their position cannot be easily obtained analytically. Fig. \ref{fig:scaling_dim_nr} in the main text depicts critical exponents accurate to $\mathcal{O}(1/k)$ at these two fixed points. As already mentioned, due to operator mixing, the biggest of the two eigenvalues must be selected to compute $\Delta_{\epsilon}$. The limit $\Tilde{\delta} \rightarrow 4C_F$ (when the stable dissipative fixed point approaches the WZW fixed point) is interesting since $\Delta_{\epsilon}$ at the stable fixed point seemingly approaches $2$, accurate to $\mathcal{O}(1/k)$. This may seem contradictory with the fact that $\Delta_{\epsilon} = 2 + \frac{2N}{k}$ at the WZW fixed point. The resolution of this is as follows: as $\Tilde{\delta} \rightarrow 4C_F$, the overlap between the energy density operator $\epsilon$ (associated with coupling $x$), and the scaling operator with the dominant eigenvalue (i.e. $\mathcal{O}_{+}$ in our notation) approaches zero, and exactly at $\delta = 4 C_F$, $\epsilon = \mathcal{O}_-$. Therefore, only at $\Tilde{\delta} = 4C_F$, $\Delta_{\epsilon} = 1 + z - e_- = 2 + \frac{2N}{k}$, which agrees with the expression for the scaling dimension of the energy operator at the WZW fixed point.


\subsection{Relation between $\eta$ and $z$} \label{sec:relation_eta_z_A}

One can derive the relation between $\Tilde{z}$ and $\Delta_g$ presented in the main text using $\beta(\Tilde{\gamma})$ and $\beta(h)$. Indeed, by setting $\beta(\Tilde{\gamma}) = 0$, we get

\begin{equation} \label{eq:betagt0}
0 = \Tilde{\delta} - \Tilde{z} - \frac{C_F}{2\pi} c \Tilde{\lambda} w(c^{2} \Tilde{\lambda} \Tilde{\gamma}) \, .
\end{equation}

\noindent Moreover, as illustrated in the previous section, $\beta(h)$ allows to compute the eigenvalue $e_h$, which is itself related with the scaling dimension of $g$

\begin{equation}
\Delta_g = 1+z-e_h = 2 + \frac{\Tilde{z}}{k} - 2 - \Bigg[ \frac{\Tilde{z}}{k} - \frac{C_F}{4\pi k} c \Tilde{\lambda} w(c^{2} \Tilde{\lambda} \Tilde{\gamma}) \Bigg] = \frac{C_F}{4\pi k} c \Tilde{\lambda} w(c^{2} \Tilde{\lambda} \Tilde{\gamma}) \, .
\end{equation}

\noindent By isolating $w$ and replacing in Eq. (\ref{eq:betagt0}), we arrive at the desired expression

\begin{equation}\label{eq:Relation_eta_z_1overk}
\Tilde{z} = \Tilde{\delta} - 2 k \Delta_g \, .
\end{equation}

\noindent This relation only holds at $\mathcal{O}(1/k)$. An exact expression valid to all orders can be argued for by demanding the dissipation term to be scale-invariant. By applying the rescaling $x \rightarrow b x$, $\tau \rightarrow b^{z} \tau$, the following condition must be satisfied

\begin{equation} \label{eq:CondScaleInvDiss}
0 = 1 + z (\delta-1) - 2 \Delta_g \, .
\end{equation}

\noindent Using the fact that $e_h = 1 + z - \Delta_g$ and $\eta = 1+z+2-2e_h$, where $\eta$ is the anomalous dimension of $g$, one arrives at

\begin{equation}\label{eq:Relation_eta_z}
z = \frac{2-\eta}{2-\delta} \, .
\end{equation}

\noindent Expanding Eq. (\ref{eq:CondScaleInvDiss}) (or Eq. (\ref{eq:Relation_eta_z})) to leading order $1/k$ yields Eq. (\ref{eq:Relation_eta_z_1overk}).

\section{RG analysis of the relativistic theory} \label{sec:appendixB}

This Appendix details the RG analysis for the relativistic theory. The calculation is very similar to the nonrelativistic case, so only the main differences and key points are discussed.


\subsection{Expanding in slow and fast modes} \label{sec:expand_B}

The expansion in slow and fast modes proceeds exactly as in the nonrelativistic case. Once again, the resulting action is grouped into three terms: $S[g_s g_f] = S[g_s] + S^{(2)}[W] + S^{(2)}_{\text{Int}}[g_s,W]$. The first term is the initial action evaluated at $g = g_s$

\begin{align}
\begin{split}
S[g_s] &= S_{\text{Grad}}[g_s] + S_{\text{WZ}}[g_s] + S_{\text{Dis}}[g_s] \\ &= \frac{1}{\lambda} \int d^2\vb*{r} \, \tr \Big( \partial_{\mu} g_s \partial_{\mu} g_s^{-1} \Big) + \frac{\I k}{12 \pi} \int_{B^3} \tr \Big( g_s^{-1} dg_s \wedge g_s^{-1} dg_s \wedge g_s^{-1} dg_s \Big) \\ &\hspace{0.5cm}+ k^2 \gamma \int d^2\vb*{r} d^2\vb*{r}' \, K(\vb*{r}-\vb*{r}') \, \tr \Big(\mathds{1} - g_s(\vb*{r}) g_s^{-1}(\vb*{r}')\Big) \, ,
\end{split}
\end{align}

\noindent The second term is purely quadratic in $W$,

\begin{align}
\begin{split}
S^{(2)}[W] &= S_{\text{Grad}}^{(2)}[W] + S_{\text{Dis}}^{(2)}[W] \\ &= - \frac{1}{\lambda}\int d^2\vb*{r} \, \tr \Big( \partial_{\mu} W \partial_{\mu} W \Big)  -k^2 \gamma \int d^2\vb*{r} d^2\vb*{r}' K(\vb*{r}-\vb*{r}') \tr \Bigg( \frac{W^2}{2} + \frac{W^{\prime \, 2}}{2} - W W' \Bigg) \\ &= \frac{1}{2} \int \frac{d^2\vb*{p}}{(2\pi)^2} \Tilde{\phi}^a(\vb*{p}) \Big( \Pi^{-1}(\vb*{p}) - k^2 \gamma \Tilde{K}(\vb*{p}) \Big) \Tilde{\phi}^a(-\vb*{p}) \\ &= \frac{1}{2} \int \frac{d^2\vb*{p}}{(2\pi)^2} \Tilde{\phi}^a(\vb*{p}) \Tilde{G}^{-1}(\vb*{p}) \Tilde{\phi}^a(-\vb*{p}) \, ,
\end{split}
\end{align}

\noindent where

\begin{equation}
\Pi(\vb*{p}) = \Pi(p) = \frac{\lambda}{p^2} \, , \qquad \Tilde{K}(\vb*{p}) = \Tilde{K}(p) = -\frac{1}{8\pi} p^{2-\delta} \, ,
\end{equation}

\noindent and the fast propagator is then

\begin{equation}
\Tilde{G}(\vb*{p}) = \frac{\lambda}{p^2 + \frac{k^2}{8\pi} \lambda \gamma p^{2-\delta}} \, .
\end{equation}

\noindent Note that the prime notation now stands for $W' = W(\vb*{r}')$, with $\vb*{r}' = (\tau',x')$. Finally, the interaction term is

\begin{align}
\begin{split}
S_{\text{Int}}^{(2)}[g_s,W] &= S_{\text{Int,WZW}}^{(2)}[g_s,W] + S_{\text{Int,Dis}}^{(2)}[g_s,W] \\ &= \int d^2\vb*{r} \, \tr \Big( \Phi_{\mu}(\vb*{r}) [\partial_{\mu} W,W] \Big) + k^2 \gamma \int d^2\vb*{r} d^2\vb*{r}' K(\vb*{r}-\vb*{r}') \, \tr \Bigg[ \Big(\mathds{1}-g_s^{\prime \, -1} g_s\Big) \Bigg( \frac{W^2}{2} + \frac{W^{\prime \, 2}}{2} - W W' \Bigg) \Bigg] \, ,
\end{split}
\end{align}

\noindent where 

\begin{equation}
\Phi_{\mu} = g_s^{-1} \Bigg( \frac{1}{\lambda} \partial_{\mu} - \frac{\I k}{8\pi} \epsilon_{\mu \nu} \partial_{\nu} \Bigg) g_s \, .
\end{equation}


\subsection{Fourier representation of interaction terms} \label{sec:Fourier_rep_B}

The Fourier representation of the two interaction terms is almost identical to that in the nonrelativistic case

\begin{align}
\begin{split}
S_{\text{Int,WZW}}^{(2)}[g_s,W] = \I \int_{\vb*{p}_s} \int_{\vb*{p}} (2p_{\mu} + p_{s\, \mu}) \tr \Big( \Tilde{\Phi}_{\mu}(\vb*{p}_s) \Tilde{W}(\vb*{p}) \Tilde{W}(-\vb*{p}-\vb*{p}_s) \Big) \, ,
\end{split}
\end{align}

\begin{align}
\begin{split}
S_{\text{Int,Dis}}^{(2)}[g_s,W] &= T_1 + T_2 + T_3 \\ &= k^2 \gamma \int_{\vb*{p}} \int_{\vb*{p}_s,\vb*{p}_s'} \tr \Big( \Tilde{D}_s(\vb*{p}_s,\vb*{p}_s') \Tilde{W}(\vb*{p}) \Tilde{W}(-\vb*{p}-\vb*{p}_s - \vb*{p}_s') \Big) \Bigg( \frac{1}{2} \Tilde{K}(p_S') + \frac{1}{2} \Tilde{K}(p_s) - \Tilde{K}(\vb*{p}+\vb*{p}_s) \Bigg) \, ,
\end{split}
\end{align}

\noindent where we have defined $D_s(\vb*{r},\vb*{r}') = \mathds{1} - g_s^{-1}(\vb*{r}') g_s(\vb*{r})$, while $\int_{\vb*{p}}$ is a shorthand for $\int\frac{d^2\vb*{p}}{(2\pi)^2}$.


\subsection{Integration of fast modes} \label{sec:integration_fast_B}

We proceed with the  cumulant expansion as in the nonrelativistic case.

\subsubsection{Order 1 in interaction action} \label{sec:integration_fast_order1_B}

We start with the expectation value of the interaction action. Let us focus first on the dissipative terms. The expectation values of $T_1$ and $T_2$ are essentially the same as before

\begin{align}
\begin{split}
\ev{T_1}_f = \ev{T_2}_f = - \frac{k^2\gamma}{2} C_F \, I_1 \int d^2\vb*{r} d^2\vb*{r}' \, K(\vb*{r}-\vb*{r}') \, \tr \Big( \mathds{1} - g_s^{\prime \, -1} g_s \Big) \, ,
\end{split}
\end{align}

\noindent where $I_1 = \int_{\vb*{p}} \Tilde{G}(p)$. For $T_3$, the main difference is the expansion of the mixed kernel. Expanding to quadratic order in $\vb*{p}_s$, we get

\begin{equation}
\Tilde{K}(\vb*{p}+\vb*{p}_s) \approx -\frac{1}{8\pi} \Bigg[ p^{2-\delta} + \frac{2-\delta}{2} \frac{p_s^2}{p^{\delta}} - \frac{\delta(2-\delta)}{2} \frac{(\vb*{p}\cdot\vb*{p}_s)^2}{p^{2+\delta}} \Bigg] + ... \, ,
\end{equation}

\noindent where the ellipsis denote higher order terms in $\vb*{p}_s$ as well as linear terms, which have a vanishing fast integral. As in the nonrelativistic case, the contribution to $\langle T_3 \rangle_f$ from the leading order term vanishes. Since we still need $\delta \sim 1/k$ to control the expansion, the third term is of higher order in $1/k$ and is thus dropped. In this case, we get

\begin{equation}
\ev{T_3}_f = \frac{(2-\delta)}{16\pi} C_F k^2 \gamma \int_p \frac{\Tilde{G}(p)}{p^{\delta}} \int d^2\vb*{r} \tr \Big( \partial_{\mu} g_s \partial_{\mu} g_s^{-1} \Big) \, .
\end{equation}

\noindent Naturally, we still have $\ev{S_{\text{Int,WZW}}^{(2)}[g_s,W]}_f = 0$. Hence, the expectation value of the interaction action is

\begin{align}
\begin{split}
\ev{S_{\text{Int}}^{(2)}[g_s,W]}_f &\approx -k^2\gamma C_F \, I_1 \int d^2\vb*{r} d^2\vb*{r}' \, K(\vb*{r}-\vb*{r}') \, \tr \Big( \mathds{1} - g_s g_s^{\prime \, -1} \Big) +\frac{k^2\gamma C_F}{8\pi} I_1 \int d^2\vb*{r} \tr \Big( \partial_{\mu} g_s \partial_{\mu} g_s^{-1} \Big) + \mathcal{O}(\delta) \, ,
\end{split}
\end{align}

\noindent where the higher order terms in $\delta$ have been dropped.

\subsubsection{Order 2 in interaction action} \label{sec:integration_fast_order2_B}

We now move to the expectation value of the square of the interaction action. We only need to focus on the same three contributions as in the nonrelativistic case, since all the other terms either vanish or are irrelevant. For the square of the WZW action, we get

\begin{align}
\begin{split}
\ev{(S_{\text{Int,WZW}}^{(2)})^2}_f^c &= -N \int_{\vb*{p}} p_{\mu} p_{\nu} \Tilde{G}^2(p) \, \int d^2\vb*{r} \tr \Big( \Phi_{\mu}(\vb*{r}) \Phi_{\nu}(\vb*{r}) \Big) \\ &= - \frac{N}{2} \int_{\vb*{p}} p^2 \Tilde{G}^2(p) \int d^2\vb*{r} \tr \Big( \Phi_{\mu}(\vb*{r}) \Phi_{\mu}(\vb*{r}) \Big) \\ &= - \frac{N}{2} I_2 \int d^2\vb*{r} \tr \Big( \Phi_{\mu}(\vb*{r}) \Phi_{\mu}(\vb*{r}) \Big) \, ,
\end{split}
\end{align}

\noindent where rotational invariance has been used, while $I_2 = \int_{\vb*{p}} p^2 \Tilde{G}^2(p)$. Using the expression for $\Phi_{\mu}$, the trace yields

\begin{align}
\begin{split}
\tr \Big( \Phi_{\mu} \Phi_{\mu} \Big) &= \tr \Bigg[ \Bigg( \frac{1}{\lambda} g_s^{-1} \partial_{\mu} g_s - \frac{\I k}{8\pi} \epsilon_{\mu \nu} g_s^{-1} \partial_{\nu} g_s \Bigg) \Bigg( \frac{1}{\lambda} g_s^{-1} \partial_{\mu} g_s - \frac{\I k}{8\pi} \epsilon_{\mu \rho} g_s^{-1} \partial_{\rho} g_s \Bigg) \Bigg] \\ &= \tr \Bigg[ \frac{1}{\lambda^2} g_s^{-1} \partial_{\mu} g_s  g_s^{-1} \partial_{\mu} g_s - \frac{\I k}{8\pi k} \epsilon_{\mu \nu} \Bigg( g_s^{-1} \partial_{\mu} g_s g_s^{-1} \partial_{\nu} g_s + g_s^{-1} \partial_{\nu} g_s g_s^{-1} \partial_{\mu} g_s \Bigg) \\ &\hspace{1cm} - \frac{k^2}{(8\pi)^2} \epsilon_{\mu \nu} \epsilon_{\mu \rho} g_s^{-1}\partial_{\nu} g_s g_s^{-1}\partial_{\rho} g_s \Bigg] \\ &= - \frac{1}{\lambda^2}\Bigg( 1 - \frac{k^2 \lambda^2}{(8\pi)^2} \Bigg) \tr \Big( \partial_{\mu} g_s \partial_{\mu} g_s^{-1} \Big) \, .
\end{split}
\end{align}

\noindent Hence

\begin{equation}
\ev{(S_{\text{Int,WZW}}^{(2)})^2}_f^c = \frac{N}{2\lambda^2} \Bigg( 1 - \frac{k^2 \lambda^2}{(8\pi)^2} \Bigg) I_2 \int d^2\vb*{r} \tr \Big( \partial_{\mu} g_s \partial_{\mu} g_s^{-1} \Big) \, .
\end{equation}

We now move on to the expectation value of the square of the dissipation term. As in the nonrelativistic case, only $\ev{T_3^2}_f^c$ contributes. Following the same steps as before, one finds that

\begin{align}
\begin{split} \label{T3toExpand}
\ev{T_3^2}_f^c &\approx k^4 \gamma^2 \int_{\vb*{p}_s,\vb*{p}_s',\vb*{p}_s''} \int_{\vb*{p}} \Tilde{K}(\vb*{p}+\vb*{p}_s) \Tilde{G}^2(p) \\ &\hspace{0.5cm} \times \Bigg[ \Bigg( -\frac{1}{2N} \Tilde{K}(\vb*{p}-\vb*{p}_s'') + \Big( \frac{N}{4} - \frac{1}{2N} \Big) \Tilde{K}(\vb*{p}+\vb*{p}_s'') \Bigg)  \tr \Big( \Tilde{D}_s(\vb*{p}_s,\vb*{p}_s') \Tilde{D}_s(\vb*{p}_s'',-\vb*{p}_s-\vb*{p}_s'-\vb*{p}_s'') \Big)  \Bigg] \, ,
\end{split}
\end{align}

Similarly to the nonrelativistic case, the non-vanishing contributions, when expanding in terms of the slow modes, are proportional to

\begin{align}
\begin{split}
\int_{\vb*{p}_s,\vb*{p}_s',\vb*{p}_s''} p_{s\,\mu} p_{s\,\nu}'' \tr \Big( \Tilde{D}_s(\vb*{p}_s,\vb*{p}_s') \Tilde{D}_s(\vb*{p}_s'',-\vb*{p}_s-\vb*{p}_s'-\vb*{p}_s'') \Big) = \int d^2\vb*{r} \tr \Big( \partial_{\mu} g_s \partial_{\nu} g_s^{-1} \Big) \, ,
\end{split}
\end{align}

\begin{align}
\begin{split}
\int_{\vb*{p}_s,\vb*{p}_s',\vb*{p}_s''} p_{s\,\mu}' p_{s\,\nu}'' \tr \Big( \Tilde{D}_s(\vb*{p}_s,\vb*{p}_s') \Tilde{D}_s(\vb*{p}_s'',-\vb*{p}_s-\vb*{p}_s'-\vb*{p}_s'') \Big) = -\int d^2\vb*{r} \tr \Big( \partial_{\mu} g_s \partial_{\nu} g_s^{-1} \Big) \, .
\end{split}
\end{align}

\noindent From this, we get

\begin{align}
\begin{split}
\ev{T_3^2}_f^c = \frac{N (2-\delta)^2}{4(8\pi)^2} k^4 \gamma^2 \int_{\vb*{p}} \frac{p_{\mu} p_{\nu}}{p^{2\delta}} \Tilde{G}(p) \int d^2\vb*{r} \tr \Big( \partial_{\mu} g_s \partial_{\nu} g_s^{-1} \Big) = \frac{N}{2(8\pi)^2} k^4 \gamma^2 I_2 \int d^2\vb*{r} \tr \Big( \partial_{\mu} g_s \partial_{\mu} g_s^{-1} \Big)+ \mathcal{O}(\delta) \, ,
\end{split}
\end{align}

\noindent where rotational invariance has been used.

Finally, the last contribution comes from the mixed term $2 \ev{S_{\text{Int,WZW}}^{(2)} T_3}_f^c$

\begin{align}
\begin{split}
2 \ev{S_{\text{Int,WZW}}^{(2)} T_3}_f^c = -\I \frac{N}{2} k^2 \gamma \int_{\vb*{p}} \int_{\vb*{p}_s,\vb*{p}_s'} \Tilde{K}(\vb*{p}+\vb*{p}_s) (2p_{\mu} + p_{s\, \mu} + p_{s\,\mu}') \Tilde{G}(p) \Tilde{G}(\vb*{p}+\vb*{p}_s+\vb*{p}_s') \tr \Big( \Tilde{D}_s(\vb*{p}_s,\vb*{p}_s') \Tilde{\Phi}_{\mu}(-\vb*{p}_s-\vb*{p}_s') \Big) \, .
\end{split}
\end{align}

\noindent The nonzero contributions when expanding to linear order in the slow modes are proportional to

\begin{align}
\begin{split}
\int_{\vb*{p}_s,\vb*{p}_s'} p_{s\,\nu} \tr \Big( \Tilde{D}_s(\vb*{p}_s,\vb*{p}_s') \Tilde{\Phi}_{\mu}(-\vb*{p}_s-\vb*{p}_s') \Big) = -\I \int d^2\vb*{r} \tr \Big( \partial_{\nu} g_s^{-1} g_s \Phi_{\mu} \Big) \, ,
\end{split}
\end{align}

\begin{align}
\begin{split}
\int_{\vb*{p}_s,\vb*{p}_s'} p_{s\,\nu}' \tr \Big( \Tilde{D}_s(\vb*{p}_s,\vb*{p}_s') \Tilde{\Phi}_{\mu}(-\vb*{p}_s-\vb*{p}_s') \Big) = \I \int d^2\vb*{r} \tr \Big( \partial_{\nu} g_s^{-1} g_s \Phi_{\mu} \Big) \, .
\end{split}
\end{align}

\noindent Performing the slow mode expansion then yields

\begin{align}
\begin{split}
2 \ev{S_{\text{Int,WZW}}^{(2)} T_3}_f^c = \frac{N(2-\delta)}{8\pi} k^2 \gamma \int_{\vb{p}} \frac{p_{\mu} p_{\nu}}{p^{\delta}} \Tilde{G}^2(p) \int d^2\vb*{r} \tr \Big( \partial_{\nu} g_s^{-1} g_s \Phi_{\mu} \Big) = \frac{N}{8\pi} k^2 \gamma I_2 \int d^2\vb*{r} \tr \Big( \partial_{\mu} g_s^{-1} g_s \Phi_{\mu} \Big) + \mathcal{O}(\delta) \, ,
\end{split}
\end{align}

\noindent where rotational invariance has once again been used in the fast integral. Let us simplify the trace

\begin{align}
\begin{split}
\tr \Big( \partial_{\mu} g_s^{-1} g_s \Phi_{\mu} \Big) &= \tr \Bigg[ \partial_{\mu} g_s^{-1} g_s \Bigg( \frac{1}{\lambda} g_s^{-1} \partial_{\mu} g_s - \frac{\I k}{8\pi} \epsilon_{\mu \nu} g_s^{-1} \partial_{\nu} g_s \Bigg) \Bigg] \\ &= \frac{1}{\lambda} \tr \Big( \partial_{\mu} g_s \partial_{\mu} g_s^{-1} \Big) - \frac{\I k}{8\pi} \epsilon_{\mu \nu} \tr \Big( \partial_{\mu} g_s^{-1} \partial_{\nu} g_s \Big) \\ &= \frac{1}{\lambda} \tr \Big( \partial_{\mu} g_s \partial_{\mu} g_s^{-1} \Big) \, ,
\end{split}
\end{align}

\noindent where the second term vanishes since $\tr \Big( \partial_{\mu} g_s^{-1} \partial_{\nu} g_s \Big) = \tr \Big( \partial_{\nu} g_s^{-1} \partial_{\mu} g_s \Big)$. Hence

\begin{equation}
2 \ev{S_{\text{Int,WZW}}^{(2)} T_3}_f^c = \frac{N}{8\pi} \frac{k^2 \gamma}{\lambda} I_2 \int d^2r \tr \Big( \partial_{\mu} g_s \partial_{\mu} g_s^{-1} \Big) \, .
\end{equation}

Therefore, the expectation value of the square of the interaction action is

\begin{align}
\begin{split}
\ev{(S^{(2)}_{\text{Int}}[g_s,W])^2}_f^c &= \frac{N}{2\lambda^2} \Bigg( 1 - \frac{k^2 \lambda^2}{(8\pi)^2} \Bigg) I_2 \int d^2\vb*{r} \tr \Big( \partial_{\mu} g_s \partial_{\mu} g_s^{-1} \Big) + \frac{N}{2(8\pi)^2} k^4 \gamma^2 I_2 \int d^2\vb*{r} \tr \Big( \partial_{\mu} g_s \partial_{\mu} g_s^{-1} \Big) \\ &\hspace{0.5cm} + \frac{N}{8\pi} \frac{k^2 \gamma}{\lambda} I_2 \int d^2\vb*{r} \tr \Big( \partial_{\mu} g_s \partial_{\mu} g_s^{-1} \Big) \, .
\end{split}
\end{align}

\noindent Note that in the relativistic case, the unphysical terms with mixed partial derivatives are not generated, since these would break Lorentz invariance.

\subsubsection{Effective action full expression} \label{sec:integration_fast_effective_action_B}

The effective action at one-loop is thus

\begin{align}
\begin{split}
S_{\text{Eff}}[g_s] &= \frac{1}{\lambda} \int d^2\vb*{r} \, \tr \Big(  \partial_{\mu} g_s \partial_{\mu} g_s^{-1} \Big) \\ &\hspace{0.5cm} + \frac{\I k}{12 \pi} \int_{B^3} \tr \Big( g_s^{-1} dg_s \wedge g_s^{-1} dg_s \wedge g_s^{-1} dg_s \Big) \\ &\hspace{0.5cm}+ k^2 \gamma \int d^2\vb*{r} d^2\vb*{r}' \, K(\vb*{r}-\vb*{r}') \, \tr \Big(\mathds{1} - g_s(\vb*{r}) g_s^{-1}(\vb*{r}') \Big) \\&\hspace{0.5cm}- C_F k^2\gamma  \, I_1 \int d^2\vb*{r} d^2\vb*{r}' \, K(\vb*{r}-\vb*{r}') \, \tr \Big( \mathds{1} - g_s(\vb*{r}) g_s^{-1}(\vb*{r}') \Big) \\ &\hspace{0.5cm}+\frac{ C_F k^2\gamma}{8\pi} I_1 \int d^2\vb*{r} \tr \Big( \partial_{\mu} g_s \partial_{\mu} g_s^{-1} \Big) \\&\hspace{0.5cm}- \frac{N}{4\lambda^2} \Bigg( 1 - \frac{k^2 \lambda^2}{(8\pi)^2} \Bigg) I_2 \int d^2\vb*{r} \tr \Big( \partial_{\mu} g_s \partial_{\mu} g_s^{-1} \Big) \\ &\hspace{0.5cm}- \frac{N}{4(8\pi)^2} k^4 \gamma^2 I_2 \int d^2\vb*{r} \tr \Big( \partial_{\mu} g_s \partial_{\mu} g_s^{-1} \Big) \\ &\hspace{0.5cm} - \frac{N}{16\pi} \frac{k^2 \gamma}{\lambda} I_2 \int d^2\vb*{r} \tr \Big( \partial_{\mu} g_s \partial_{\mu} g_s^{-1} \Big) \, .
\end{split}
\end{align}

\noindent As in the nonrelativistic case, higher order terms in the cumulant expansion yield irrelevant terms which can be neglected.


\subsection{$\beta$ functions calculation} \label{sec:beta_functions_B}

From the effective action, we see that $\tr \big( \partial_{\mu} g \partial_{\mu} g^{-1} \big)$ and $K(\vb*{r}-\vb*{r}') \tr \big( \mathds{1}-g(\vb*{r})g^{-1}(\vb*{r}') \big)$ will be renormalized. The $\beta$ functions for $\lambda$ and $\gamma$ are obtained by rescaling $\vb*{r} \rightarrow b \vb*{r}$, with $b = \e^{dl}$. Once again, only terms coming from $S[g_s]$ are rescaled, since the fast integrals will be proportional to $dl$. The gradient term is scale-invariant and does not pick up any factor of $b$, while the dissipation term picks up a factor of $b^{\delta} \approx 1 + \delta dl$.

\subsubsection{Fast integrals} \label{sec:beta_functions_fast_integrals_B}

The fast integrals are evaluated over a shell $b^{-1} = \e^{-dl} < p < 1$. For $I_1$, we have

\begin{align}
\begin{split}
I_1 = \int \frac{d^2\vb*{p}}{(2\pi)^2} \frac{\lambda}{p^2 + \frac{k^2}{8\pi} \lambda \gamma p^2} + \mathcal{O}(\delta) = \frac{\lambda}{4\pi^2} \int_0^{2\pi} d\theta \int_{\e^{-dl}}^1 dp \frac{p}{p^2 + \frac{k^2}{8\pi} \lambda \gamma p^2} = \frac{\lambda}{2\pi} \frac{1}{1+\frac{k^2}{8\pi} \lambda \gamma} dl = \frac{\lambda}{2\pi} F(\lambda \gamma) dl \, ,
\end{split}
\end{align}

\noindent where $F(\lambda \gamma) = \frac{1}{1+\frac{k^2}{8\pi} \lambda \gamma}$. On the other hand, $I_2$ yields

\begin{align}
\begin{split}
I_2 = \int \frac{d^2\vb*{p}}{(2\pi)^2} \frac{\lambda^2 p^2}{\big( p^2 + \frac{k^2}{8\pi} \lambda \gamma p^2 \big)^2} + \mathcal{O}(\delta) = \frac{\lambda^2}{4\pi^2} \int_0^{2\pi} d\theta \int_{\e^{-dl}}^1 dp \frac{p^3}{\big( p^2 + \frac{k^2}{8\pi} \lambda \gamma p^2 \big)^2} = \frac{\lambda^2}{2\pi} \frac{1}{\big( 1+\frac{k^2}{8\pi} \lambda \gamma \big)^2} dl = \frac{\lambda^2}{2\pi} F^2(\lambda \gamma) dl \, .
\end{split}
\end{align}

\subsubsection{$\beta$ functions} \label{sec:beta_functions_beta_functions_B}

The $\beta$ functions for the two couplings are  obtained following the same procedure as in the nonrelativistic case.  It will again be useful to introduce $\mathcal{O}(k^0)$ couplings $\Tilde{\lambda} = k \lambda$ and $\Tilde{\gamma} = k \gamma$.  One finds,

\begin{align} \label{eq:betalt_relativistic}
\begin{split}
\beta(\Tilde{\lambda}) = \frac{1}{k} \Bigg[ \frac{N \Tilde{\lambda}^2}{8\pi} \Bigg( 1 - \frac{\Tilde{\lambda}^2}{(8\pi)^2} \Bigg) F^2(\Tilde{\lambda} \Tilde{\gamma}) - \frac{C_F}{16\pi^2} \Tilde{\lambda}^3 \Tilde{\gamma} F(\Tilde{\lambda} \Tilde{\gamma}) + \frac{N}{(8\pi)^3} \Tilde{\lambda}^4 \Tilde{\gamma}^2 F^2(\Tilde{\lambda} \Tilde{\gamma}) + \frac{N}{32\pi^2} \Tilde{\lambda}^3 \Tilde{\gamma} F^2(\Tilde{\lambda} \Tilde{\gamma}) \Bigg] + \mathcal{O}(1/k^2) \, ,
\end{split}
\end{align}

\begin{equation} \label{eq:betagt_relativistic}
\beta(\Tilde{\gamma}) = \frac{1}{k} \Bigg[ \Tilde{\delta} \Tilde{\gamma} - \frac{C_F}{2\pi} \Tilde{\lambda} \Tilde{\gamma} F(\Tilde{\lambda} \Tilde{\gamma}) \Bigg] + \mathcal{O}(1/k^2) \, ,
\end{equation}

\noindent with $F(\Tilde{\lambda} \Tilde{\gamma}) = \frac{1}{1 + \frac{1}{8\pi} \Tilde{\lambda} \Tilde{\gamma}}$.


\subsection{Fixed point analysis} \label{sec:RG_fp_analysis_B}

\subsubsection{Solving for fixed points} \label{sec:RG_fp_analysis_solving_fp_B}

Solving $\beta(\Tilde{\lambda}) = \beta(\Tilde{\gamma}) = 0$, we find three fixed points. There is the trivial Gaussian fixed point at $\Tilde{\lambda} = \Tilde{\gamma} = 0$ and the WZW fixed point at $\Tilde{\lambda} = 8\pi$ and $\Tilde{\gamma} = 0$. The fixed point of our main interest is the dissipative fixed point located at

\begin{align}
\begin{split}
\Tilde{\lambda} = \frac{128 \pi C_F^2  \Tilde{\delta}}{64 C_F^3 - 16 C_F^2 N + N \Tilde{\delta}^2} \, , \qquad \Tilde{\gamma} = \frac{(16C_F^2 - \tilde{\delta}^2)N}{16C_F^2 \Tilde{\delta}} \, .
\end{split}
\end{align}

\noindent Note that this fixed point only exists for $\Tilde{\delta} < 4C_F$. When $\Tilde{\delta} > 4C_F$, the WZW fixed point becomes unstable, similar to the nonrelativistic theory, the main difference being that now it becomes unstable towards a fixed point at $\Tilde{\gamma} \rightarrow \infty$ in contrast to the nonrelativistic case, where it became unstable towards the fixed point corresponding to the stable, dissipative phase (see Fig. \ref{fig:rglow_nonrel}(b) of the main text).

\subsubsection{Scaling dimensions and critical exponents} \label{sec:RG_fp_analysis_scaling_dim_B}

The calculation of $\Delta_g$ is once again done by adding a magnetic field to the action. The resulting $\beta$ function for $h$ is

\begin{equation}
\beta(h) = 2h - \frac{C_F}{4\pi k} \Tilde{\lambda} h F(\Tilde{\lambda} \Tilde{\gamma}) + \mathcal{O}(1/k^2) \, ,
\end{equation}

\noindent from which we get the magnetic field eigenvalue

\begin{equation}
e_h = 2 - \frac{C_F}{4\pi k} \Tilde{\lambda} F(\Tilde{\lambda} \Tilde{\gamma}) \, ,
\end{equation}

\noindent which must be evaluated at the various fixed points. For the scaling dimension of the energy density operator $\epsilon = \tr \big( \partial_{\mu} g \partial_{\mu} g^{-1} \big)$, we must obtain the eigenvalues of the following $2\times2$ matrix

\begin{equation}
M_{\Tilde{\lambda} \Tilde{\gamma}} = \begin{pmatrix} \partial_{\Tilde{\lambda}} \beta(\Tilde{\lambda}) & \partial_{\Tilde{\gamma}} \beta(\Tilde{\lambda}) \\ 
\partial_{\Tilde{\lambda}} \beta(\Tilde{\gamma}) & \partial_{\Tilde{\gamma}} \beta(\Tilde{\gamma}) \end{pmatrix}\Big|_{(\Tilde{\lambda},\Tilde{\gamma})=(\Tilde{\lambda}^*,\Tilde{\gamma}^*)} \, ,
\end{equation}

\noindent At the two relativistic fixed points (Gaussian and WZW), the two scaling dimensions are identical as in the nonrelativistic theory. However, for the dissipative fixed point, we can this time obtain closed-form expressions. For the scaling dimension of $g$, we get

\begin{equation}
\Delta_g = \frac{\Tilde{\delta}}{2k} \, ,
\end{equation}

\noindent while the eigenvalues of the above $2\times2$ matrix are

\begin{equation}
e_{\pm} = \frac{-N \Tilde{\delta}^3 \pm \Tilde{\delta} \sqrt{N(1024C_F^5 - 64 C_F^3 \Tilde{\delta}^2 + N \Tilde{\delta}^4)}}{64C_F^3 k} \, .
\end{equation}

\noindent As in the nonrelativistic case, the eigenvalue contributing to the scaling dimension of the energy density operator is the biggest, that is $e_+$. Therefore, we find

\begin{equation}
\Delta_{\epsilon} = 2  -e_+ = 2 + \frac{\Tilde{\delta}}{64 C_F^3 k} \Bigg[ N \Tilde{\delta}^2 - \sqrt{ N (1024 C_F^5 - 64 C_F^3 \Tilde{\delta}^2 + N \Tilde{\delta}^4)} \Bigg] \, .
\end{equation}

\noindent Once again, as $\Tilde{\delta} \rightarrow 4C_F$, we see that $\Delta_{\epsilon}$ approaches $2 \neq 2 + \frac{2N}{k}$, the value at the WZW fixed point. The reason is identical to the nonrelativistic case:  as $\Tilde{\delta} \rightarrow 4C_F$, the overlap between the energy density operator and the scaling operator with the dominant eigenvalue approaches zero, and therefore, at $\Tilde{\delta} = 4C_F$ the scaling dimension of the energy operator matches with what is expected for the WZW CFT, namely, $2 + \frac{2N}{k}$.


\twocolumngrid	
\bibliographystyle{apsrev4-2}

%


\end{document}